\UseRawInputEncoding
\documentclass[preprint,aps,12pt,showpacs,nofootinbib,tightenlines]{revtex4}
\usepackage{mathrsfs}
\usepackage{amsmath}
\usepackage{amssymb}
\usepackage{epsfig}
\usepackage{graphicx}
\usepackage{subfigure}
\usepackage{booktabs}
\def\be{\begin{eqnarray}}
\def\en{\end{eqnarray}}
\def\non{\nonumber\\}

\def\prd{{Phys. Rev. D}~}
\def\prl{{ Phys. Rev. Lett.}~}
\def\plb{{ Phys. Lett. B}~}

\def\jpg{{ J. Phys. G}~}

\begin{document}
\title{Covariant light-front approach for $B_c$ decays into charmonium: implications on form factors and branching ratios}

\author{Zhi-Qing Zhang
\footnote{Electronic address: zhangzhiqing@haut.edu.cn}, Zhi-Jie Sun, Yan-Chao Zhao, You-Ya Yang, Zi-Yu Zhang} 
\affiliation{\it \small  Institute of Theoretical Physics, School of Sciences, Henan University
of Technology, Zhengzhou450052, Henan, China } 
\date{\today}

\begin{abstract}
In this work, we investigate the form factors of $B_c$ decays into
$J/\Psi, \psi(2S,3S),\eta_c,$ $ \eta_c(2S,3S), \chi_{c0}, \chi_{c1}, h_c$ and $X(3872)$ mesons
in the covariant light-front quark model (CLFQM). For the purpose of the branching ratio
calculation, the form factors of $B_c\to D^{(*)}, D^{(*)}_s$ transitions are also included. In order to obtain the
form factors for the physical transition
processes, we need to extend these form factors from the space-like region to the time-like region. The $q^2$-dependence for each transition
form factor is also plotted. Then, using the factorization method, we calculate
the branching ratios of 80 $B_c$ decay channels with a charmonium involved in each mode. Most of our predictions are comparable to
the results given by other approaches. As to the decays with the radially excited-state S-wave charmonia involved, such as $\psi(2S,3S)$ and $\eta_c(2S,3S)$, two sets of parameters for their light-front wave functions, corresponding to scenario I (SI) and scenario II (SII), are adopted
to calculate the branching ratios. By comparing with the
future experimental data, one can discriminate which parameters are more favored.
\end{abstract}

 \vspace{1cm}

\maketitle
{\centering\section{Introduction}\label{sec£ºintro}}

During the period from the 1970s to 1980s, the light-front quark model (LFQM) was developed \cite{terent,chung} to deal with nonperturbative
physical quantities such as decay constants, transition form factors, and so on. This relativistic quark model is
based on a light-front formalism \cite{dirac} and quantum chromodynamics (QCD) light-front quantization \cite{brodsky}. The LFQM can provide a relativistic
treatment of the hadron momentum and fully treatment of the quark spin by using the so-called Melosh rotation. At the same time,
the light-front wave functions are independent of the hadron momentum and thus are manifestly Lorentz invariant. Equipped with these advantages,
the LFQM becomes a convenient approach and has been employed to calculate decay constants and form factors \cite{jaus90,jaus91,ji92,cheng97,cheng98}.
While under the LFQM, the constituent quark and antiquark in a bound state are required to be on their mass shells, which makes
degrees of freedom of light-front momentum become three and the Lorentz covariance of the matrix elements to be lost. The usual practice is
only taking the plus component ($\mu=+$) of the current matrix elements, which will miss the zero-mode contributions.
However, lacking the zero-mode contributions sometimes affects the calculation
accuracy. Unfortunately, such conventional LFQM approach with defect is powerless to calculate the zero-mode contributions.
At the end of the twentieth century, Jaus put forward the covariant Light-Front quark model (CLFQM) \cite{jaus}.
The previous LFQM is usually called the standard Light-Front quark model (SLFQM) \cite{jaus}. The CLFQM is more
convincing than the SLFQM. In the CLFQM approach \cite{jaus}, when evaluating the light-front matrix element from the momentum loop integral by a light-front
decomposition to the internal momentum and carrying out the integration over the minus component ($p^-=p^0-p^3$) by
means of contour methods, one will encounter additional spurious contribution proportional to the light-like
vector $\omega^\mu=(0,2,0_\bot)$, which violates the covariance. While this spurious contribution is just canceled
by the zero-mode contribution, at the same time the covariance of the current matrix elements is restored, and all the problems can be
resolved.
Since the popular CLFQM was proposed, it has been widely used to study the form factors and the decay constants of the
ground-state S-wave and low-lying P-wave mesons, and is further applied to
phenomenological studies about $B_c$ decays \cite{hycheng,wwang,wwang1,wwang2,Ke13}.
Certainly, there still exist some discussions about the self-consistency of the CLFQM, for example, the decay constant of the
vector meson, which is different as a result of extracting from different polarization (longitudinal and transverse) states \cite{choi14,chang18,chang20,chang21}.

$B_c$ meson decays  have received
extensive attention because of its unique structure in the Standard Model. The $B_c$ meson is the only heavy meson composed of
two heavy quarks with different flavors (b and c), which cannot annihilate into gluons (photon) via strong
(electromagnetic) interaction. Decays of the $B_c$ meson occur only via weak interaction,
which includes three types at the quark level, the $b\to c(u)$, $c\to s(d)$ transitions, and the weak annihlilations.
Although the phase space of the $c$ quark decays is much smaller than
that of the $b$ quark decays, the Cabibbo-Kobayashi-Maskawa (CKM) matrix elements
are greatly in favor of the $c$ quark decays (i.e., $|V_{cs}|\gg |V_{cb}|, |V_{cd}|\gg |V_{ub}|$), which provide about $70\%$ of the $B_c$ decay
width, while the $b$ quark decays and the weak annihilations only amount
to about $20\%$ and $10\%$, respectively \cite{gouz}.  On the experimental side,
since the $B_c$ meson was first discovered by the Collider Detector at Fermilab (CDF) collaboration via
the decay of $B_c\to J/\Psi l\nu$ in 1.8 TeV $p\bar p$ collisions at
the Fermilab Tevatron, many $B_c$ decay channels have been observed by the Large Hadron Collider beauty (LHCb) collaboration, such as
$B_c^+\to J/\Psi \pi^+\pi^-\pi^+$ \cite{aajj1}, $B_c^+\to J/\Psi \pi^+$ \cite{aajj11}, $B_c^+\to J/\Psi K^+$ \cite{aajj2}, $B_c^+\to J/\Psi D_s^{(*)+}$ \cite{aajj3},
$B_c^+\to J/\Psi K^+K^-\pi^+$ \cite{aajj4}, and $B_c^+\to B^0_s \pi^+$ \cite{aajj5}.
The inclusive production cross section of the $B_c$ meson
at the LHC is estimated to be at a level of $1 \mu b$ for $\sqrt{14}$ TeV. This means that around $\mathcal{O}(10^9)$ $B_c$ events with
a luminosity of $1 fb^{-1}$ can be provided \cite{gao10}, which are sufficient for studying the $B_c$ meson decays.
On the theoretical side,
many theoretical methods have been used to study the
$B_c$ meson decays to charmonium states, such as the perturbation QCD (PQCD) approach
\cite{Zhu,Rui:2016opu}, the generalized
factorization (GF) approach \cite{Hsiao:2016pml}, the QCD factorization  (QCDF) approach
\cite{chang182}, the QCD sum rule (QCDSR) approach \cite{V}, the Bethe-Salpter Equation approach \cite{chang}, the relativistic quark
model (RQM) \cite{Ivanov, kang}, and nonrelativistic QCD approach(NRQCD) \cite{Qiao}.

The development of the theoretical and experimental aspects of the $B_c$ meson physics motivates us to investigate the $B_c$ weak decays
with a charmonium involved in each mode. In Sect.2, we recapitulate the CLFQM, including the definitions of the
decay constants and the relevant formulae
of $B_c$ to charmonium or charmed meson transition form factors.
In Sect.3, after determining the shape parameters $\beta'$ using the corresponding decay constants, we provide the numerical results of
 $B_c$ transition form factors and their $q^2$ dependence.
Then, using the transition form factors, we calculate the branching ratios of $B_c$ decays with a charmonium meson involved in each mode.
In addition, detailed data analysis and discussion, including a comparison with the other model calculations,
are carried out. The conclusions are presented in the final part.

\begin{table}
\caption{Feynman rules for the vertices $i\Gamma^\prime_M$ of the incoming meson-quark-antiquark, where $p^\prime_1$ and $p_2$ are the
quark and antiquark momenta, respectively.}
\begin{center}
\begin{tabular}{c|c}
\hline\hline $M\left({ }^{2 S+1} L_{J}\right)$ & $i \Gamma_{M}^{\prime}$ \\
\hline
$\text { pseudoscalar }\left({ }^{1} S_{0}\right) $&$H_{P}^{\prime} \gamma_{5}$\\
$\text { vector }\left({ }^{3} S_{1}\right) $&$ i H_{V}^{\prime}\left[\gamma_{\mu}-\frac{1}{W_{V}^{\prime}}\left(p_{1}^{\prime}-p_{2}\right)_{\mu}\right]$\\
$\text { scalar }\left({ }^{3} P_{0}\right) $&$-i H_{S}^{\prime}$\\
$\text { axial }\left({ }^{3} P_{1}\right) $&$-i H_{3_{A}}^{\prime}\left[\gamma_{\mu}+\frac{1}{W_{3_{A}}^{\prime}}\left(p_{1}^{\prime}-p_{2}\right)_{\mu}\right] \gamma_{5}$\\
$\text { axial }\left({ }^{1} P_{1}\right)  $&$-i H_{1_{A}}^{\prime}\left[\frac{1}{W_{1_{A}}^{\prime}}\left(p_{1}^{\prime}-p_{2}\right)_{\mu}\right] \gamma_{5}$\\
 \hline\hline
\end{tabular}\label{tab0}
\end{center}
\end{table}

{\centering\section{Formalism}\label{form}}
{\centering\subsection{Covariant light-front quark model}}

\begin{figure}[htbp]
\centering \subfigure{
\begin{minipage}{5cm}
\centering
\includegraphics[width=5cm]{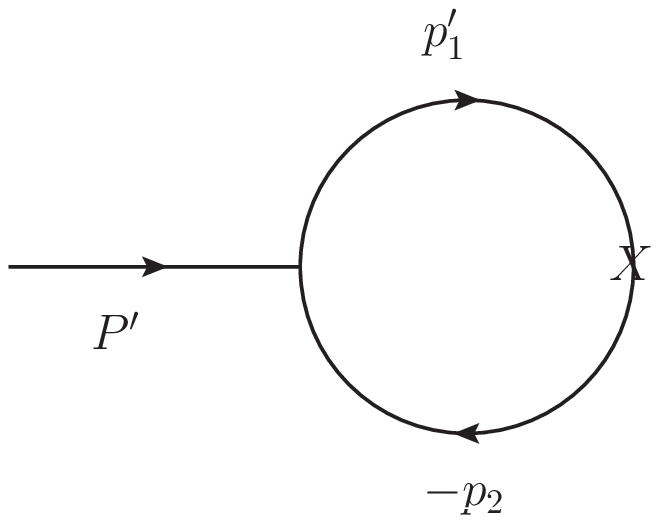}
\end{minipage}}
\subfigure{
\begin{minipage}{6cm}
\centering
\includegraphics[width=6cm]{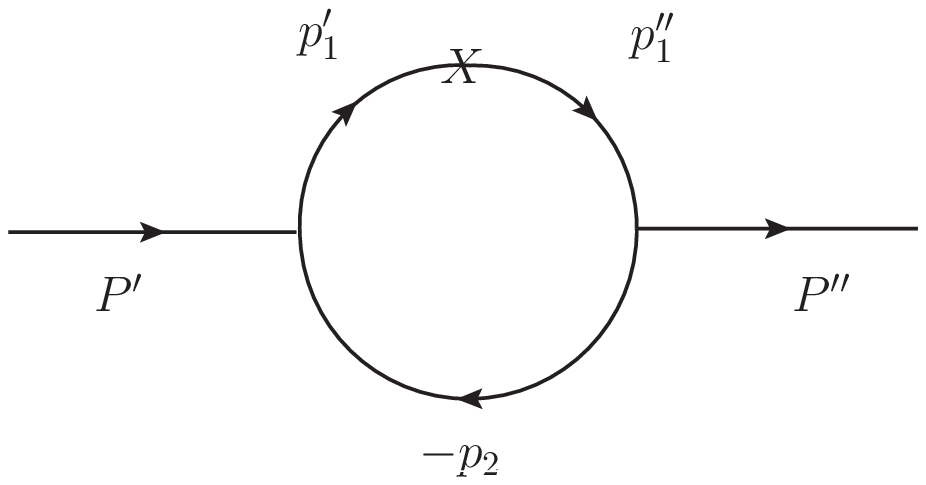}
\end{minipage}}
\caption{Feynman diagrams for $B_c$ decay (left) and transition
(right) amplitudes, where $P^{\prime(\prime\prime)}$ is the
incoming (outgoing) meson momentum, $p^{\prime(\prime\prime)}_1$
is the quark momentum, $p_2$ is the antiquark momentum and X
denotes the vector or axial vector transition vertex.}
\label{feyn}
\end{figure}

Under the covariant light-front quark model, the light-front coordinates of a momentum $p$ are used, $p=(p^-,p^+,p_\perp)$, with
$p^\pm=p^0\pm p_z,$ and $p^2=p^+p^--p^2_\perp$.
The Feynman diagrams for $B_c$ meson decay and transition amplitudes are shown in Fig. \ref{feyn}.
The incoming (outgoing) meson has mass $M^\prime(M^{\prime\prime})$
with the momentum $P^\prime=p_1^\prime+p_2 (P^{\prime\prime}=p_1^{\prime\prime}+p_2)$, where $p_{1}^{\prime(\prime\prime)} $
and $p_{2}$ are the momenta of the quark and antiquark
inside the incoming (outgoing) meson with mass $m_{1}^{\prime(\prime\prime)}$and $m_{2}$, respectively. Here, we use the same notation
as those in Refs. \cite{jaus,hycheng} and $M^\prime=m_{B_c}$ for $B_c$ meson decays.
These momenta can be expressed in terms of the internal variables $(x_{i},p{'}_{\perp})$ as
\be
p_{1,2}^{\prime+}=x_{1,2} P^{\prime+}, \quad p_{1,2 \perp}^{\prime}=x_{1,2} P_{\perp}^{\prime} \pm p_{\perp}^{\prime}
\en
with $x_{1}+x_{2}=1$. Using these internal variables,
we can define some quantities for the incoming meson which will be used in the following calculations:
\be
M_{0}^{\prime 2} &=&\left(e_{1}^{\prime}+e_{2}\right)^{2}=\frac{p_{\perp}^{\prime 2}+m_{1}^{\prime 2}}{x_{1}}
+\frac{p_{\perp}^{2}+m_{2}^{2}}{x_{2}}, \quad \widetilde{M}_{0}^{\prime}=\sqrt{M_{0}^{\prime 2}-\left(m_{1}^{\prime}-m_{2}\right)^{2}}, \non
e_{i}^{(\prime)} &=&\sqrt{m_{i}^{(\prime) 2}+p_{\perp}^{\prime 2}+p_{z}^{\prime 2}}, \quad \quad p_{z}^{\prime}
=\frac{x_{2} M_{0}^{\prime}}{2}-\frac{m_{2}^{2}+p_{\perp}^{\prime 2}}{2 x_{2} M_{0}^{\prime}},\en
where $M'_0$ is the kinetic invariant mass of the incoming meson and can be expressed as the energies of the quark and the antiquark
$e^{(\prime)}_i$. It is similar to the case of the outgoing meson.

To calculate the amplitudes for the transition form factors,
we need the Feynman rules for the meson-quark-antiquark vertices ($i\Gamma'_M(M=P,V,A,S)$), which are listed in Tab. \ref{tab0}.
It is noted that for the outgoing meson, we should use $i(\gamma_0\Gamma^{\prime\dag}_M\gamma_0)$ for the relevant vertices. The
$B_c\to M$
($M$ denotes a pseudoscalar (P), a vector (V), an axial-vector (A) or a scalar (S) meson) form factors induced by vector and aixal-vector currents
are defined as
\be
\left\langle P\left(P^{\prime
\prime}\right)\left|V_{\mu}\right|
B_{c}\left(P^{\prime}\right)\right\rangle&=&f_+(q^2)P_\mu+f_-(q^2)q_\mu,\\
\left\langle V\left(P^{\prime
\prime},\varepsilon\right)\left|V_{\mu}\right|
B_{c}\left(P^{\prime}\right)\right\rangle&=&\epsilon_{\mu\nu\alpha\beta}\varepsilon^{*\nu}P^\alpha q^\beta g(q^2),\label{formv1}\\
\left\langle V\left(P^{\prime
\prime},\varepsilon\right)\left|A_{\mu}\right|
B_{c}\left(P^{\prime}\right)\right\rangle&=&-i\left\{\varepsilon_\mu^*f(q^2)+\varepsilon^*\cdot P\left[P_\mu a_+(q^2)+q_\mu a_-(q^2)\right]\right\},\label{formv2}\\
\left\langle \;^1A\left(P^{\prime
\prime},\varepsilon\right)\left|A_{\mu}\right|
B_{c}\left(P^{\prime}\right)\right\rangle&=-&q^{\;^1A}(q^2)\epsilon_{\mu\nu\alpha\beta}\varepsilon^{*\nu}P^\alpha q^\beta,\\
\left\langle \;^1A\left(P^{\prime
\prime},\varepsilon\right)\left|V_{\mu}\right|
B_{c}\left(P^{\prime}\right)\right\rangle&=&i\left\{l^{\;^1A}(q^2)\varepsilon_\mu^*+\varepsilon^*\cdot P\left[P_\mu c^{\;^1A}_+(q^2)+q_\mu c^{\;^1A}_-(q^2)\right]\right\},\\
\left\langle \;^3A\left(P^{\prime
\prime},\varepsilon\right)\left|A_{\mu}\right|
B_{c}\left(P^{\prime}\right)\right\rangle&=-&q^{\;^3A}(q^2)\epsilon_{\mu\nu\alpha\beta}\varepsilon^{*\nu}P^\alpha q^\beta,\\
\left\langle \;^3A\left(P^{\prime
\prime},\varepsilon\right)\left|V_{\mu}\right|
B_{c}\left(P^{\prime}\right)\right\rangle&=&i\left\{l^{\;^3A}(q^2)\varepsilon_\mu^*+\varepsilon^*\cdot P\left[P_\mu c^{\;^3A}_
+(q^2)+q_\mu c^{\;^3A}_-(q^2)\right]\right\},\\
\left\langle S\left(P^{\prime
\prime}\right)\left|A_{\mu}\right|
B_{c}\left(P^{\prime}\right)\right\rangle&=&i\left[u_+(q^2)P_\mu+u_-(q^2)q_\mu\right].
\en
In calculations, the Bauer-Stech-Wirbel (BSW) \cite{bsw} form factors for the $B_c\to M$ transition are more frequently used and defined by
\be
\left\langle P\left(P^{\prime
\prime}\right)\left|V_{\mu}\right|
B_{c}\left(P^{\prime}\right)\right\rangle&=&\left(P_{\mu}-\frac{m_{B_{c}}^{2}-m_{P}^{2}}{q^{2}}
q_{\mu}\right) F_{1}^{B_{c}
P}\left(q^{2}\right)+\frac{m_{B_{c}}^{2}-m_{P}^{2}}{q^{2}} q_{\mu}
F_{0}^{B_{c} P}\left(q^{2}\right),\;\;\;\;\;\;\\
\left\langle
V\left(P^{\prime \prime}, \varepsilon^{\mu
*}\right)\left|V_{\mu}\right|
B_{c}\left(P^{\prime}\right)\right\rangle&=&-\frac{1}{m_{B_{c}}+m_{V}}
\epsilon_{\mu \nu \alpha \beta} \varepsilon^{ * \nu}
P^{\alpha} q^{\beta} V^{B_{c} V}\left(q^{2}\right),\\
\left\langle V\left(P^{\prime \prime}, \varepsilon^{\mu *}\right)\left|A_{\mu}\right| B_{c}\left(P^{\prime}\right)\right\rangle&=& i\left\{\left(m_{B_{c}}
+m_{V}\right) \varepsilon_{\mu}^{ *} A_{1}^{B_{c} V}\left(q^{2}\right)-\frac{\varepsilon^{ *}
\cdot P}{m_{B_{c}}+m_{V}} P_{\mu} A_{2}^{B_{c} V}\left(q^{2}\right)\right.\non &&
\left.-2 m_{V} \frac{\varepsilon^{ *} \cdot
P}{q^{2}} q_{\mu}\left[A_{3}^{B_{c}
V}\left(q^{2}\right)-A_{0}^{B_{c}
V}\left(q^{2}\right)\right]\right\},\label{bctov2}\\
\left\langle A\left(P^{\prime \prime}, \varepsilon^{\mu *}\right)\left|V_{\mu}\right| B_{c}\left(P^{\prime}\right)\right\rangle
&=&-i\left\{\left(m_{B_{c}}-m_{A}\right) \varepsilon_{\mu}^{ *} V_{1}^{B_{c} A}\left(q^{2}\right)
-\frac{\varepsilon^{ *} \cdot P}{m_{B_{c}}-m_{A}} P_{\mu} V_{2}^{B_{c} A}\left(q^{2}\right)\right.\non &&
\left.-2 m_{A} \frac{\varepsilon^{ *} \cdot
P}{q^{2}} q_{\mu}\left[V_{3}^{B_{c}
A}\left(q^{2}\right)-V_{0}^{B_{c}
A}\left(q^{2}\right)\right]\right\},\label{bctoa1}\\
\left\langle A\left(P^{\prime \prime}, \varepsilon^{\mu*}\right)\left|A_{\mu}\right| B_{c}\left(P^{\prime}\right)\right\rangle
&=&-\frac{1}{m_{B_{c}}-m_{A}} \epsilon_{\mu \nu \alpha \beta} \varepsilon^{ * \nu} P^{\alpha} q^{\beta} A^{B_{c} A}\left(q^{2}\right),
\en
\be
\left\langle S\left(P^{\prime
\prime}\right)\left|A_{\mu}\right|
B_{c}\left(P^{\prime}\right)\right\rangle&=&\left(P_{\mu}-\frac{m_{B_{c}}^{2}-m_{S}^{2}}{q^{2}}
q_{\mu}\right) F_{1}^{B_{c}
S}\left(q^{2}\right)+\frac{m_{B_{c}}^{2}-m_{S}^{2}}{q^{2}} q_{\mu}
F_{0}^{B_{c} S}\left(q^{2}\right),\;\;\;\;\;\;\;\;
\en
where $P=P'+P'', q=P'-P''$, and the convention $\epsilon_{0123}=1$ is adopted.

To smear the singularity at $q^2=0$ in Eq.(\ref{bctov2}) and (\ref{bctoa1}), the relations $V^{B_cA}_3(0)=V^{B_cA}_0(0),A^{B_cV}_3(0)=A^{B_cV}_0(0)$ are required, and
\be
V^{B_cA}_3(q^2)&=&\frac{m_{B_c}-m_A}{2m_A}V^{B_cA}_1(q^2)-\frac{m_{B_c}+m_A}{2m_A}V^{B_cA}_2(q^2),\\
A^{B_cV}_3(q^2)&=&\frac{m_{B_c}+m_V}{2m_V}A^{B_cV}_1(q^2)-\frac{m_{B_c}-m_V}{2m_V}A^{B_cV}_2(q^2).
\en
These two kinds of form factors are related to each other via
\be
F^{B_cP}_1(q^2)&=&f_+(q^2),F^{B_cP}_0(q^2)=f_+(q^2)+\frac{q^2}{q\cdot P}f_-(q^2),\label{relationp}\\
V^{B_cV}(q^2)&=&-(m_{B_c}+m_V)g(q^2), A^{B_cV}_1(q^2)=-\frac{f(q^2)}{m_{B_c}+m_V}, \label{relationv1}\\
A^{B_cV}_2(q^2)&=&(m_{B_c}+m_V)a_+(q^2),A^{B_cV}_3(q^2)-A^{B_cV}_0(q^2)=\frac{q^2}{2m_V}a_-(q^2),\label{relationv2}\\
A^{B_cA}(q^2)&=&-(m_{B_c}-m_A)q(q^2), V^{B_cA}_1(q^2)=-\frac{l(q^2)}{m_{B_c}-m_A},\label{relationa1}\\
V^{B_cA}_2(q^2)&=&(m_{B_c}-m_A)c_+(q^2),V^{B_cA}_3(q^2)-V^{B_cA}_0(q^2)=\frac{q^2}{2m_A}c_-(q^2),\label{relationa2}\\
F^{B_cS}_1(q^2)&=&-u_+(q^2),F^{B_cS}_0(q^2)=-u_+(q^2)-\frac{q^2}{q\cdot P}u_-(q^2).\label{relations}
\en

{\centering\subsection{Wave functions and decay constants}\label{decaycons}}
In order to calculate the form factors, we need to specify the light-front wave functions. In principle, one can obtain them by solving the
relativistic Schr$\ddot{o}$dinger equation. But it is difficult to obtain the exact solution in many cases. Therefore, phenomenological wave functions are usually employed to describe the hadronic structure.
In the present work, we shall use the phenomenological Gaussian-type wave functions
\be
\varphi^{\prime} &=&\varphi^{\prime}\left(x_{2}, p_{\perp}^{\prime}\right)=4\left(\frac{\pi}{\beta^{\prime 2}}\right)^{\frac{3}{4}}
\sqrt{\frac{d p_{z}^{\prime}}{d x_{2}}} \exp \left(-\frac{p_{z}^{\prime 2}+p_{\perp}^{\prime 2}}{2 \beta^{\prime 2}}\right),\non
\varphi_{p}^{\prime} &=&\varphi_{p}^{\prime}\left(x_{2}, p_{\perp}^{\prime}\right)=\sqrt{\frac{2}{\beta^{\prime 2}}} \varphi^{\prime},
\quad \frac{d p_{z}^{\prime}}{d x_{2}}=\frac{e_{1}^{\prime} e_{2}}{x_{1} x_{2} M_{0}^{\prime}},\label{betap}
\en
where the parameter $\beta'$ describes the momentum distribution and is approximately of order $\Lambda_{QCD}$.
It is usually determined by the decay constants through the analytic expressions in the conventional light-front approach, which
are given as follows \cite{jaus,hycheng}:
\be
f_{P}&=&\frac{N_{c}}{16 \pi^{3}} \int d x_{2} d^{2} p_{\perp}^{\prime} \frac{h^\prime_P}{x_{1} x_{2} (M^{\prime2}-M^{\prime2}_0)}
4\left(m_{1}^{\prime} x_{2}+m_{2} x_{1}\right),\\
f_{V}&=&\frac{N_{c}}{4 \pi^{3} M^{\prime}} \int d x_{2} d^{2} p_{\perp}^{\prime} \frac{h_{V}^{\prime}}{x_{1} x_{2}\left(M^{\prime 2}-M_{0}^{\prime 2}\right)}\non &&
\times\left[x_{1} M_{0}^{\prime 2}-m_{1}^{\prime}\left(m_{1}^{\prime}-m_{2}\right)-p_{\perp}^{\prime 2}+\frac{m_{1}^{\prime}+m_{2}}{w_{V}^{\prime}} p_{\perp}^{\prime 2}\right],\\
f_{\;^3A}&=&-\frac{N_{c}}{4 \pi^{3} M^{\prime}}  \int d x_{2} d^{2} p_{\perp}^{\prime} \frac{h_{\;^3A}^{\prime}}{x_{1} x_{2}\left(M^{\prime 2}-M_{0}^{\prime 2}\right)}\non &&
\times\left[x_{1} M_{0}^{\prime 2}-m_{1}^{\prime}\left(m_{1}^{\prime}+m_{2}\right)-p_{\perp}^{\prime 2}-\frac{m_{1}^{\prime}-m_{2}}{w_{\;^3A}^{\prime}} p_{\perp}^{\prime 2}\right],\\
f_{\;^1A}&=&\frac{N_{c}}{4 \pi^{3} M^{\prime}} \int d x_{2} d^{2} p_{\perp}^{\prime} \frac{h_{\;^3A}^{\prime}}{x_{1} x_{2}\left(M^{\prime 2}
-M_{0}^{\prime 2}\right)}\left(\frac{m_{1}^{\prime}-m_{2}}{w_{\;^1A}^{\prime}} p_{\perp}^{\prime 2}\right),\\
f_{S}&=&\frac{N_{c}}{16 \pi^{3}} \int d x_{2} d^{2} p_{\perp}^{\prime} \frac{h_{S}^{\prime}}{x_{1} x_{2}\left(M^{\prime 2}-M_{0}^{\prime 2}\right)}
4\left(m_{1}^{\prime} x_{2}-m_{2} x_{1}\right),
\en
where $m_{1}^{\prime}$ and $m_{2}$ are the constituent quarks of meson $M (M=P,V,\;^{3}A,\;^{1}A,S)$. By the way, a tensor meson
($\;^3P_2$ state)
cannot be produced through $(V\pm A)$ or tensor current, so we should not define its decay constant. The explicit forms of $h'_{M}$  are given by \cite{hycheng}
\be
h_{P}^{\prime} &=&h_{V}^{\prime}=\left(M^{\prime 2}-M_{0}^{\prime 2}\right) \sqrt{\frac{x_{1} x_{2}}{N_{c}}} \frac{1}{\sqrt{2} \widetilde{M}_{0}^{\prime}} \varphi^{\prime},\label{hp}\\
h_{S}^{\prime} &=&\sqrt{\frac{2}{3}} h_{\;^3A}^{\prime}=\left(M^{\prime 2}-M_{0}^{\prime 2}\right) \sqrt{\frac{x_{1} x_{2}}{N_{c}}} \frac{1}{\sqrt{2} \widetilde{M}_{0}^{\prime}}
\frac{\widetilde{M}_{0}^{\prime 2}}{2 \sqrt{3} M_{0}^{\prime}} \varphi_{p}^{\prime},\label{hs3a}\\
h_{\;^1A}^{\prime} &=&\left(M^{2 \prime}-M_{0}^{\prime 2}\right) \sqrt{\frac{x_{1} x_{2}}{N_{c}}} \frac{1}{\sqrt{2} \widetilde{M}_{0}^{\prime}} \varphi_{p}^{\prime}.
\label{h1a}\en

It is easy to see that the decay constants of the scalar meson and $\;^1A$ type of axial meson are zero for $m^\prime_1=m_2$, which satisfy
the $SU(N)$ flavor constrain. The other nontrivial decay constants can be obtained through the experimental results for the purely leptonic decays or the lattice QCD calculations.
The constituent quark masses used in the calculations will be listed in the next section.

{\centering\subsection{Form factors}}
One important difference between the conventional light-front
quark approach and the covariant one lies in the treatment of the constituent quarks.
In the conventional light-front framework, the constituent
quarks are required to be on their mass shells, and the physical quantities, such as decay constant and form factor, can be extracted
from the plus component of the corresponding current matrix elements. However, this framework misses the zero-mode contributions and renders
the matrix elements non-covariant. In order to resolve this problem, the covariant light-front approach was proposed by Jaus \cite{jaus},
which provides a systematical way to deal with the zero-mode contributions by including the so-called Z-diagram contributions. Then physical
quantities can be calculated in terms of Feynman momentum loop integrals in a manifestly covariant way. As a result, the constituent
quarks of the meson will be off-shell.
For the general $B_c\rightarrow P$ transition, the decay amplitude for the lowest order is
\be
\mathcal{B}_{\mu}^{B_c P}=-i^{3} \frac{N_{c}}{(2 \pi)^{4}} \int d^{4} p_{1}^{\prime} \frac{H_{B_c}^{\prime}\left(H_{P}^{\prime\prime}\right)}
{N_{1}^{\prime} N_{1}^{\prime \prime} N_{2}} S_{\mu}^{B_c P},
\en
where $N_{1}^{\prime(\prime \prime)}=p_{1}^{\prime(\prime \prime) 2}-m_{1}^{\prime (\prime\prime) 2}, N_{2}=p_{2}^{2}-m_{2}^{2} $ arise
from the quark propagators, and
the trace $S_{\mu}^{B_cP}$ can be directly obtained by using the Lorentz contraction,
\be
S_{\mu}^{B_c P}&=&\operatorname{Tr}\left[\gamma_{5}\left(\not p_{1}^{\prime \prime}+m_{1}^{\prime \prime}\right) \gamma_{\mu}\left(\not p_{1}^{\prime}
+m_{1}^{\prime}\right) \gamma_{5}\left(-\not p_{2}+m_{2}\right)\right] \non
&=& 2 p_{1 \mu}^{\prime}\left[M^{\prime 2}+M^{\prime \prime 2}-q^{2}-2 N_{2}-\left(m_{1}^{\prime}-m_{2}\right)^{2}-\left(m_{1}^{\prime \prime}
-m_{2}\right)^{2}+\left(m_{1}^{\prime}-m_{1}^{\prime \prime}\right)^{2}\right]\non
&&+q_{\mu}\left[q^{2}-2 M^{\prime 2}+N_{1}^{\prime}-N_{1}^{\prime \prime}+2 N_{2}+2\left(m_{1}^{\prime}-m_{2}\right)^{2}-\left(m_{1}^{\prime}
-m_{1}^{\prime \prime}\right)^{2}\right] \non
&&+P_{\mu}\left[q^{2}-N_{1}^{\prime}-N_{1}^{\prime \prime}-\left(m_{1}^{\prime}-m_{1}^{\prime \prime}\right)^{2}\right].
\label{ptop}
\en
In practice, we use the light-front decomposition of the Feynman loop momentum and integrate out
the minus component through the contour method. If the covariant vertex functions are not singular when performing integration,
the transition amplitudes will
pick up the singularities in the antiquark propagators. The integration then leads to
\be
N_{1}^{\prime(\prime \prime)} &\rightarrow& \hat{N}_{1}^{\prime(\prime \prime)}=x_{1}\left(M^{\prime(\prime \prime 2}-M_{0}^{\prime(\prime \prime) 2}\right),\non
H_{M}^{\prime(\prime\prime)} &\rightarrow& h_{M}^{\prime(\prime \prime)}\non
W_{M}^{\prime \prime} &\rightarrow& w_{M}^{\prime \prime} \non
\int \frac{d^{4} p_{1}^{\prime}}{N_{1}^{\prime} N_{1}^{\prime \prime} N_{2}} H_{B_c}^{\prime} H_{M}^{\prime \prime} S^{B_cM} & \rightarrow&-i \pi \int \frac{d x_{2} d^{2}
p_{\perp}^{\prime}}{x_{2} \hat{N}_{1}^{\prime} \hat{N}_{1}^{\prime \prime}} h_{B_c}^{\prime} h_{M}^{\prime \prime} \hat{S}^{B_cM},
\en
where
\be
M_{0}^{\prime \prime 2}=\frac{p_{\perp}^{\prime \prime 2}+m_{1}^{\prime \prime 2}}{x_{1}}+\frac{p_{\perp}^{\prime \prime 2}+m_{2}^{2}}{x_{2}},
\label{vertex}
\en
with $p''_\perp=p'_\perp-x_2q_\perp$, and $M$ in the subscript and superscript denotes a
pseudoscalar (P), a vector (V), an axial-vector (A) or a scalar (S) meson. The explicit forms of $h^{\prime(\prime\prime)}_{M}$ have been given in Eq.(\ref{hp})-Eq.(\ref{h1a}).
For the $B_c\to V,A$ transitions, the $\omega''_M (M=V,A)$ in the corresponding vertex operators listed in Tab. \ref{tab0} are given as
\be
w_{V}^{\prime\prime} &=&M_{0}^{\prime\prime}+m_{1}^{\prime\prime}+m_{2}, \quad w_{3 A}^{\prime\prime}=\frac{\widetilde{M}_{0}^{\prime\prime 2}}{m_{1}^{\prime\prime}-m_{2}}, \quad w_{1 A}^{\prime\prime}=2,
\label{vertexva}\en
where $\widetilde{M}_{0}^{\prime\prime}=\sqrt{M_{0}^{\prime\prime 2}-\left(m_{1}^{\prime\prime}-m_{2}\right)^{2}}$.

After performing the integration with the contour method, we will be confronted with additional spurious contributions proportional to the
light-like four-vector $\tilde{\omega}=(0,2,\bf{0}_\perp)$. These undesired spurious contributions can be eliminated by inclusion of the
zero-mode contributions, which amount to performing the $p^-$ integration in a proper way. The specific rules under the
$p^-$ integration have been derived in Refs. \cite{jaus,hycheng}, and the relevant ones are collected in the Appendix.

Using Eqs.(\ref{ptop})-(\ref{vertex}) and taking the integration rules given in Refs \cite{jaus,hycheng},
we obtain the $B_c\to P$ form factors,
\be
f_{+}\left(q^{2}\right)=\frac{N_{c}}{16 \pi^{3}} \int d x_{2} d^{2} p_{\perp}^{\prime} \frac{h_{B_c}^{\prime}
h_{P}^{\prime \prime}}{x_{2} \hat{N}_{1}^{\prime} \hat{N}_{1}^{\prime \prime}}\left[x_{1}\left(M_{0}^{\prime 2}+M_{0}^{\prime \prime 2}\right)+x_{2} q^{2}\right.\notag\\
\left.-x_{2}\left(m_{1}^{\prime}-m_{1}^{\prime \prime}\right)^{2}-x_{1}\left(m_{1}^{\prime}-m_{2}\right)^{2}-x_{1}\left(m_{1}^{\prime \prime}-m_{2}\right)^{2}\right],
\en
\be
f_{-}\left(q^{2}\right)=\frac{N_{c}}{16 \pi^{3}} & \int d x_{2} d^{2} p_{\perp}^{\prime} \frac{2 h_{B_c}^{\prime}
h_{P}^{\prime \prime}}{x_{2} \hat{N}_{1}^{\prime} \hat{N}_{1}^{\prime \prime}}\left\{-x_{1} x_{2} M^{\prime 2}
-p_{\perp}^{\prime 2}-m_{1}^{\prime} m_{2}+\left(m_{1}^{\prime \prime}-m_{2}\right)\left(x_{2} m_{1}^{\prime}
+x_{1} m_{2}\right)\right.\notag\\
&+2 \frac{q \cdot P}{q^{2}}\left(p_{\perp}^{\prime 2}+2 \frac{\left(p_{\perp}^{\prime} \cdot q_{\perp}\right)^{2}}{q^{2}}\right)
+2 \frac{\left(p_{\perp}^{\prime} \cdot q_{\perp}\right)^{2}}{q^{2}}-\frac{p_{\perp}^{\prime} \cdot q_{\perp}}{q^{2}}\left[M^{\prime \prime 2}-x_{2}\left(q^{2}+q \cdot P\right)\right.\notag\\
&\left.\left.-\left(x_{2}-x_{1}\right) M^{\prime 2}+2 x_{1} M_{0}^{\prime 2}-2\left(m_{1}^{\prime}-m_{2}\right)\left(m_{1}^{\prime}+m_{1}^{\prime \prime}\right)\right]\right\} .
\en
It is similar for the $B_c\to V$ transition amplitudes, which are given by \cite{hycheng}
\be
\mathcal{B}_{\mu}^{B_c V}=-i^{3} \frac{N_{c}}{(2 \pi)^{4}} \int d^{4} p_{1}^{\prime} \frac{H_{B_c}^{\prime}\left(i H_{V}^{\prime \prime}\right)}{N_{1}^{\prime} N_{1}^{\prime \prime} N_{2}}
 S_{\mu \nu}^{B_c V} \varepsilon^{*\nu},
\en
where \be
S_{\mu \nu}^{B_c V}&=&\left(S_{V}^{B_c V}-S_{A}^{B_c V}\right)_{\mu \nu}\non
&=&\operatorname{Tr}\left[\left(\gamma_{\nu}-\frac{1}{W_{V}^{\prime \prime}}\left(p_{1}^{\prime \prime}-p_{2}\right)_{\nu}\right)\left(p_{1}^{\prime \prime}
+m_{1}^{\prime \prime}\right)\left(\gamma_{\mu}-\gamma_{\mu} \gamma_{5}\right)\left(\not p_{1}^{\prime}+m_{1}^{\prime}\right) \gamma_{5}\left(-\not p_{2}
+m_{2}\right)\right] \non
&=&-2 i \epsilon_{\mu \nu \alpha \beta}\left\{p_{1}^{\prime \alpha} P^{\beta}\left(m_{1}^{\prime \prime}-m_{1}^{\prime}\right)
+p_{1}^{\prime \alpha} q^{\beta}\left(m_{1}^{\prime \prime}+m_{1}^{\prime}-2 m_{2}\right)+q^{\alpha} P^{\beta} m_{1}^{\prime}\right\}\non &&
+\frac{1}{W_{V}^{\prime \prime}}\left(4 p_{1 \nu}^{\prime}-3 q_{\nu}-P_{\nu}\right) i \epsilon_{\mu \alpha \beta \rho} p_{1}^{\prime \alpha} q^{\beta} P^{\rho}\nonumber
\en
\be
&&
+2 g_{\mu \nu}\left\{m_{2}\left(q^{2}-N_{1}^{\prime}-N_{1}^{\prime \prime}-m_{1}^{\prime 2}-m_{1}^{\prime \prime 2}\right)
-m_{1}^{\prime}\left(M^{\prime \prime 2}-N_{1}^{\prime \prime}-N_{2}-m_{1}^{\prime \prime 2}-m_{2}^{2}\right)\right.\nonumber
\non &&\left.-m_{1}^{\prime \prime}\left(M^{\prime 2}-N_{1}^{\prime}-N_{2}-m_{1}^{\prime 2}-m_{2}^{2}\right)
-2 m_{1}^{\prime} m_{1}^{\prime \prime} m_{2}\right\} \non &&
+8 p_{1 \mu}^{\prime} p_{1 \nu}^{\prime}\left(m_{2}-m_{1}^{\prime}\right)-2\left(P_{\mu} q_{\nu}
+q_{\mu} P_{\nu}+2 q_{\mu} q_{\nu}\right) m_{1}^{\prime}+2 p_{1 \mu}^{\prime} P_{\nu}\left(m_{1}^{\prime}-m_{1}^{\prime \prime}\right)\non &&
+2 p_{1 \mu}^{\prime} q_{\nu}\left(3 m_{1}^{\prime}-m_{1}^{\prime \prime}-2 m_{2}\right)
+2 P_{\mu} p_{1 \nu}^{\prime}\left(m_{1}^{\prime}+m_{1}^{\prime \prime}\right)+2 q_{\mu} p_{1 \nu}^{\prime}\left(3 m_{1}^{\prime}+m_{1}^{\prime \prime}-2 m_{2}\right)\non &&
+\left\{2 p_{1 \mu}^{\prime}\left[M^{\prime 2}
+M^{\prime \prime 2}-q^{2}-2 N_{2}+2\left(m_{1}^{\prime}-m_{2}\right)\left(m_{1}^{\prime \prime}+m_{2}\right)\right]\right.\non&&
+q_{\mu}\left[q^{2}-2 M^{\prime 2}+N_{1}^{\prime}-N_{1}^{\prime \prime}+2 N_{2}-\left(m_{1}+m_{1}^{\prime \prime}\right)^{2}+2\left(m_{1}^{\prime}-m_{2}\right)^{2}\right]\non&&
\left.+P_{\mu}\left[q^{2}-N_{1}^{\prime}-N_{1}^{\prime \prime}-\left(m_{1}^{\prime}+m_{1}^{\prime \prime}\right)^{2}\right]\right\}\frac{1}{2 W_{V}^{\prime \prime}}\left(4 p_{1 \nu}^{\prime}-3 q_{\nu}-P_{\nu}\right) .
\label{sptov}\en
From the above equation, we can get the expressions for $B_c\to V$ form factors defined in Eqs.(\ref{formv1}) and (\ref{formv2}) \cite{hycheng}
\be
g(q^{2})&=&-\frac{N_{c}}{16 \pi^{3}} \int d x_{2} d^{2} p_{\perp}^{\prime} \frac{2 h_{B_c}^{\prime}
 h_{V}^{\prime \prime}}{x_{2} \hat{N}_{1}^{\prime} \hat{N}_{1}^{\prime \prime}}\left\{x_{2} m_{1}^{\prime}
 +x_{1} m_{2}+\left(m_{1}^{\prime}-m_{1}^{\prime \prime}\right) \frac{p_{\perp}^{\prime} \cdot q_{\perp}}{q^{2}}\right.\non &&\left.
 +\frac{2}{w_{V}^{\prime \prime}}\left[p_{\perp}^{\prime 2}+\frac{\left(p_{\perp}^{\prime} \cdot q_{\perp}\right)^{2}}{q^{2}}\right]\right\},\\
f(q^{2})&=& \frac{N_{c}}{16 \pi^{3}} \int d x_{2} d^{2} p_{\perp}^{\prime} \frac{h_{B_c}^{\prime} h_{V}^{\prime \prime}}{x_{2}
\hat{N}_{1}^{\prime}
\hat{N}_{1}^{\prime \prime}}\left\{2 x_{1}\left(m_{2}-m_{1}^{\prime}\right)\left(M_{0}^{\prime 2}+M_{0}^{\prime \prime 2}\right)
-4 x_{1} m_{1}^{\prime \prime} M_{0}^{\prime 2}+2 x_{2} m_{1}^{\prime} q \cdot P\right.\non
&&\left.+2 m_{2} q^{2}-2 x_{1} m_{2}\left(M^{\prime 2}+M^{\prime \prime 2}\right)+2\left(m_{1}^{\prime}-m_{2}\right)\left(m_{1}^{\prime}
+m_{1}^{\prime \prime}\right)^{2}+8\left(m_{1}^{\prime}-m_{2}\right) \right.\non &&
\left. \times\left[p_{\perp}^{\prime 2}+\frac{\left(p_{\perp}^{\prime}
\cdot q_{\perp}\right)^{2}}{q^{2}}\right]+2\left(m_{1}^{\prime}+m_{1}^{\prime \prime}\right)\left(q^{2}+q \cdot P\right) \frac{p_{\perp}^{\prime} \cdot q_{\perp}}{q^{2}}
-4 \frac{q^{2} p_{\perp}^{\prime 2}+\left(p_{\perp}^{\prime} \cdot q_{\perp}\right)^{2}}{q^{2} w_{V}^{\prime \prime}}
\right.\non && \left.\times\left[2 x_{1}\left(M^{\prime 2}+M_{0}^{\prime 2}\right)-q^{2}-q \cdot P-2\left(q^{2}+q \cdot P\right) \frac{p_{\perp}^{\prime} \cdot q_{\perp}}{q^{2}}
\right.\right.\non &&
\left.\left.-2\left(m_{1}^{\prime}-m_{1}^{\prime \prime}\right)\left(m_{1}^{\prime}-m_{2}\right)\right]\right\},\\
a_{+}(q^{2})&=& \frac{N_{c}}{16 \pi^{3}} \int d x_{2} d^{2} p_{\perp}^{\prime} \frac{2 h_{B_c}^{\prime} h_{V}^{\prime \prime}}{x_{2} \hat{N}_{1}^{\prime}
\hat{N}_{1}^{\prime \prime}}\left\{\left(x_{1}-x_{2}\right)\left(x_{2} m_{1}^{\prime}+x_{1} m_{2}\right)-\left[2 x_{1} m_{2}
+m_{1}^{\prime \prime}+\left(x_{2}-x_{1}\right) m_{1}^{\prime}\right] \right.\non &&
\left.\times \frac{p_{\perp}^{\prime} \cdot q_{\perp}}{q^{2}}-2 \frac{x_{2} q^{2}+p_{\perp}^{\prime} \cdot q_{\perp}}{x_{2} q^{2} w_{V}^{\prime \prime}}\left[p_{\perp}^{\prime} \cdot p_{\perp}^{\prime \prime}
+\left(x_{1} m_{2}+x_{2} m_{1}^{\prime}\right)\left(x_{1} m_{2}-x_{2} m_{1}^{\prime \prime}\right)\right]\right\},\\
a_{-}(q^{2})&=& \frac{N_{c}}{16 \pi^{3}} \int d x_{2} d^{2} p_{\perp}^{\prime} \frac{h_{B_c}^{\prime} h_{V}^{\prime \prime}}{x_{2} \hat{N}_{1}^{\prime}
\hat{N}_{1}^{\prime \prime}}\left\{2\left(2 x_{1}-3\right)\left(x_{2} m_{1}^{\prime}+x_{1} m_{2}\right)-8\left(m_{1}^{\prime}-m_{2}\right)
\right.\non &&\times\left[\frac{p_{\perp}^{\prime 2}}{q^{2}}
+2 \frac{\left(p_{\perp}^{\prime} \cdot q_{\perp}\right)^{2}}{q^{4}}\right]-\left[\left(14-12 x_{1}\right) m_{1}^{\prime}-2 m_{1}^{\prime \prime}-\left(8-12 x_{1}\right) m_{2}\right] \frac{p_{\perp}^{\prime} \cdot q_{\perp}}{q^{2}} \non
&&+\frac{4}{w_{V}^{\prime \prime}}\left(\left[M^{\prime 2}+M^{\prime \prime 2}-q^{2}+2\left(m_{1}^{\prime}-m_{2}\right)\left(m_{1}^{\prime \prime}
+m_{2}\right)\right]\left(A_{3}^{(2)}+A_{4}^{(2)}-A_{2}^{(1)}\right)\right.\non
&&+Z_{2}\left(3 A_{2}^{(1)}-2 A_{4}^{(2)}-1\right)+\frac{1}{2}\left[x_{1}\left(q^{2}+q \cdot P\right)
-2 M^{\prime 2}-2 p_{\perp}^{\prime} \cdot q_{\perp}-2 m_{1}^{\prime}\left(m_{1}^{\prime \prime}+m_{2}\right)\right.\non
&&\left.\left.\left.-2 m_{2}\left(m_{1}^{\prime}-m_{2}\right)\right]\left(A_{1}^{(1)}+A_{2}^{(1)}-1\right) q \cdot P\left[\frac{p_{\perp}^{\prime 2}}{q^{2}}
+\frac{\left(p_{\perp}^{\prime} \cdot q_{\perp}\right)^{2}}{q^{4}}\right]\left(4 A_{2}^{(1)}-3\right)\right)\right\},\;\;\;
\en
where the functions $A^{(1)}_1,A^{(1)}_2,A^{(2)}_3,A^{(2)}_4$ and $Z_2$ are listed in the Appendix.
and the physical form factors $V^{B_cV}(q^2), A_0^{B_cV}(q^2), A_1^{B_cV}(q^2),$ and $A_2^{B_cV}(q^2)$ can be related to the above formulae
through Eqs. (\ref{relationv1}) and (\ref{relationv2}).

The extension to $B_c\to A$ transitions is straightforward and their form factors have similar expressions as those in the $B_c\to V$ transitions case.
The $B_c\to \;^{3}A, \;^{1}A$ transition amplitudes are defined as \cite{hycheng}
\be
\mathcal{B}_{\mu}^{B_c \;^{1} A}&=&-i^{3} \frac{N_{c}}{(2 \pi)^{4}} \int d^{4} p_{1}^{\prime} \frac{H_{B_c}^{\prime} H_{{ }^{1} A}^{\prime \prime}}{N_{1}^{\prime} N_{1}^{\prime \prime}
N_{2}} S_{\mu \nu}^{B_c \;^{1} A} \varepsilon^{\prime \prime * \nu},\\
\mathcal{B}_{\mu}^{B_c\;^{3} A}&=&-i^{3} \frac{N_{c}}{(2 \pi)^{4}} \int d^{4} p_{1}^{\prime} \frac{H_{B_c}^{\prime} H_{{ }^{3} A}^{\prime \prime}}{N_{1}^{\prime} N_{1}^{\prime \prime}
N_{2}} S_{\mu \nu}^{B_c \;^{3} A} \varepsilon^{\prime \prime * \nu},
\en
where the traces $S_{\mu \nu}^{B_c\;^{i}A}(i=1,3)$
\be
S_{\mu \nu}^{B_c \;^{3}A} &=&\left(S_{V}^{B_c \;^{3}A}-S_{A}^{B_c \;^{3}A}\right)_{\mu \nu} \non
&=&\operatorname{Tr}\left[\left(\gamma_{\nu}-\frac{1}{W_{\;^{3}A}^{\prime \prime}}\left(p_{1}^{\prime \prime}-p_{2}\right)_{\nu}\right) \gamma_{5}\left(\not p_{1}^{\prime \prime}
+m_{1}^{\prime \prime}\right)\left(\gamma_{\mu}-\gamma_{\mu} \gamma_{5}\right)\left(\not p_{1}^{\prime}+m_{1}^{\prime}\right) \gamma_{5}\left(-\not p_{2}+m_{2}\right)\right] \non
&=&\operatorname{Tr}\left[\left(\gamma_{\nu}-\frac{1}{W_{\;^{3}A}^{\prime \prime}}\left(p_{1}^{\prime \prime}
-p_{2}\right)_{\nu}\right)\left(\not p_{1}^{\prime \prime}-m_{1}^{\prime \prime}\right)\left(\gamma_{\mu} \gamma_{5}-\gamma_{\mu}\right)\left(\not p_{1}^{\prime}
+m_{1}^{\prime}\right) \gamma_{5}\left(-\not p_{2}+m_{2}\right)\right],\label{btoa3}\non\\
S_{\mu \nu}^{B_c\;^{1}A} &=&\left(S_{V}^{B_c \;^{1}A}-S_{A}^{B_c\;^{1}A}\right)_{\mu \nu} \non
&=&\operatorname{Tr}\left[\left(-\frac{1}{W_{\;^{1}A}^{\prime \prime}}\left(p_{1}^{\prime \prime}-p_{2}\right)_{\nu}\right) \gamma_{5}\left(\not p_{1}^{\prime \prime}
+m_{1}^{\prime \prime}\right)\left(\gamma_{\mu}-\gamma_{\mu} \gamma_{5}\right)\left(\not p_{1}^{\prime}+m_{1}^{\prime}\right) \gamma_{5}\left(-\not p_{2}+m_{2}\right)\right] \non
&=&\operatorname{Tr}\left[\left(-\frac{1}{W_{\;^{1}A}^{\prime \prime}}\left(p_{1}^{\prime \prime}
-p_{2}\right)_{\nu}\right)\left(\not p_{1}^{\prime \prime}-m_{1}^{\prime \prime}\right)\left(\gamma_{\mu} \gamma_{5}-\gamma_{\mu}\right)\left(\not p_{1}^{\prime}
+m_{1}^{\prime}\right) \gamma_{5}\left(-\not p_{2}+m_{2}\right)\right].\;\;\label{btoa1}
\en
By comparing Eq.(\ref{sptov}) and Eqs.(\ref{btoa3}),(\ref{btoa1}), we have $S^{B_c\;^{i}A}_{V(A)}=S^{B_cV}_{A(V)}(i=1,3)$ with the replacements
$m^{\prime\prime}_1\to -m^{\prime\prime}_1, W^{\prime\prime}_V\to W^{\prime\prime}_{^{3}A,^{1}A}$.
Note that only the term $1/W''$ is kept in $S_{\mu \nu}^{B_c \;^{1}A}$.  Thus the $B_c\to \;^{i}A(i=1,3)$ form factors
can be related
to the $B_c\to V$ form factors through the following replacements:
\be
l^{^{3}A,^{1}A}(q^2)&=&f(q^2), \;\;\;\; \text{with}\;\;\;\; \non &&m^{\prime\prime}_1\to -m^{\prime\prime}_1, h^{\prime\prime}_V\to h^{\prime\prime}_{^{3}A,^{1}A},
w^{\prime\prime}_V\to w^{\prime\prime}_{^{3}A,^{1}A},\\
q^{^{3}A,^{1}A}(q^2)&=&g(q^2), \;\;\;\; \text{with}\;\;\;\; \non &&m^{\prime\prime}_1\to -m^{\prime\prime}_1, h^{\prime\prime}_V\to h^{\prime\prime}_{^{3}A,^{1}A},
w^{\prime\prime}_V\to w^{\prime\prime}_{^{3}A,^{1}A},\\
c^{^{3}A,^{1}A}_{\pm}(q^2)&=&a_{\pm}(q^2), \;\; \text{with}\;\;\;\; \non &&m^{\prime\prime}_1\to -m^{\prime\prime}_1, h^{\prime\prime}_V\to h^{\prime\prime}_{^{3}A,^{1}A},
w^{\prime\prime}_V\to w^{\prime\prime}_{^{3}A,^{1}A},
\en
where the replacement of $m^{\prime\prime}_1\to -m^{\prime\prime}_1$ is not applied to $m''_1$ in $w''$ and $h''$, because they arise from the
propagator and quark-antiquark-meson coupling vertex. The physical form
factors $A^{B_cA}(q^2), V_0^{B_cA}(q^2), V_1^{B_cA}(q^2), V_2^{B_cA}(q^2)$ can be related to the above formulae
through Eqs.(\ref{relationa1}) and (\ref{relationa2}).

we finally turn to the $B_c\to S$ transition amplitude, which is given as \cite{hycheng}
\be
\mathcal{B}^{B_{c} S}=-i^{3} \frac{N_{c}}{(2 \pi)^{4}} \int d^{4} p_{1}^{\prime} \frac{H_{B_c}^{\prime}\left(H_{S}^{\prime \prime}\right)}
{N_{1}^{\prime} N_{1}^{\prime \prime} N_{2}} S_{\mu}^{B_{c} S},
\en
where the trace $S_{\mu}^{B_{c} S}$
\be
S_{\mu}^{B_{c} S} &=& Tr\left[\left(\not p_{1}^{\prime \prime}+m_{1}^{\prime \prime}\right) \gamma_{\mu} \gamma_{5}\left(\not p_{1}^{\prime}
+m_{1}^{\prime}\right) \gamma_{5}\left(-\not p_{2}+m_{2}\right)\right]\notag\\
&=&2 p_{1 \mu}^{\prime}\left[M^{\prime 2}+M^{\prime \prime 2}-q^{2}-2 N_{2}-\left(m_{1}^{\prime}-m_{2}\right)^{2}
-\left(m_{1}^{\prime \prime}+m_{2}\right)^{2}+\left(m_{1}^{\prime}+m_{1}^{\prime \prime}\right)^{2}\right]\non &&
+q_{\mu}\left[q^{2}-2 M^{\prime 2}+N_{1}^{\prime}-N_{1}^{\prime \prime}+2 N_{2}+2\left(m_{1}^{\prime}-m_{2}\right)^{2}
-\left(m_{1}^{\prime}+m_{1}^{\prime \prime}\right)^{2}\right]\non &&
+P_{\mu}\left[q^{2}-N_{1}^{\prime}-N_{1}^{\prime
\prime}-\left(m_{1}^{\prime}+m_{1}^{\prime
\prime}\right)^{2}\right]. \en
Using the formulae above and the integration rules obtained in Refs. \cite{jaus,hycheng}, we have the $B_c\to S$ form factors
\be
F_{1}^{B_{c} S}\left(q^{2}\right)  &=& \frac{N_{c}}{16 \pi^{3}} \int d x_{2} d^{2} p_{\perp}^{\prime} \frac{h_{B_c}^{\prime}
 h_{S}^{\prime \prime}}{x_{2} \hat{N}_{1}^{\prime} \hat{N}_{1}^{\prime \prime}}\left[x_{1}\left(M_{0}^{\prime 2}+M_{0}^{\prime \prime 2}\right)+x_{2} q^{2}\right.\non &&
\left.-x_{2}\left(m_{1}^{\prime}+m_{1}^{\prime
\prime}\right)^{2}-x_{1}\left(m_{1}^{\prime}-m_{2}\right)^{2}-x_{1}\left(m_{1}^{\prime
\prime}+m_{2}\right)^{2}\right],\\
F_{0}^{B_{c} S}\left(q^{2}\right) &=& F_{1}^{B_{c} S}\left(q^{2}\right)+\frac{q^{2}}{q \cdot P} \frac{N_{c}}{16 \pi^{3}} \int d x_{2} d^{2} p_{\perp}^{\prime}
\frac{2 h_{B_c}^{\prime} h_{S}^{\prime \prime}}{x_{2} \hat{N}_{1}^{\prime} \hat{N}_{1}^{\prime \prime}}\left\{-x_{1} x_{2} M^{\prime 2}
-p_{\perp}^{\prime 2}-m_{1}^{\prime} m_{2}\right.\non &&\left.
-\left(m_{1}^{\prime \prime}+m_{2}\right)\left(x_{2} m_{1}^{\prime}+x_{1} m_{2}\right)
+2 \frac{q \cdot P}{q^{2}}\left(p_{\perp}^{\prime 2}+2 \frac{\left(p_{\perp}^{\prime} \cdot q_{\perp}\right)^{2}}{q^{2}}\right)
+2 \frac{\left(p_{\perp}^{\prime} \cdot q_{\perp}\right)^{2}}{q^{2}}\right.\non &&\left.
-\frac{p_{\perp}^{\prime} \cdot q_{\perp}}{q^{2}}\left[M^{\prime \prime 2}-x_{2}\left(q^{2}+q \cdot P\right)
-\left(x_{2}-x_{1}\right) M^{\prime 2}+2 x_{1} M_{0}^{\prime 2}\right.\right.\non &&\left.\left.-2\left(m_{1}^{\prime}-m_{2}\right)
\left(m_{1}^{\prime}-m_{1}^{\prime \prime}\right)\right]\right\}.
\en
{\centering\section{Numerical results and discussions} \label{results}}
Equipped with explicit expressions of the form factors $f_+(q^2),f_-(q^2)$ for $B_c\to P$ transitions, $g(q^2), f(q^2), a_+(q^2), a_-(q^2)$
for $B_c\to V$ transitions, $l^{\;^{i}A}(q^2), q^{\;^{i}A}(q^2), c^{\;^{i}A}_+(q^2), c^{\;^{i}A}_-(q^2)$
for $B_c\to\;^iA(i=1,3)$ transitions, and $F_1^{B_cS}(q^2), F_0^{B_cS}(q^2)$ for $B_c\to S$ transitions, we now proceed to perform
numerical studies using the CLFQM. In the earlier works \cite{wwang,wwang1}, the form factors of $B_c$ decays into the
ground-state charmonia and charmed mesons
were calculated. In this work, besides updating the transition form factors of $B_c$ decays to these ground-state
charmonia and charmed mesons, we also study the results of $B_c$ transitions to some excited-state charmonia.
With these form factors, we then calculate the branching ratios of 80 $B_c$ decays with a charmonium
involved in each channel.

As mentioned earlier, the shape parameter $\beta'$ in the wave function describes the momentum distribution and can be calculated using
the meson's decay constant under the CLFQM. The analytic expressions for the calculations are listed in Sect.2.2. The decay constant for the $B_c$ meson is
employed by the result provided by the lattice QCD \cite{chiu1}
\be
f_{B_c}=(489\pm4\pm3)\;\text{MeV},
\en
which is larger than the value used in Refs. \cite{wwang,wwang1}. The decay constant of $J/\Psi$ can be determined by the leptonic decay
width
\be
\Gamma_{ee}\equiv\Gamma(J\Psi\to e^+e^-)=\frac{4\pi\alpha^2_{em}Q^2_cf^3_{J\Psi}}{3m_{J\Psi}},
\en
with the electric charge of the charm quark $Q_c=\frac{2}{3}$, $\alpha_{em}$ being a fine-structure constant. Using the updated measured
result for the electronic width of $J/\Psi$ given in PDG22 \cite{pdg22} $\Gamma_{ee}=(5.53\pm0.10)$ keV, one can obtain the decay constant
of $J/\Psi$
\be
f_{J/\Psi}=(431.0\pm4.3)\;\text{MeV},
\en
which is different from the previous value $f_{J/\Psi}=(416\pm5)$ MeV \cite{wwang}.
Similarly, using the measured result $\Gamma(\psi(2S)\to e^+e^-)=2.36\pm0.04$ keV, we obtain the decay constant of the radially excited
meson $\psi(2S)$,
$f_{\psi(2S)}=296^{+3}_{-2}$ MeV. The decay constant of the radially excited state $\psi(3S)$ is determined as $f_{\psi(3S)}=(187\pm8)$ MeV by using the
data $\Gamma_{\psi(3S)\to ee}=(5.53\pm0.10)$ keV. As for the decay constant $f_{\eta_c}$, we use the the lattice QCD result given in Ref. \cite{damir}
\be
f_{\eta_c}=(387\pm7\pm2)\;\text{MeV},
\en
which is a little larger than the value $f_{\eta_c}=340.9^{+16.3}_{-16.6}$ MeV extracted from the data of $\eta_c\to\gamma\gamma$ decay.
The decay constant $f_{\eta_c(2S)}$ can be determined by the double photon decay of $\eta_c(2S)$ as
\be
f_{\eta_c(2S)}=\sqrt{\frac{81m_{\eta_c(2S)}\Gamma_{\eta_c(2S)\to\gamma\gamma}}{64\pi\alpha^2_{em}}}.
\en

By using the measured results of the branching ratio $Br(\eta_c(2S)\to\gamma\gamma)=(1.9\pm1.3)\times10^{-4}$ and
$\Gamma_{\eta_c(2S)}=11.3^{+3.2}_{-2.9}$ MeV \cite{pdg22}, we can obtain the decay constant
\be
f_{\eta_c(2S)}=(243^{+79}_{-111})\;\text{MeV}.\label{etac2s}
\en
However, there is no calculation for the decay constant of $\eta_c(3S)$ or the data on $\eta_c(3S)\to \gamma\gamma$ decay
used to extract it
from experiment. We can fix the decay constant $f_{\eta_c(3S)}$ through the assumption
$\frac{f_{\eta_c(3S)}}{f_{\eta_c}}=\frac{f_{\psi(3S)}}{f_{J/\Psi}}$ \cite{ymwang,cmeng} and obtain it as
\be
f_{\eta_c(3S)}=(170.0\pm8.0)\;\text{MeV}.
\en

To determine the shape parameter of $\chi_{c1}$, we use the decay constant $f_{\chi_{c1}}=185$ MeV evaluated from the light-cone QCD sum
rules at the scale $\mu=m_c$ \cite{olpak}. This value is much smaller than $f_{\chi_{c1}}=340^{+119}_{-101}$ MeV given in Ref. \cite{wwang1}.
So the corresponding shape parameter $\beta^{\prime}_{\chi_{c1}}=(0.536\pm0.023)$ GeV is smaller than the value
$\beta^{\prime}_{\chi_{c1}}=(0.7\pm0.1)$ GeV obtained in Ref. \cite{wwang1}. For the charmonia $\chi_{c0}$ and $h_{c}$, we
will assume the same values and introduce an uncertainty of $10\%$ to the shape parameters to compensate the different Lorentz structures, that is
$\beta'_{\chi_{c0}}=\beta'_{h_{c}}=(0.536\pm0.023)$ GeV. The decay constant of $X(3872)$ is determined using the branching fractions
$Br(B^-\to J/\Psi K^-)=(1.026\pm0.031)\times10^{-3}$ and $Br(B^-\to X(3872) K^-)=(2.3\pm0.9)\times10^{-4}$ and is obtained as
\be
f_{X(3872)}=(234\pm52) \;\text{MeV},
\en
which is lower than $f_{X(3872)}=329^{+111}_{-95}$ MeV used in the previous CLFQM calculations \cite{wwang2}.
The experimental results for the decay constants of charmed mesons are given as \cite{pdg22}
\be
f_D=(204.6\pm5.0)\;\text{MeV},\;\;\;f_{D_s}=(257.5\pm4.6)\;\text{MeV}.
\en
As for the decay constants of the vector charmed meson $D^*$ and $D^*_s$, we used the lattice QCD results $f_{D^*}=(245\pm20^{+3}_{-2})$ MeV,
\;$f_{D^*_s}=(272\pm16^{+3}_{-20})$ MeV \cite{Boucaud}. \footnote{It is noticed that all the charmed mesons appeared in this paper are positively charged.
In some place, we will omit the sign of charge for simplicity. } Using these decay constants and the masses of the
constituent quarks and mesons given in Tab. \ref{mass}, we can obtain the values of the shape parameters $\beta'$ for
our considered mesons which are listed in Tab. \ref{beta}.

\begin{table}
\begin{center}
\caption{The masses (GeV) of the constituent quarks and mesons \cite{ymwang,pdg22}.}
\begin{tabular}{ccccc}
\hline\hline
$m_{u}$& $m_{d}$ & $m_{s}$&$ m_{c}$&$m_{b}$\\
$0.25$&$0.25$&$0.37$&$1.4$&$4.8$\\
\hline
$m_{B_c}$&$m_{D^+}$&$m_{D^+_{s}}$&$m_{D^{\ast+}}$&$m_{D^{\ast+}_{s}}$\\
$6.27447$&$1.86966$&$1.96835$&$2.01026$&$2.1122$\\
\hline
$m_{J/\Psi}$&$m_{\psi(2S)}$&$m_{\psi(3S)}$&$m_{\chi_{c0}}$&$m_{\chi_{c1}}$\\
$3.09690$&$3.68610$&$4.039$&$3.41471$&$3.51067$\\
\hline
$m_{\eta_{c}}$&$m_{\eta_{c}(2S)}$&$m_{\eta_{c}(3S)}$&$m_{h_{c}}$&$m_{X(3872)}$\\
$2.9839$&$3.6375$&$3.940$&$3.52538$&$3.87165$\\
\hline\hline
\end{tabular}\label{mass}
\end{center}
\end{table}
\begin{table}
\begin{center}
\caption{The shape parameters $\beta'$ (in units of GeV) in the Gaussian-type light-front wave functions defined in Eq.(\ref{betap}), and the uncertainties are from the decay constants.}
\begin{tabular}{ccccc}
\hline\hline
$\beta'_{B_{c}}$&$ \beta'_{D} $ & $\beta'_{D_{s}} $&$\beta'_{D^{\ast}} $&$\beta'_{D^{\ast}_{s}} $ \\
$1.058^{+0.009}_{-0.010}$&$0.464^{+0.011}_{-0.014}$&$0.497^{+0.032}_{-0.028}$&$0.409^{+0.021}_{-0.022}$&$0.438^{+0.016}_{-0.027}$\\
\hline
$\beta'_{J/\Psi}$&$\beta'_{\psi(2S)}$&$\beta'_{\psi(3S)}$&$\beta'_{\eta_{c}}$&$\beta'_{\eta_{c}(2S)}$\\
$0.646^{+0.041}_{-0.041}$&$0.566^{+0.004}_{-0.003}$&$0.449^{+0.012}_{-0.013}$&$0.754\pm0.014$&$0.488^{+0.140}_{-0.187}$\\
\hline
$\beta'_{\eta_{c}(3s)}$&$\beta'_{\chi_{c0}} $&$\beta'_{\chi_{c1}}$&$\beta'_{h_c}$&$\beta'_{X(3872)}$\\
$0.382^{+0.045}_{-0.054}$&$0.536\pm0.023$&$0.536\pm0.023$&$0.536\pm0.023$&$0.62^{+0.057}_{-0.064}$\\
\hline\hline
\end{tabular}\label{beta}
\end{center}
\end{table}

\begin{table}[!htbp]
\caption{$B_c\to D, D^*, D_s, D_s^*, \eta_c, \eta_c{(2S,3S)}, J/\Psi, \psi(2S,3S)$ form factors in the CLFQM.
The uncertainties are from the decay constants of $B_c$ and final-state meson.}
\begin{center}
\begin{tabular}{c|c|c|c|c}
\hline\hline F& F(0)&$F(q^{2}_{max})$&$a$&$b$\\
\hline
$F^{B_{c}\eta_{c}}_{1} $&$0.60^{+0.00+0.01}_{-0.00-0.01}$  &$ 1.06^{+0.00+0.03}_{-0.00-0.03}$    &$1.95^{+0.01+0.03}_{-0.01-0.03}$        & $0.48^{+0.00+0.01}_{-0.00-0.01}$           \\
$F^{B_{c}\eta_{c}}_{0} $&$0.60^{+0.00+0.01}_{-0.01-0.00}$   &$ 0.85^{+0.00+0.02}_{-0.01-0.02}$   &$1.44^{+0.00+0.03}_{-0.00-0.03}$        & $-0.62^{+0.02+0.02}_{-0.02-0.03}$           \\
$F^{B_{c}\eta_{c}(2S)}_{1} $&$0.37^{+0.00+0.12}_{-0.00-0.18}$  &$ 0.48^{+0.00+0.28}_{-0.00-0.31}$    &$1.44^{+0.00+0.92}_{-0.00-0.66}$        & $0.15^{+0.02+0.50}_{-0.02-0.34}$           \\
$F^{B_{c}\eta_{c}(2S)}_{0} $&$0.37^{+0.00+0.12}_{-0.00-0.18}$    &$ 0.41^{+0.00+0.28}_{-0.01-0.28}$  &$0.73^{+0.01+0.99}_{-0.01-0.95}$        & $-0.81^{+0.02+0.34}_{-0.02-0.28}$           \\
$F^{B_{c}\eta_{c}(3S)}_{1} $&$0.29^{+0.00+0.04}_{-0.00-0.05}$   &$ 0.36^{+0.00+0.07}_{-0.00-0.08}$   &$1.53^{+0.00+0.29}_{-0.00-0.23}$        & $0.23^{+0.01+0.13}_{-0.01-0.13}$           \\
$F^{B_{c}\eta_{c}(3S)}_{0} $&$0.29^{+0.00+0.04}_{-0.00-0.05}$   &$ 0.32^{+0.00+0.07}_{-0.00-0.08}$   &$0.85^{+0.01+0.44}_{-0.01-0.31}$        & $-0.74^{+0.01+0.05}_{-0.00-0.24}$           \\
$F^{B_{c} D}_{1}$&$0.17^{+0.00+0.01}_{-0.00-0.01}$    &$0.97^{+0.01+0.10}_{-0.01-0.08}$  &$3.09^{+0.02+0.07}_{-0.02-0.05}$        & $0.91^{+0.00+0.02}_{-0.00-0.01}$           \\
$F^{B_{c} D}_{0}$&$0.17^{+0.00+0.01}_{-0.00-0.01}$  &$ 0.30^{+0.01+0.04}_{-0.01-0.04}$    &$2.32^{+0.01+0.08}_{-0.01-0.06}$        & $-2.42^{+0.06+0.11}_{-0.06-0.16}$           \\
$F^{B_{c} D_{s}}_{1}$&$0.21^{+0.00+0.01}_{-0.00-0.01}$  &$ 1.09^{+0.01+0.07}_{-0.01-0.06}$    &$2.68^{+0.02+0.04}_{-0.01-0.04}$        & $0.79^{+0.00+0.01}_{-0.00-0.01}$           \\
$F^{B_{c} D_{s}}_{0}$&$0.21^{+0.00+0.01}_{-0.00-0.01}$   &$ 0.45^{+0.01+0.03}_{-0.01-0.03}$   &$1.91^{+0.04+0.01}_{-0.04-0.01}$        & $-1.55^{+0.04+0.06}_{-0.04-0.06}$           \\
\hline
$V^{B_{c} J/\Psi}$&$0.76^{+0.00+0.04}_{-0.00-0.04}$    &$ 1.37^{+0.00+0.11}_{-0.00-0.10}$       &$2.16^{+0.01+0.09}_{-0.01-0.08}$              & $0.53^{+0.00+0.01}_{-0.00-0.01}$                          \\
$A^{B_{c} J/\Psi}_{0}$&$0.55^{+0.00+0.03}_{-0.00-0.04}$   &$ 0.76^{+0.00+0.06}_{-0.00-0.07}$        &$1.22^{+0.02+0.07}_{-0.02-0.07}$              & $0.16^{+0.00+0.00}_{-0.00-0.00}$                          \\
$A^{B_{c} J/\Psi}_{1}$&$0.53^{+0.00+0.02}_{-0.03-0.00}$    &$ 0.78^{+0.01+0.02}_{-0.01-0.05}$       &$1.45^{+0.03+0.09}_{-0.01-0.09}$              & $0.29^{+0.00+0.02}_{-0.00-0.00}$                          \\
$A^{B_{c} J/\Psi}_{2}$&$0.49^{+0.00+0.00}_{-0.00-0.01}$     &$ 0.84^{+0.00+0.03}_{-0.00-0.00}$      &$1.97^{+0.01+0.11}_{-0.01-0.11}$              & $0.43^{+0.00+0.03}_{-0.00-0.03}$                          \\
$V^{B_{c} \psi(2S)}$&$0.57^{+0.00+0.01}_{-0.00-0.00}$  &$ 0.67^{+0.00+0.02}_{-0.00-0.00}$       &$1.01^{+0.01+0.01}_{-0.01-0.02}$      & $-0.16^{+0.03+0.01}_{-0.03-0.02}$                         \\
$A^{B_{c}\psi(2S)}_{0}$&$0.41^{+0.00+0.00}_{-0.00-0.00}$   &$ 0.44^{+0.00+0.00}_{-0.00-0.00}$      &$0.39^{+0.01+0.01}_{-0.01-0.01}$      & $-0.15^{+0.02+0.01}_{-0.02-0.01}$                         \\
$A^{B_{c} \psi(2S)}_{1}$&$0.35^{+0.00+0.00}_{-0.00-0.00}$     &$ 0.35^{+0.00+0.00}_{-0.00-0.00}$    &$0.08^{+0.01+0.02}_{-0.02-0.03}$      & $-0.69^{+0.03+0.01}_{-0.04-0.02}$                         \\
$A^{B_{c} \psi(2S)}_{2}$&$0.17^{+0.00+0.00}_{-0.00-0.00}$      &$ 0.12^{+0.00+0.00}_{-0.00-0.00}$   &$-1.53^{+0.07+0.09}_{-0.09-0.13}$      & $-3.67^{+0.14+0.13}_{-0.19-0.21}$                         \\
$V^{B_{c} \psi(3S)}$&$0.46^{+0.00+0.02}_{-0.00-0.02}$ &$ 0.53^{+0.00+0.03}_{-0.00-0.03}$&$1.14^{+0.03+0.04}_{-0.03-0.03}$          & $-0.01^{+0.01+0.01}_{-0.01-0.01}$           \\
$A^{B_{c} \psi(3S)}_{0}$&$0.31^{+0.00+0.01}_{-0.00-0.01}$&$ 0.33^{+0.00+0.01}_{-0.00-0.01}$&$0.49^{+0.03+0.03}_{-0.02-0.03}$           & $-0.04^{+0.01+0.01}_{-0.01-0.01}$                         \\
$A^{B_{c}\psi(3S)}_{1}$&$0.27^{+0.00+0.01}_{-0.00-0.01}$&$ 0.28^{+0.00+0.02}_{-0.00-0.01}$ &$0.25^{+0.02+0.04}_{-0.02-0.03}$         & $-0.45^{+0.01+0.00}_{-0.01-0.00}$                         \\
$A^{B_{c}\psi(3S)}_{2}$&$0.14^{+0.00+0.01}_{-0.00-0.01}$ &$ 0.12^{+0.00+0.02}_{-0.00-0.02}$ &$-1.01^{+0.02+0.04}_{-0.02-0.04}$     & $-2.73^{+0.01+0.00}_{-0.01-0.00}$                         \\
$V^{B_{c}D^{\ast}}$&$0.20^{+0.03+0.00}_{-0.03-0.00}$&$ 1.18^{+0.19+0.10}_{-0.19-0.07}$ &$3.39^{+0.02+0.15}_{-0.02-0.12}$       & $0.99^{+0.01+0.04}_{-0.01-0.03}$                         \\
$A^{B_{c}D^{\ast}}_{0}$&$0.14^{+0.00+0.02}_{-0.00-0.02}$&$0.35^{+0.01+0.07}_{-0.01-0.06}$ &$1.88^{+0.03+0.12}_{-0.03-0.10}$     & $0.18^{+0.01+0.01}_{-0.01-0.01}$                         \\
$A^{B_{c}D^{\ast}}_{1}$&$0.13^{+0.02+0.00}_{-0.02-0.00}$  &$ 0.44^{+0.07+0.04}_{-0.07-0.03}$ &$2.38^{+0.02+0.16}_{-0.02-0.13}$   & $0.52^{+0.00+0.05}_{-0.00-0.04}$                         \\
$A^{B_{c}D^{\ast}}_{2}$&$0.12^{+0.01+0.00}_{-0.00-0.01}$&$ 0.55^{+0.05+0.05}_{-0.01-0.09}$  &$2.98^{+0.02+0.17}_{-0.02-0.15}$    & $0.68^{+0.00+0.06}_{-0.00-0.06}$                         \\
$V^{B_{c}D^{\ast}_{s}}$&$0.25^{+0.00+0.00}_{-0.00-0.00}$ &$ 1.24^{+0.01+0.08}_{-0.01-0.05}$ &$3.22^{+0.02+0.16}_{-0.02-0.09}$    & $0.94^{+0.01+0.04}_{-0.00-0.02}$                         \\
$A^{B_{c}D^{\ast}_{s}}_{0}$&$0.18^{+0.02+0.00}_{-0.03-0.00}$ &$ 0.41^{+0.05+0.02}_{-0.07-0.01}$           &$1.77^{+0.02+0.13}_{-0.02-0.07}$       & $0.19^{+0.01+0.01}_{-0.01-0.00}$                         \\
$A^{B_{c}D^{\ast}_{s}}_{1}$&$0.16^{+0.00+0.01}_{-0.02-0.00}$   &$ 0.47^{+0.00+0.07}_{-0.06-0.02}$         &$2.25^{+0.02+0.17}_{-0.02-0.09}$        & $0.50^{+0.00+0.05}_{-0.00-0.03}$                         \\
$A^{B_{c}D^{\ast}_{s}}_{2}$&$0.15^{+0.01+0.00}_{-0.01-0.00}$  &$ 0.60^{+0.05+0.06}_{-0.03-0.04}$          &$2.85^{+0.02+0.19}_{-0.02-0.10}$   & $0.67^{+0.00+0.06}_{-0.00-0.04}$                         \\
 \hline\hline
\end{tabular}\label{btopv}
\end{center}
\end{table}

From Tab. \ref{btopv}, we can find that the form factors of $B_c$ transitions to charmed mesons ($D, D^*,D_s, D^*_s$) at the maximally
recoiling point ($q^2=0$)
are smaller than those of $B_c$ transitions to ground-state charmonia. This is because the initial charm quark in the $B_c$
decays to charmed mesons is almost
at rest, and its momentum is of order $m_c$, while the charmed mesons in the final states move very fast, and the final charm quark tends to have a very
large momentum of order $m_b$. So the overlaps of the initial and final states' light-front wave functions in these transitions are limited,
which induce small values for the
form factors. In the $B_c$ transitions to charmonia, both the spectator charm quark and the charm antiquark generated from the weak vertex
are heavy, and the light-front wave functions of the charmonia have a maximum near $E\sim m_c$. It is expected that the overlaps of the $B_c$ and
charmonium's light-front wave functions become large, which induce larger form factors. Thus it is easy to understand that for the $B_c$ decays to the charmonium and charmed meson, for example $B_c\to J/\Psi D$, the Feynman amplitudes associated with $B_c$ transitions to charmonia are
much more important than that associated with $B_c$ transitions to charmed mesons. Furthermore, the SU(3) symmetry breaking effects between the form factors of
$B_c\to D$ and $B_c\to D_s$ transitions are large, since the decay constant of $D_s$ meson is larger than that of the $D$ meson. It is similar between the
form factors of $B_c\to D^*$ and $B_c\to D^*_s$ transitions. These can be checked by future experiments.
The uncertainties from the decay constant of $\eta_c(2S)$ shown in Eq.(\ref{etac2s}) are very large, so there are relevant large uncertainties in the
$B_c\to \eta_c(2S)$ transition form factors.
It is noted that if evaluating  the form factors at $q^2>0$ region in the frame of $q_\perp=0$, we must include the non-valence
configuration (the so-called Z-graph contribution)
arising from quark pair creation from the vacuum, which is difficult for us to calculate reliably. While if one calculates
in the frame of $q^+=0$, such
non-valence contribution vanishes automatically. Because of the condition $q^+=0$ imposed in the course of calculation, the
form factors are obtained only for space-like momentum transfer
$q^2=-q^2_\perp\leq0$, while the physical transition processes are relevant for the time-like form factors.
Many authors \cite{jaus,hycheng,wwang} have proposed parametrization of form factors by using some
explicit functions of $q^2$ in the space-like region, then extending them to the time-like region. Here, we will adopt the
parametrization form given in
Ref. \cite{wwang}:
\be
F(q^2)=F(0)\exp(a\frac{q^2}{m^2_{B_c}}+b\frac{q^4}{m^4_{B_c}}).
\en

\begin{table}
\begin{center}
\caption{Comparison of the $B_c\to \eta_c$ and $B_c\to J/\Psi$ transition form factors at $q^2=0$ between this work and other literature.}
\begin{tabular}{c|c c c c c }
\hline\hline
&$F^{B_{c}\eta_{c}}_{1}=F^{B_{c}\eta_{c}}_{0}$&$V^{B_{c}J/\psi}$&$A^{B_{c} J/\psi}_{0}$&$A^{B_{c} J/\psi}_{1}$&$A^{B_{c}J/\psi}_{2}$\\
\hline
This work&$0.60$&$0.76$&$0.55$&$0.53$&$0.49$\\
\hline
\cite{wwang}&$0.61$&$0.74$&$0.53$&$0.50$&$0.44$\\
\hline
\cite{Nobes}&$0.5359$&$0.736$&$0.532$&$0.524$&$0.509$\\
\hline
\cite{Santorelli}&$0.61$&$0.83$&$0.57$&$0.56$&$0.54$\\
\hline
\cite{sun}$^1$&$0.66$&--&$0.655$&$0.578$&$0.427$\\
\hline
\cite{Verma}&$0.58$&$0.91$&$0.58$&$0.63$&$0.74$\\
\hline
\cite{Onishchenko}&$0.66$&$1.03$&$0.60$&$0.63$&$0.69$\\
\hline
\cite{P}&$0.76$&$0.96$&$0.69$&$0.68$&$0.66$\\
\hline
\cite{Zuo,T}&$0.87$&$1.69$&$0.27$&$0.75$&$1.69$\\
\hline
\cite{Likhoded,Kiselev}$^2$&$0.66[0.7]$&$1.03[0.94]$&$0.60[0.66]$&$0.63[0.66]$&$0.69[0.66]$\\
\hline
\cite{D.s}&$0.420$&$0.591$&$0.408$&$0.416$&$0.431$\\
\hline
\cite{Galkin}&$0.47$&$0.49$&$0.40$&$0.50$&$0.73$\\
\hline
\cite{Hernandez}&$0.49$&$0.61$&$0.45$&$0.49$&$0.56$\\
\hline
\cite{Colangelo}&$0.20$&$0.38$&$0.26$&$0.27$&$0.28$\\
\hline
\cite{and}&$0.23$&$0.33$&$0.21$&$0.21$&$0.23$\\
\hline\hline
\end{tabular}\label{formetacpsi}
\end{center}
{\footnotesize $^1$ Here the results with $\omega=0.8$ GeV are quoted. }\\
{\footnotesize $^2$ The results out (in) the brackets are evaluated in sum rules (potential) model.}\\
\end{table}
\begin{table}[htbp]
\begin{center}
\caption{Comparison of the $B_c\to D, D^*, D_s, D_s^*$ transition form factors at $q^2=0$ between this work and other literature.}
\begin{tabular}{c|c c c c c }
\hline\hline
&$F^{B_{c}D}_{1}=F^{B_{c} D}_{0}$&$A_0^{B_{c}D^{\ast}}$&$A^{B_{c}D^{\ast}}_{1}$&$A^{B_{c}D^{\ast}}_{2}$&$V^{B_{c} D^{\ast}}$\\
\hline
This work&$0.17$&$0.20$&$0.14$&$0.13$&$0.12$\\
\hline
\cite{Colangelo}&$0.13$&$0.25$&$0.05$&$0.11$&$0.17$\\
\hline
\cite{Galkin}&$0.14$&$0.18$&$0.14$&$0.17$&$0.19$\\
\hline
\cite{Nobes}&$0.1446$&$0.175$&$0.094$&$0.100$&$0.105$\\
\hline
 \cite{D.s}$^1$ &$0.154$&$0.224$&$0.156$&$0.145$&$0.134$\\
\hline
\cite{wwang}&$0.16$&$0.13$&$0.09$&$0.08$&$0.07$\\
\hline
\cite{Likhoded, Kiselev}$^2$&$0.32[0.29]$&$1.66[1.74]$&$0.35[0.37]$&$0.43[0.43]$&$0.51[0.50]$\\
\hline
\cite{Zuo, T}&$0.35$&$0.57$&$0.05$&$0.32$&$0.57$\\
\hline
\cite{P}&$0.69$&$0.98$&$0.47$&$0.56$&$0.64$\\
\hline
\cite{Verma}&$0.075$&$0.16$&$0.081$&$0.095$&$0.11$\\
\hline\hline
&$F^{B_{c} D_{s}}_{1}=F^{B_{c}D_{s}}_{0}$&$A^{B_{c} D^{\ast}_{s}}$&$V^{B_{c}D^{\ast}_{s}}_{0}$&$V^{B_{c} D^{\ast}_{s}}_{1}$&$V^{B_{c} D^{\ast}_{s}}_{2}$\\
\hline
This work&$0.21$&$0.25$&$0.18$&$0.16$&$0.15$\\
\hline
\cite{Verma}&$0.15$&$0.29$&$0.16$&$0.18$&$0.20$\\
\hline
\cite{wwang}&$0.28$&$0.23$&$0.17$&$0.14$&$0.12$\\
\hline
\cite{Likhoded, Kiselev}$^2$&$0.45[0.43]$&$2.02[2.27]$&$0.47[0.52]$&$0.56[0.56]$&$0.65[0.60]$\\
\hline\hline
\end{tabular}\label{compD}
\end{center}
{\footnotesize $^1$ Here the results with $\omega=0.6$ GeV are quoted.}\\
{\footnotesize $^2$ The results out (in) the brackets are evaluated in sum rules (potential) model.}\\
\end{table}

\begin{figure}[htbp]
\centering
\begin{minipage}[t]{0.45\linewidth}
\includegraphics[scale=0.5]{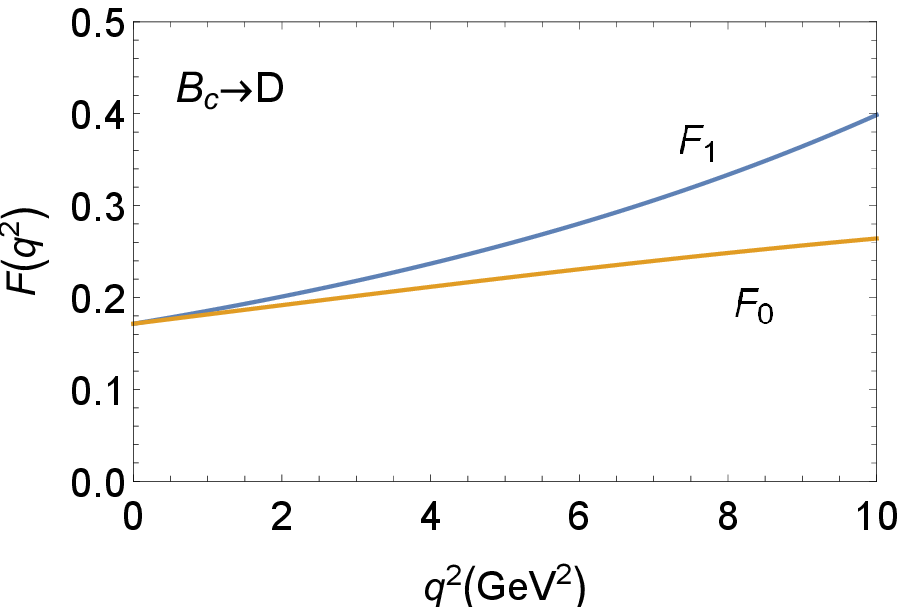}\caption{Form factors $F_{1}(q^2), F_{0}(q^2)$ for the $B_{c}\to D$ transition.}
\label{BD}
\end{minipage}
\hspace{0.4cm}
\begin{minipage}[t]{0.45\linewidth}
\includegraphics[scale=0.5]{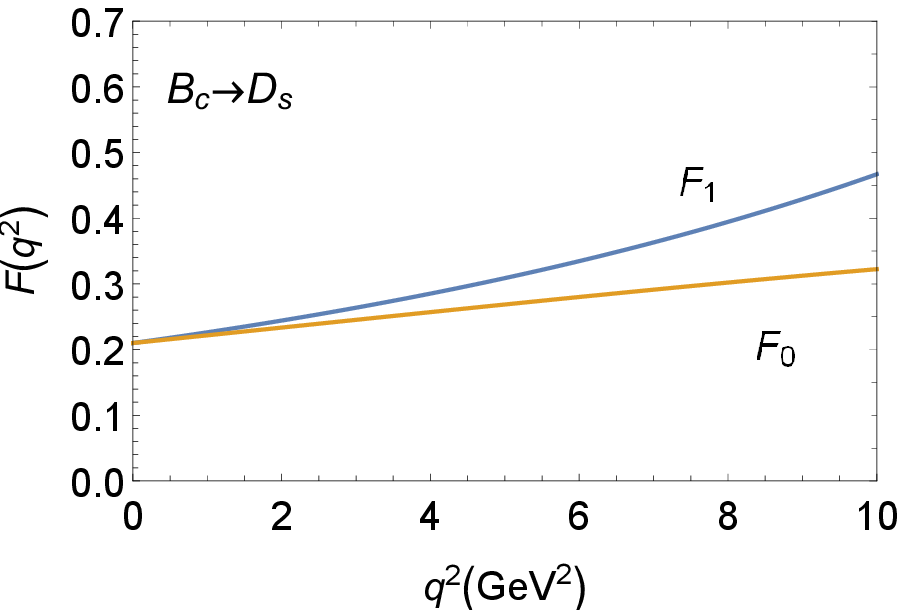}\caption{Form factors $F_{1}(q^2), F_{0}(q^2)$ for the $B_{c^{}}\to D_{s}$ transition.}
\label{BDs}
\end{minipage}
\end{figure}
\begin{figure}[htbp]
\centering
\begin{minipage}[t]{0.45\linewidth}
\includegraphics[scale=0.55]{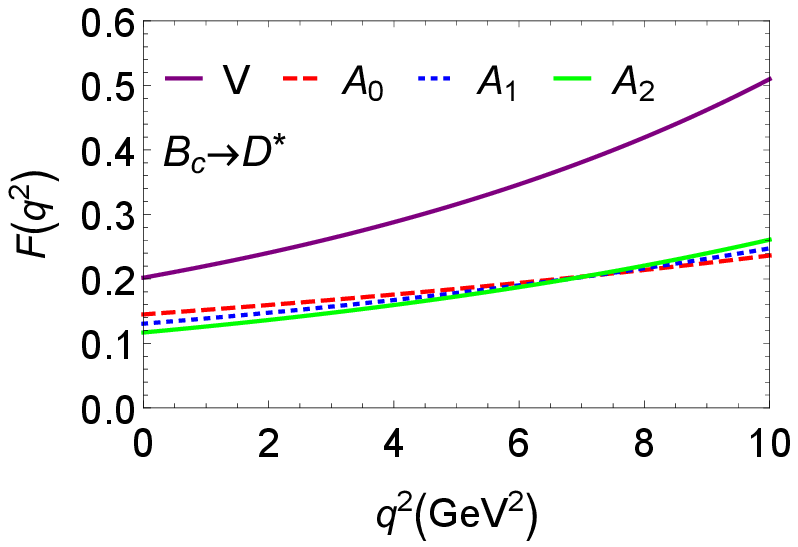}\caption{Form factors $V(q^2)$, $A_{0}(q^2)$, $A_{1}(q^2)$ and $A_{2}(q^2)$ for the $B_{c}\to D^{\ast}$ transition.}
\label{BDst}
\end{minipage}
\hspace{0.4cm}
\begin{minipage}[t]{0.45\linewidth}
\includegraphics[scale=0.55]{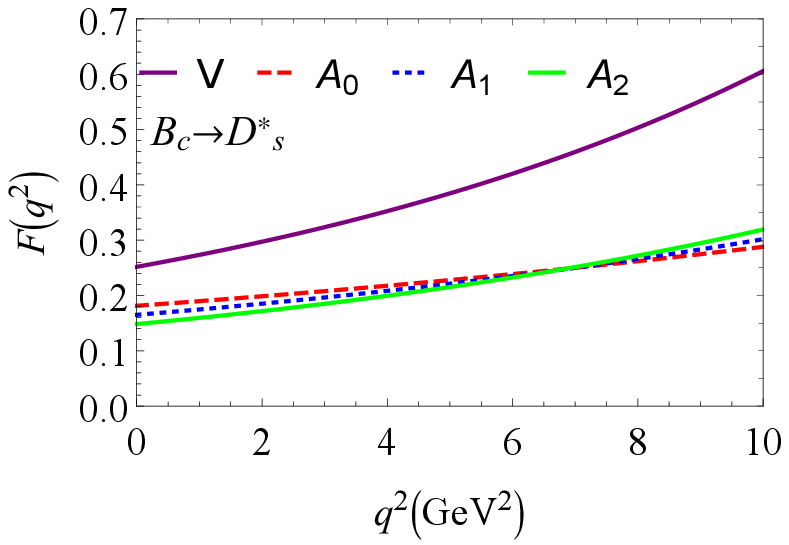}\caption{Form factors $V(q^2)$, $A_{0}(q^2)$, $A_{1}(q^2)$ and $A_{2}(q^2)$ for the $B_{c}\to D^{\ast}_s$ transition.}
\label{BDsst}
\end{minipage}
\end{figure}
\begin{figure}[htbp]
\centering
\subfigure{
\begin{minipage}{4.8cm}
\centering
\includegraphics[width=4.8cm]{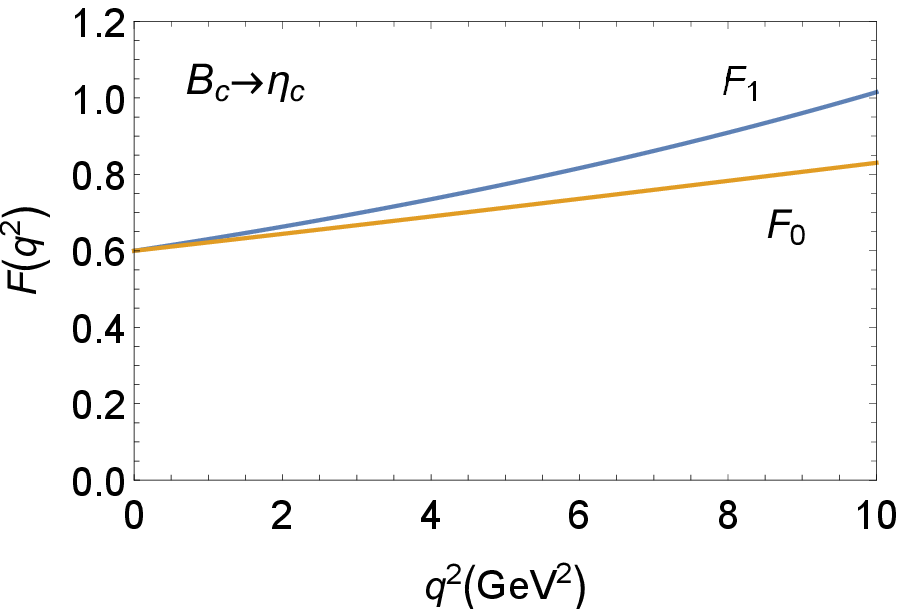}
\end{minipage}}
\subfigure{
\begin{minipage}{4.8cm}
\centering
\includegraphics[width=4.8cm]{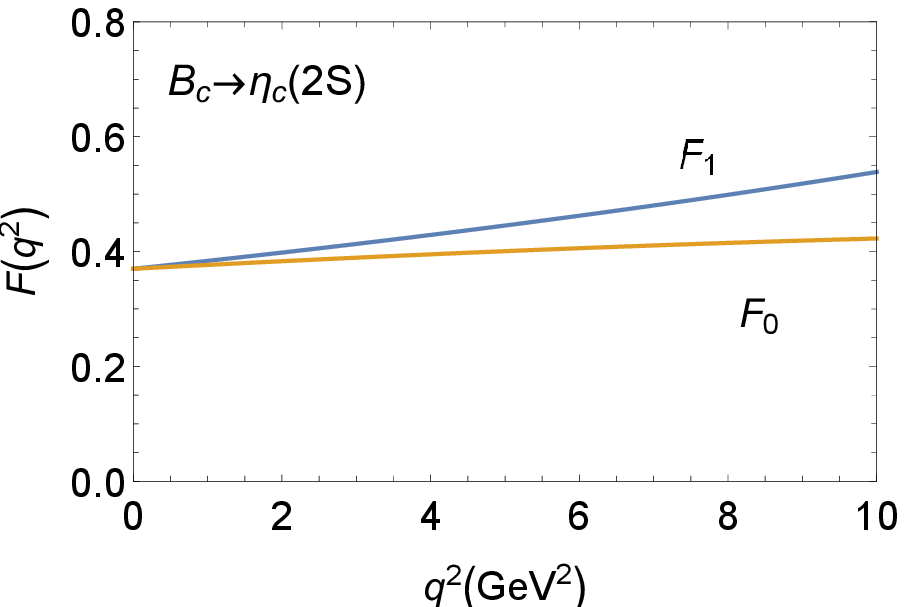}
\end{minipage}}
\subfigure{
\begin{minipage}{4.8cm}
\centering
\includegraphics[width=4.8cm]{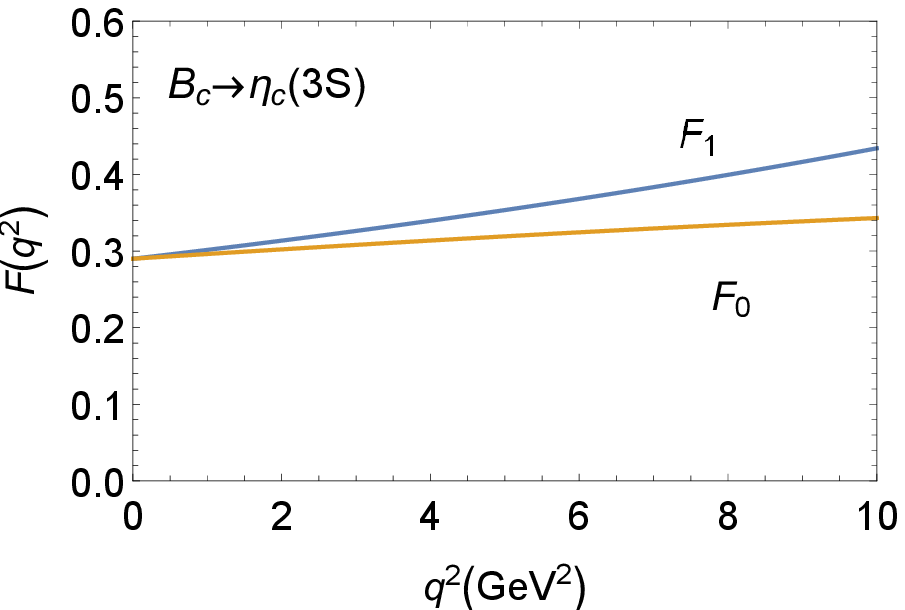}
\end{minipage}}
\caption{Form factors $F_{1}(q^2), F_{0}(q^2)$ for the $B_{c}\to \eta_{c}$ (left), $B_{c}\to \eta_{c}(2S)$ (center), $B_{c}\to \eta_{c}(3S)$ (right)  transitions.}
\label{etac13}
\end{figure}
\begin{figure}[htbp]
\centering
\subfigure{
\begin{minipage}{4.8cm}
\centering
\includegraphics[width=4.8cm]{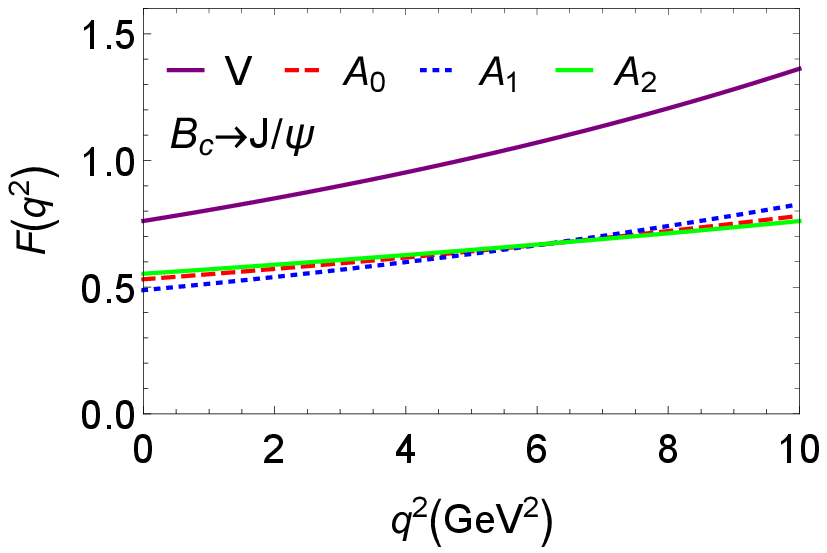}
\label{jpsi}
\end{minipage}}
\subfigure{
\begin{minipage}{4.8cm}
\centering
\includegraphics[width=4.8cm]{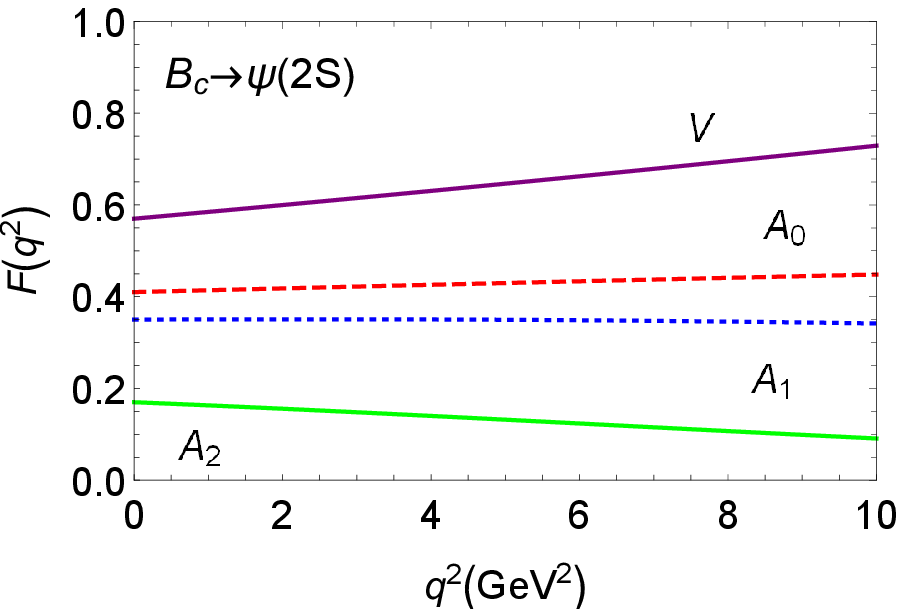}
\label{jpsi2}
\end{minipage}}
\subfigure{
\begin{minipage}{4.8cm}
\centering
\includegraphics[width=4.8cm]{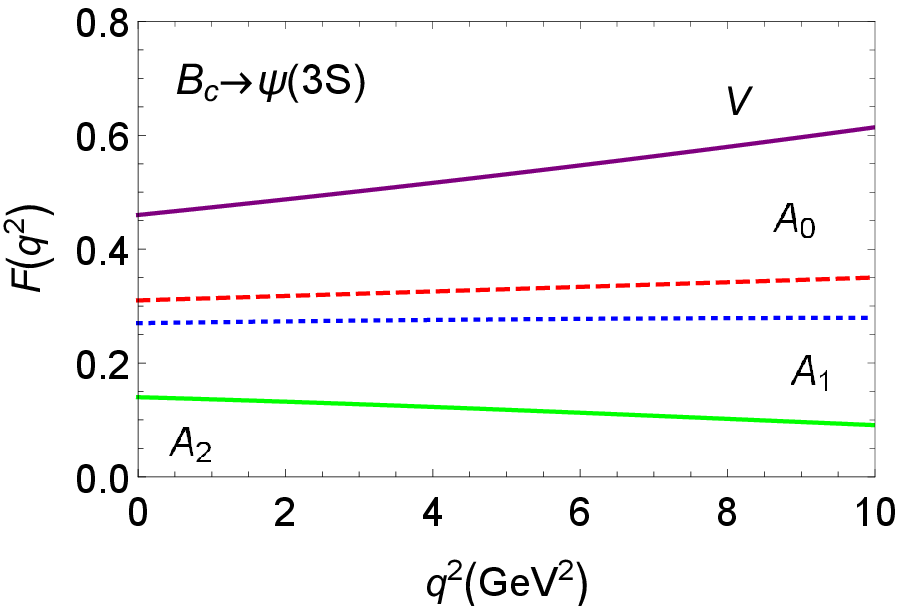}
\label{jpsi3}
\end{minipage}}
\caption{Form factors $V(q^2)$, $A_{0}(q^2),$ $A_{1}(q^2)$ and $A_{2}(q^2)$ for the  $B_{c}\to J/\Psi$ (left), $B_{c}\to \psi(2S)$ (center) and $B_{c}\to \psi(3S)$ (right) transitions.}
\label{jpsi13}
\end{figure}
\begin{figure}[htbp]
\centering
\begin{minipage}[t]{0.45\linewidth}
\includegraphics[scale=0.5]{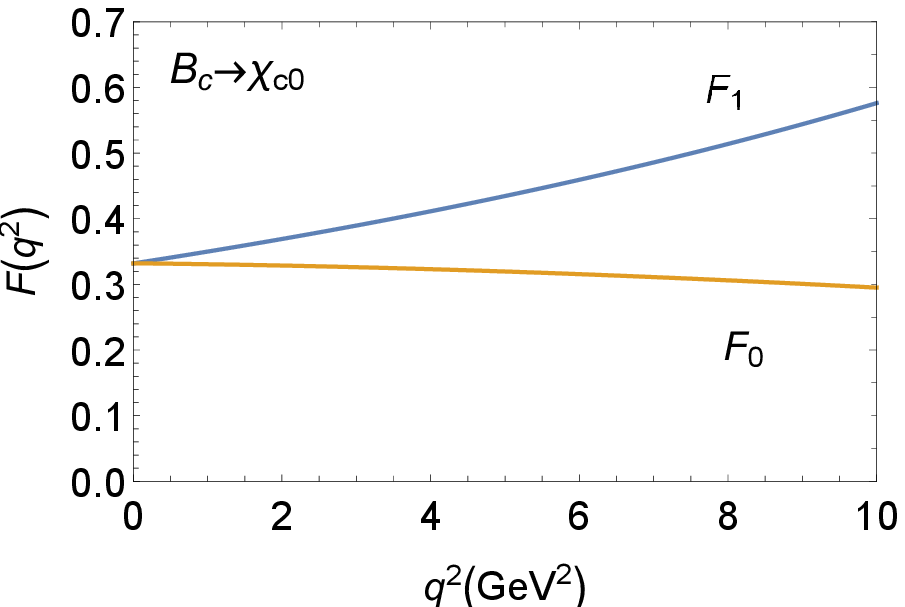}\caption{Form factors$ F_{1}(q^2), F_{0}(q^2)$ for the $B_{c}\to \chi_{c0}$ transition.}
\label{chic0}
\end{minipage}
\hspace{0.4cm}
\begin{minipage}[t]{0.45\linewidth}
\includegraphics[scale=0.5]{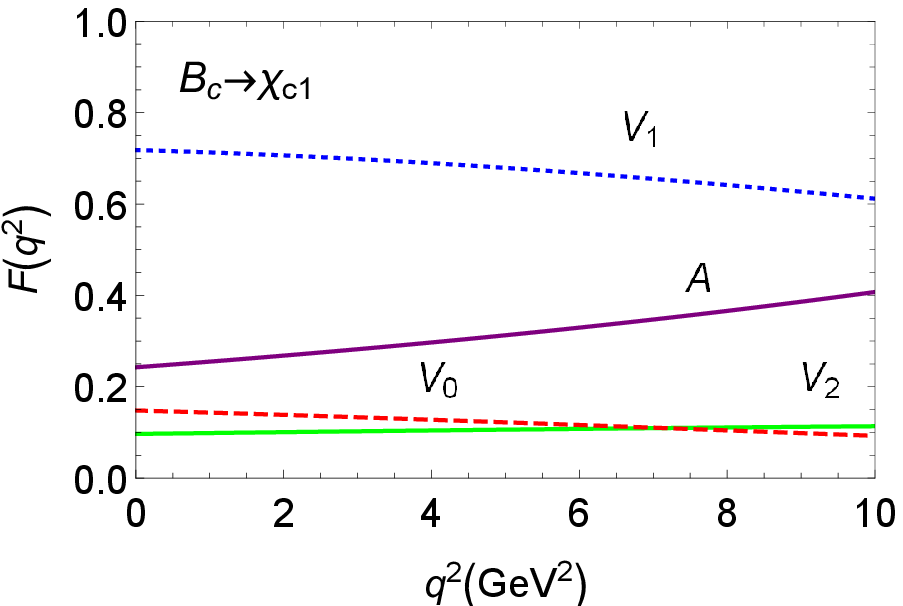}\caption{Form factors $A(q^2), V_{0}(q^2),V_{1}(q^2)$ and $V_{2}(q^2)$ for the $B_{c}\to\chi_{c1}$ transition.}
\label{chic1}
\end{minipage}
\end{figure}
\begin{figure}[htbp]
\centering
\begin{minipage}[t]{0.45\linewidth}
\includegraphics[scale=0.5]{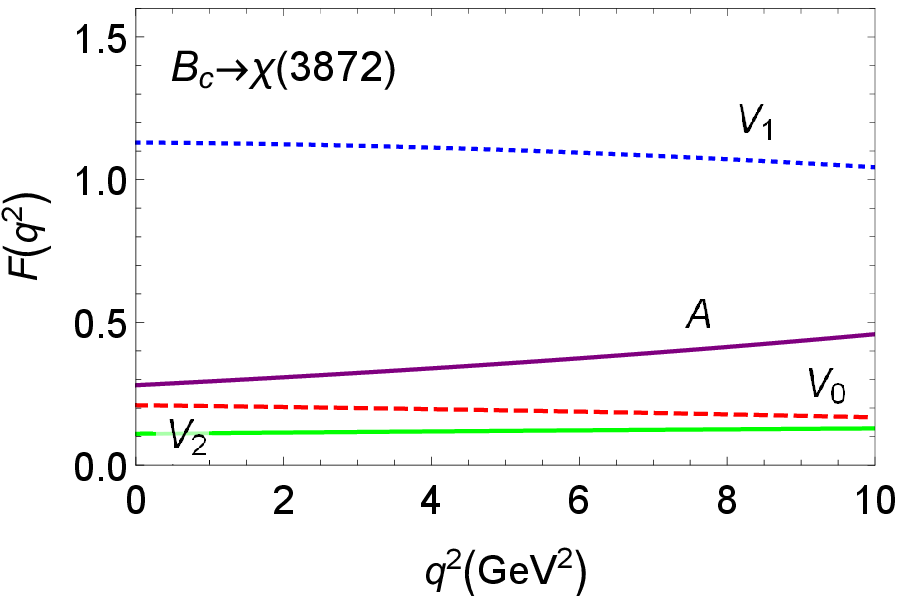}\caption{Form factors $A(q^2)$, $V_{0}(q^2)$, $V_{1}(q^2)$ and $V_{2}(q^2)$ for the $B_{c}\to X(3872)$ transition.}
\label{3872}
\end{minipage}
\hspace{0.4cm}
\begin{minipage}[t]{0.45\linewidth}
\includegraphics[scale=0.5]{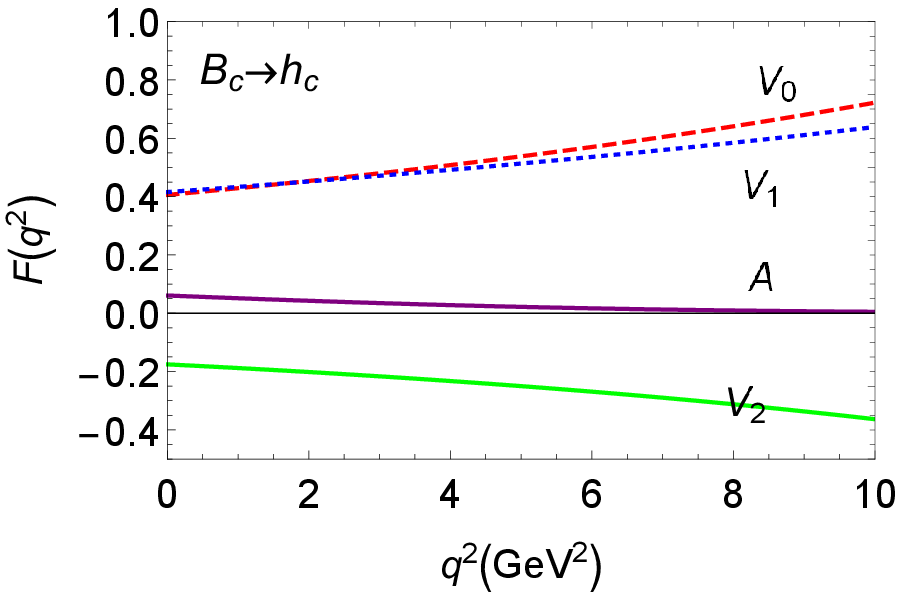}\caption{Form factors $A(q^2)$, $V_{0}(q^2)$, $V_{1}(q^2)$ and $V_{2}(q^2)$ for the $B_{c}\to h_{c}$ transition.}
\label{hc}
\end{minipage}
\end{figure}
The parameters $a$ and $b$ will be fitted in the space-like region ($-10$ GeV$^2\leq q^2\leq 0$).
The $q^2$-dependence of  form factors in the time-like region are plotted in Figs. \ref{BD}, \ref{BDs}, \ref{BDst}, \ref{BDsst}, \ref{etac13}, \ref{jpsi13}, \ref{chic0}, \ref{chic1}, \ref{3872} and \ref{hc}.
In general the slope parameters $a, b$ are very sensitive to the values of $\beta'$,
while the form factors at $q^2=0$ are less sensitive to the variation in $\beta'$ values.

{\centering\subsection{Transition form factors}}
In Tab. \ref{formetacpsi}, we can find that the values of the $B_c\to \eta_c$ transition form factors $F^{B_{c}\eta_{c}}_{0,1}$ predicted
by many works \cite{wwang,Nobes, Santorelli, sun, Verma, sharm, Dhir, Onishchenko, P, Zuo, T, Likhoded, Kiselev} are larger than $0.5$, only a few of values
\cite{D.s, Galkin, Hernandez, Colangelo, and} are less than $0.5$. Similarly, many predictions  for the form factor
$V^{B_{c}J/\psi}$ are larger than $0.6$ except for a few of values. As for the form factors $A^{B_{c} J/\psi}_{0,1,2}$, most of their values lie in the range
of $0.5\sim0.7$.

In Tab. \ref{compD}, we compare our results of the $B_c\to D, D^*, D_s, D_s^*$ transition form factors at $q^2=0$ with other calculations.
One can find that our results are consistent with those calculated using the relativistic quark model (RQM) \cite{D.s, Galkin},
while they are too large for the results given by the relativistic constituent quark model (RCQM) \cite{P},
the light-cone sum rule (LCSR) \cite{Zuo, T}, and the QCD sum rules \cite{Likhoded, Kiselev}. In Refs. \cite{Likhoded, Kiselev}, the form factors have a threefold
enhancement by including the Coulomb-like $\alpha_s/v$ corrections for the heavy quarkonium $B_c$. It seems too small for the values of
$F^{B_{c}D}_{0,1}$ predicted using the Bauer-Stech-Wirbel (BSW) relativistic quark model \cite{Verma}. Compared with the previous
CLFQM calculations \cite{wwang}, our predictions for the form factors $V^{B_cD^*}, A^{B_cD^*}_{0,1,2}$ have a significant enhancement
by using a larger decay constant $f_{D^*}$, while the influence from the difference values for the decay constant $f_{B_c}$ is small.

For the $B_c$ transition to a ground-state vector meson, which is either a charmed meson or a charmonium, the form factor $V$ is the largest one, and
$A_{0,1,2}$ are close to each other. It is easy to find this character in Figs. \ref{BDst} and \ref{BDsst} and the first pannel of Fig. \ref{jpsi13}. On the other hand, if the
final-state meson is a radially excited meson $\psi(2S)$ or $\psi(3S)$, the form factors
show a hierarchy $V>A_2>A_1>A_0$, which can be found in the last two pannels of Fig.\ref{jpsi13}. Nevertheless, $V$ is always the largest one among the form factors of
$B_c$
transition to either a ground state or a radially excited charmonium.
There also exits another hierarchy for the $B_c\to \eta_c, \eta_c(2S), \eta_c(3S)$ transitions, $F_{0,1}^{B_c\to\eta_c}>F_{0,1}^{B_c\to\eta_c(2S)}
>F_{0,1}^{B_c\to\eta_c(3S)}$.
The $q^2$ dependence of the $B_c\to \chi_{c1}$ transition is shown in Fig.\ref{chic1}, where $V_1$ is much larger than other form
factors $A,V_{0,2}$. It is very like the case of the $B_c\to X(3872)$ transition shown in Fig.\ref{3872}. Thus it is a natural assignment of this state as the first radial excitation
of a 1P charmonium state $\chi_{c1}$. Because both $\chi_{c1}$ and $X(3872)$ have the same quantum numbers $J^{PC}=1^{++}$, they should have similar
properties in $B_c$ decays, while it is very different for the $B_c$ transition to another type of axial vector meson $h_c$ with $J^{PC}=1^{+-}$,
where the value of $V_0$ is large and close
to that of $V_1$ as shown in Fig.\ref{hc}. Certainly, the values of the form factor $V_1$ for both of $B_c$ transitions to these two types of axial vector
charmonia are large.
By comparing with Fig.\ref{3872} and \ref{hc}, one can find that both of the values of $V_2$ in the $B_c\to X(3873)$ and $B_c\to h_c$ transition
form factors are the smallest;
especially, $V_2$ for the $B_c\to h_c$ transition becomes negative.
 As we know, there is sill no definite answer about the internal properties of the $X(3872)$. From Fig.\ref{3872}, one can find that
  the form factors of the $B_c \to X(3872)$ transition are
almost flat in their $q^2$ behaviors except for $A^{B_cX(3872)}$. A comparison of these values with experimental measurements for the $B_c \to X(3872)$ transition form factors will provide unique insight into the mysterious inner structure of $X(3872)$. The form factors for the $B_c$ to these P-wave charmonium
transitions are listed in Tab. \ref{pwaveform}.
\begin{table}
\begin{center}
\caption{Results for the $B_c\to \chi_{c0}, \chi_{c1}, h_c, X(3872)$ transition form factors and the fitted parameters $a$ and $b$.
The uncertainties are from the decay constants of $B_c$ and final state mesons.}
\begin{tabular}{c|c|c|c|c}
\hline\hline F& F(0)&$F(q^{2}_{max})$&a&b\\
\hline
$F^{B_{c} \chi_{c0}}_{1}$&$0.33^{+0.00+0.02}_{-0.02-0.00}$&$ 0.52^{+0.00+0.04}_{-0.00-0.03}$     &$2.07^{+0.02+0.04}_{-0.02-0.04}$        & $0.39^{+0.01+0.01}_{-0.01-0.02}$           \\
$F^{B_{c}\chi_{c0}}_{0}$&$0.33^{+0.02+0.02}_{-0.00-0.02}$  &$ 0.30^{+0.02+0.02}_{-0.00-0.02}$    &$-0.14^{+0.01+0.06}_{-0.01-0.06}$        & $-1.29^{+0.01+0.03}_{-0.01-0.03}$           \\
\hline
$A^{B_{c}\chi_{c1}}$&$0.24^{+0.00+0.01}_{-0.00-0.01}$&$ 0.36^{+0.00+0.02}_{-0.00-0.02}$&$1.96^{+0.02+0.03}_{-0.02-0.03}$ & $0.33^{+0.02+0.01}_{-0.01-0.02}$      \\
$V^{B_{c}\chi_{c1}}_{0}$&$0.15^{+0.00+0.00}_{-0.00-0.00}$&$ 0.11^{+0.00+0.01}_{-0.00-0.00}$&$-1.19^{+0.01+0.20}_{-0.20-0.01}$& $-2.60^{+0.05+0.24}_{-0.05-0.25}$      \\
$V^{B_{c}\chi_{c1}}_{1}$&$0.72^{+0.00+0.02}_{-0.00-0.02}$&$ 0.65^{+0.00+0.02}_{-0.00-0.03}$&$-0.25^{+0.02+0.10}_{-0.02-0.09}$ & $-1.51^{+0.00+0.06}_{-0.00-0.06}$        \\
$V^{B_{c}\chi_{c1}}_{2}$&$0.10^{+0.00+0.00}_{-0.00-0.00}$&$ 0.11^{+0.00+0.00}_{-0.00-0.00}$ &$0.83^{+0.04+0.02}_{-0.03-0.02}$ & $-0.77^{+0.05+0.03}_{-0.03-0.05}$        \\
\hline
$A^{B_{c}h_{c}}$&$0.06^{+0.00+0.00}_{-0.00-0.00}$ &$ 0.01^{+0.00+0.00}_{-0.00-0.00}$ &$-6.24^{+0.08+0.60}_{-0.07-0.66}$ & $-14.41^{+0.12+1.46}_{-0.11-1.62}$            \\
$V^{B_{c}h_{c}}_{0}$&$0.41^{+0.03+0.00}_{-0.03-0.00}$ &$ 0.63^{+0.05+0.01}_{-0.05-0.01}$&$2.17^{+0.02+0.05}_{-0.02-0.05}$ & $0.38^{+0.01+0.01}_{-0.01-0.02}$          \\
$V^{B_{c} h_{c}}_{1}$&$0.42^{+0.02+0.00}_{-0.02-0.00}$ &$ 0.58^{+0.03+0.01}_{-0.03-0.01}$ &$1.64^{+0.02+0.05}_{-0.02-0.0.05}$  & $0.22^{+0.00+0.01}_{-0.00-0.01}$        \\
$V^{B_{c} h_{c}}_{2}$&$-0.18^{+0.00+0.01}_{-0.00-0.01}$ &$ -0.31^{+0.01+0.00}_{-0.00-0.01}$ &$2.71^{+0.02+0.04}_{-0.02-0.04}$  & $0.61^{+0.01+0.01}_{-0.01-0.02}$        \\
\hline\hline
$A^{B_{c} X(3872)}$&$0.28^{+0.00+0.02}_{-0.00-0.03}$   &$ 0.37^{+0.00+0.03}_{-0.00-0.04}$       &$1.85^{+0.02+0.09}_{-0.02-0.08}$    & $0.38^{+0.01+0.01}_{-0.01-0.03}$       \\
$V^{B_{c}X(3872)}_{0}$&$0.21^{+0.00+0.01}_{-0.00-0.01}$  &$ 0.19^{+0.00+0.02}_{-0.00-0.02}$     &$-0.52^{+0.01+0.38}_{-0.01-0.32}$      & $-1.45^{+0.02+0.36}_{-0.03-0.32}$      \\
$V^{B_{c}X(3872)}_{1}$&$1.13^{+0.00+0.01}_{-0.00-0.03}$  &$ 1.10^{+0.00+0.05}_{-0.00-0.06}$     &$-0.05^{+0.01+0.24}_{-0.01-0.20}$      & $-1.03^{+0.00+0.15}_{-0.00-0.12}$     \\
$V^{B_{c}X(3872)}_{2}$&$0.11^{+0.00+0.01}_{-0.01-0.01}$   &$ 0.12^{+0.00+0.01}_{-0.01-0.01}$    &$0.77^{+0.03+0.04}_{-0.03-0.04}$      & $-0.61^{+0.02+0.08}_{-0.02-0.12}$    \\
\hline\hline
\end{tabular}\label{pwaveform}
\end{center}
\end{table}
\begin{table}
\begin{center}
\caption{Comparison of the $B_c\to \eta_c(2S,3S), \psi(2S,3S), \chi_{c1}, h_c, X(3872)$ transition form factors at $q^2=0$ between
this work and other literature.}
\scriptsize
\begin{tabular}{c|c c c c }
\hline\hline
&$F^{B_{c}\eta_{c}(2S)}_{1}(0)=F^{B_{c}\eta_{c}(2S)}_{0}(0)$&$F^{B_{c}\eta_{c}(3S)}_{1}(0)=F^{B_{c}\eta_{c}(3S)}_{0}(0)$&$-$&$-$\\
\hline
This work&$0.37$&$0.29$&$-$&$-$\\
\hline
\cite{Rui:2016opu}&$1.04$&$0.78$&$-$&$-$\\
\hline
&$V^{B_{c}\psi(2S)}$&$A^{B_{c} \psi(2S)}_{0}$&$A^{B_{c} \psi(2S)}_{1}$&$A^{B_{c}\psi(2S)}_{2}$\\
\hline
This work&$0.57$&$0.41$&$0.35$&$0.17$\\
\hline
\cite{Ke13}&$0.525$&$0.452$&$0.335$&$0.102$\\
\hline
\cite{Rui:2016opu}&$1.71$&$0.80$&$0.87$&$1.22$\\
\hline
&$V^{B_{c} \psi(3S)}$&$A^{B_{c} \psi(3S)}_{0}$&$A^{B_{c} \psi(3S)}_{1}$&$A^{B_{c}\psi(3S)}_{2}$\\
\hline
This work&$0.46$&$0.31$&$0.27$&$0.14$\\
\hline
\cite{Rui:2016opu}&$1.07$&$0.41$&$0.41$&$0.66$\\
\hline
&$A^{B_{c}\chi_{c1}}$&$V^{B_{c}\chi_{c1}}_{0}$&$V^{B_{c}\chi_{c1}}_{1}$&$V^{B_{c}\chi_{c1}}_{2}$\\
\hline
This work&$0.24$&$0.15$&$0.72$&$0.10$\\
\hline
\cite{wwang1}&$0.36$&$0.13$&$0.85$&$0.15$\\
\hline
&$A^{B_{c} h_{c}}$&$V^{B_{c} h_{c}}_{0}$&$V^{B_{c} h_{c}}_{1}$&$V^{B_{c} h_{c}}_{2}$\\
\hline
This work&$0.06$&$0.41$&$0.42$&$-0.18$\\
\hline
\cite{wwang1}&$0.07$&$0.64$&$0.50$&$-0.32$\\
\hline
&$A^{B_{c} X(3872)}$&$V^{B_{c} X(3872)}_{0}$&$V^{B_{c} X(3872)}_{1}$&$V^{B_{c} X(3872)}_{2}$\\
\hline
This work&$0.28$&$0.21$&$1.13$&$0.11$\\
\hline
\cite{wwang2}&$0.36$&$0.18$&$1.15$&$0.13$\\
\hline\hline
\end{tabular}\label{etajsiex1}
\end{center}
\end{table}

From Tab. \ref{etajsiex1}, we can find that the form factors of $B_c\to \eta_c(2S,3S), \psi(2S,3S)$ transitions calculated in the PQCD
approach \cite{Rui:2016opu} may be more than twice as larger as those predicted in the CLFQM, which will induce large differences for the branching ratios of
some correlative decay channels given by these two approaches. In the PQCD appraoch,
the form factors are sensitive to the formulae of the $B_c$ wave functions. In Ref. \cite{Rui:2016opu}, the authors argued that
the $B_c$ wave function in the light-cone
formula is broader in shape than that of the traditional zero-point one, which is $\propto \delta(x-r_c)$, so the overlap between
the initial and final
states' wave functions becomes larger by using the light-cone wave function for the $B_c$ meson, which induces larger form factors.
Our predictions for
the $B_c$ to vector charmonium $J/\Psi, \psi(2S)$ transition form factors are close to the results given by the LFQM calculations \cite{Ke13}
except for $A^{B_{c}\psi(2S)}_{2}$. For the $B_c$ to the axial vector charmonium transition form factors, our results
are also consistent with the previous CLQM calculations \cite{wwang1,wwang2}.

\subsection{Branching ratios}
Besides the masses of the constituent quarks and mesons listed in Table. \ref{mass}, other inputs, such as the $B_c$ meson
lifetime $\tau_{B_{c}}$,
the Wilson coefficients $a_1, a_2$, and
the Cabibbo-Kobayashi-Maskawa (CKM) matrix elements, are listed as \cite{pdg22,buras}
\be
\tau_{B_{c}}&=&(0.510\pm0.009)\times10^{-12}s, a_1=1.07,a_2=0.234, \\
V_{cd}&=&0.221\pm0.004,V_{cb}=(40.8\pm1.4)\times10^{-3}, V_{cs}=0.975\pm0.006 \\
V_{us}&=&0.2243\pm0.0008, V_{ud}=0.97373\pm0.00031 .
\en

First, we consider the branching ratios of the decays $B_c\to \eta_c(J/\Psi)P(V)$,
which can be calculated through the formula
\be
\mathcal{B}r(B_c\to \eta_c(J/\Psi)P(V))=\frac{\tau_{B_c}}{\hbar}\Gamma(B_c\to \eta_c(J/\Psi)P(V)),
\en
where the decay width $\Gamma(B_c\to \eta_c(J/\Psi)P(V))$ for each channel is given as following
\be
\Gamma\left(B_{c} \to \eta_{c} P(V)\right) & =
& \frac{\left|G_{F} V_{c b} V_{u q}^{*} a_{1} f_{P(V)}
m_{B_{c}}^{2} F_{0}^{B_{c}\eta_c}(m^2_{P(V)})\right|^{2}}{32 \pi
m_{B_{c}}}\left(1-r_{\eta_{c}}^{2}\right),\\
\Gamma\left(B_{c} \to J /\Psi P\right) & =
& \frac{\left|G_{F} V_{c b} V_{u q}^{*} a_{1} f_{P}
m_{B_{c}}^{2} A_{0}^{B_{c} J /\Psi}(m^2_P)\right|^{2}}{32 \pi
m_{B_{c}}}\left(1-r_{J / \psi}^{2}\right),
\en
with the subscript $q=d (s)$ in the CKM element $V_{uq}$ for the decays with $\pi, \rho (K, K^*)$ involved.
For the decays $B_c\to J/\Psi V$, the corresponding decay width is the summation of the three polarizations
\be
\Gamma\left(B_{c} \to J /\Psi V\right) & = & \frac{|\vec{p}|}{8 \pi m_{B_{c}}^{2}}\left(
|\mathcal{A}_{L}(B_{c}\to J/\Psi V)|^2+2\left|\mathcal{A}_{N}\left(B_{c}\to J
/\Psi V\right)\right|^{2}\right.\non &&\left.+2\left|\mathcal{A}_{T}\left(B_{c}\to J/\psi V\right)\right|^{2}\right),
\en
where $\vec{p}$ is the three-momentum of either of the two final states in the $B_c$ rest frame
\be
|\vec{p}|=\frac{\sqrt{\left(m_{B_{c}}^{2}-\left(m_{J / \psi}+m_{V}\right)^{2}\right)\left(m_{B_{c}}^{2}-
\left(m_{J / \psi}-m_{V}\right)^{2}\right)}}{2 m_{B_{c}}},
\en
and the three polarization amplitudes $\mathcal{A}_{L}, \mathcal{A}_{N}, and \mathcal{A}_{T}$ are given as
\be
\mathcal{A}_{L}\left(B_{c}\to J/\Psi V\right) &=& \frac{G_F}{\sqrt2}V^*_{cb}V_{uq}a_1f_Vm^2_{B_c}\frac{1}{2r_{J/\Psi}}
\left[\frac{\lambda(1,r^2_{J/\Psi},r^2_V)}{1+r_{J/\Psi}}A^{B_cJ/\Psi}_2(m^2_V) \right.\non && \left.
-(1-r^2_{J/\Psi}-r^2_{V})(1+r_{J/\Psi})A^{B_cJ/\Psi}_1(m^2_V)\right],\\
\mathcal{A}_{N}\left(B^{+}_{c} \to J/\psi V\right)&=& -\frac{G_{F}}{\sqrt{2}} V^{*}_{cb} V_{uq}a_{1} f_{V}
m_{B_{c}}^{2} r_{V}\left(1+r_{J /\Psi}\right) A_{1}^{B_{c}J/\Psi}\left(m_{V}^{2}\right), \\
\mathcal{A}_{T}\left(B_{c} \to J/\Psi V\right)&=&  -\frac{G_{F}}{\sqrt{2}} V^{*}_{cb} V_{uq} a_{1} f_{V}
m_{B_{c}}^{2} r_{V} \frac{\sqrt{\lambda\left(1, r_{J /\Psi}^{2}, r_{V}^{2}\right)}}{1+r_{J/\Psi}} V^{B_{c}J/\Psi}\left(m_{V}^{2}\right),
\en
with $\lambda\left(1, r_{J /\Psi}^{2}, r_V^{2}\right)=\left(1+r_{J / \psi}^{2}-r_V^{2}\right)^{2}-4r_{J/\Psi}^{2}$.

\begin{table}
\caption{ The CLFQM predictions for branching ratios $(10^{-3})$ of $ B_{c}$
 decays to final states containing a ground-state S-wave
charmonium ($\eta_c$ or $J/\Psi$) and a light pseudoscalar or vector meson. The first error is induced by the $B_c$ meson life time,
and the second and third uncertainties are from the decay constants of $B_c$ and charmonia.}
\setlength{\tabcolsep}{1mm}{
\begin{center}
\scriptsize
\begin{tabular}{c|c|c|c|c|c|c|c|c|c|c|c|c|c}
\hline\hline mode& This work &\cite{Qiao}&\cite{Ebert}&\cite{chang}&\cite{V}&\cite{Ivanov}&\cite{Naimuddin}&\cite{Colangelo1}&\cite{Abd}&\cite{Rui:2014tpa}&\cite{Rui:2016opu}&\cite{wwang2}&\cite{Ke13}\\
\hline
$B^{+}_{c}\to J/\psi\pi^+$&$1.97^{+0.04+0.00+0.24}_{-0.04-0.00-0.25}    $&$2.91(2.22)$&$0.61$&$2.1$&$1.3$&$1.7$&$0.34$&$1.3$&$1.1$&$2.33$&$2.6$ &$2.0$&$0.664$\\
$B^{+}_{c}\to J/\psi K^+$&$0.16^{+0.00+0.00+0.02}_{-0.00-0.00-0.02}    $&$0.22(0.16)$&$0.05$&$0.16$&$0.11$&$0.13$&$0.11$&$0.07$&$0.08$&$0.19$  &$-$  &$0.16$&$0.0527$  \\
$B^{+}_{c}\to J/\psi\rho^+$&$5.34^{+0.09+0.07+0.73}_{-0.09-0.07-0.63}  $&$8.08(6.03)$&$1.6$&$6.5$&$4.0$&$4.9$&$1.8$&$3.7$&$3.1$&$8.20$ &$-$  &$5.0$ &$-$  \\
$B^{+}_{c}\to J/\psi K^{*+}$&$0.31^{+0.01+0.00+0.04}_{-0.01-0.00-0.04}   $&$0.43(0.32)$&$0.10$&$0.35$&$0.22$&$0.28$&$0.09$&$0.2$&$0.18$&$0.48$ &$-$   &$0.29$ &$0.109$   \\
\hline \hline
$B^+_c\to \eta_{c}\pi^+$   &$2.36^{+0.04+0.00+0.06}_{-0.04-0.00-0.06}  $&$5.19(2.95)$&$0.85$&$2.2$&$2.0$&$1.9$&$0.34$&$0.26$&$1.4$&$2.98$   &$5.2$&$-$&$-$ \\
$B^+_c\to \eta_{c} K^+$   &$0.19^{+0.00+0.00+0.00}_{-0.00-0.00-0.01} $&$0.38(0.21)$&$0.07$&$0.17$&$0.13$&$0.15$&$0.03$&$0.02$&$0.11$&$0.24$  &$-$ &$-$&$-$  \\
$B^+_c\to \eta_{c}\rho^+$   &$6.01^{+0.11+0.00+0.16}_{-0.11-0.01-0.16}   $&$14.5(7.89)$&$2.1$&$5.9$&$4.2$&$4.5$&$1.06$&$0.67$&$3.3$&$9.83$   &$-$ &$-$ &$-$   \\
$B^+_c\to \eta_{c} K^{*+}$   &$0.34^{+0.01+0.00+0.01}_{-0.01-0.00-0.01} $&$0.77(0.41)$&$0.11$&$0.31$&$0.20$&$0.25$&$0.06$&$0.04$&$0.18$&$0.57$   &$-$    &$-$  &$-$   \\
\hline\hline
\end{tabular}\label{swave}
\end{center}}
\end{table}
From Tab. \ref{swave}, one can find that our predictions are consistent with the results given by the QCD sum rules \cite{V}, the
relativistic constituent quark model (RCQM) \cite{Ivanov} and the Bethe-Salpeter equation approach under the
so-called instantaneous nonrelativistic approximation \cite{chang}.
In Ref. \cite{Qiao},
the authors calculated these decays in the nonrelativistic QCD (NRQCD) approach at the next-to-leading-order (NLO) in the QCD coupling $\alpha_s$.
It is interesting that the leading-order (LO) results for these channels, except for the decay $B^+_c\to J/\Psi\rho^+$ are in
agreement with our predictions,
while the branching ratios obtain substantial enhancement after including the NLO QCD correction, which provides a large factor $K$.
We wonder whether these results will still be stable with the higher order corrections, such as the next-to-next-to-leading-order (NNLO) contributions, involved.
In Ref. \cite{Ebert}, the branching ratios are calculated in the relativistic quark model using $v/c$ expansion for $B_c$ and the
charmonium, and the obtained results are smaller than most of other predictions, including ours; for example, their results are only about one third of our predictions in most cases.
Certainly, the results given in the QCD relativistic (potential) models \cite{Colangelo1,Abd} are also small. As mentioned earlier, the results
calculated using the PQCD approach are sensitive to the types of wave functions for $B_c$ meson (the traditional zero-point wave function
 and the light-cone wave function).
For example, if taking the light-cone wave function for the $B_c$ meson, the branching ratio of the decay $B^+_c\to \eta_c\pi^+$ will reach
$(5.2^{+2.6}_{-1.4})\times10^{-3}$ \cite{Rui:2016opu}, which is much larger than $(2.98^{+1.24}_{-1.05})\times10^{-3}$ obtained using the
 traditional zero-point one. From Tab. \ref{swave}, one can find that the ratio of the branching fractions
$R_{K/\pi}\equiv\frac{Br(B_c^+\to J/\Psi K^+)}{Br(B_c^+\to J/\Psi \pi^+)}=0.081\pm0.011$, which is
consistent with the value $R_{K/\pi}=0.079\pm0.007\pm0.003$ given by the LHCb collaboration \cite{lhcb1}.

If replacing $P(V)$ with $D(D^{*})$, the branching ratios of the corresponding decays $B_c\to \eta_c(J/\Psi)D(D^{*})$ can be obtained
by their decay widths:
\be
\Gamma\left(B_{c} \to \eta_{c}D \right)  &=&
 \frac{\left(G_{F} V^{*}_{c b} V_{c d}m_{B_{c}}^{2}\right)^{2}\left(1-r_{\eta_{c}}^{2}-r_{D}^{2}\right)}{32 \pi m_{B_{c}}}\non &&
 \times|a_{1} f_{D}
 F_{0}^{B_{c} \eta_{c}}(m_{D}^2)+ a_{2} f_{\eta_{c}}
F_{0}^{B_{c} D}(m_{\eta_c}^2)|^2, \label{bcetacd}\\
\Gamma\left(B_{c} \to \eta_{c}D^{\ast} \right)  &=&
 \frac{\left(G_{F} V^{*}_{c b} V_{c d}m_{B_{c}}^{2}\right)^{2}\left(1-r_{\eta_{c}}^{2}-r_{D^{\ast}}^{2}\right)}{32 \pi m_{B_{c}}}\non &&
 \times|a_{1} f_{D^{\ast}}F_{0}^{B_{c} \eta_{c}}(m_{D^{*}}^2)+ a_{2} f_{\eta_{c}}
 A_{0}^{B_{c} D^{\ast}}(m_{\eta_c}^2)|^2,\\
\Gamma\left(B_{c} \to J /\Psi D \right) & =&
 \frac{\left(G_{F} V^{*}_{c b} V_{c d} m_{B_{c}}^{2}\right)^{2}\left(1-r_{J / \psi}^{2}-r_{D}^{2}\right)}{32 \pi m_{B_{c}}} \non &&
\times |a_{1} f_{D}A_{0}^{B_{c}J/\Psi}(m^2_D)+ a_{2} f_{J/\Psi}
 F_{0}^{B_{c}D}(m^2_{J/\Psi})|^{2}.
\en

For the decay $B_c\to J/\Psi D^*$, the corresponding decay width is the summation of the three polarizations:
\be
\Gamma\left(B_{c} \to J /\Psi D^*\right) & = & \frac{|\vec{p}|}{8 \pi m_{B_{c}}^{2}}\left(
|\mathcal{A}_{L}(B_{c}\to J/\Psi D^*)|^2+2\left|\mathcal{A}_{N}\left(B_{c}\to J
/\Psi D^*\right)\right|^{2}\right.\non &&\left.+2\left|\mathcal{A}_{T}\left(B_{c}\to J/\Psi D^*\right)\right|^{2}\right),
\en
where the three polarization amplitudes $\mathcal{A}_{L}, \mathcal{A}_{N}$ and $\mathcal{A}_{T}$ are given as
\be
\mathcal{A}_{L}\left(B_{c} \to  J / \Psi D^{\ast}\right)  &=&
\frac{G_F}{\sqrt2}V^*_{cb}V_{cd}m^2_{B_c}
\left\{-(1-r^2_{J/\Psi}-r^2_{D^*})\right.\non && \left.
\times\left[\frac{a_1f_{D^*}(1+r_{J/\Psi})}{2r_{J/\Psi}}A^{B_cJ/\Psi}_1(m^2_{D^*})+\frac{a_2f_{J/\Psi}(1+r_{D^*})}{2r_{D^*}}A^{B_cD^*}_1(m^2_{J/\Psi})\right]\right.\non && \left.
+\frac{a_1\lambda_1f_{D^*}}{2r_{J/\Psi}(1+r_{J/\Psi})}A^{B_cJ/\Psi}_2(m^2_{D^*})
     +\frac{a_2\lambda_2f_{J/\Psi}}{2r_{D^*}(1+r_{D^*})}A^{B_cD^*}_2(m^2_{J/\Psi})\right\},\non \\
 \mathcal{A}_{N}\left(B_{c} \rightarrow  J / \psi D^{\ast}\right)&=&  -\frac{ G_{F}}{\sqrt{2}}V^{*}_{c b} V_{c d} m_{B_{c}}^{2}[ a_{1} f_{D^{\ast}}
 r_{D^{\ast}}\left(1+r_{J / \Psi}\right) A_{1}^{B_{c} J /\Psi}\left(m_{D^{\ast}}^{2}\right)\non &&
+ a_{2} f_{J / \Psi} r_{J / \Psi}\left(1+r_{D^{\ast}}\right) A_{1}^{B_{c} D^{\ast}}\left(m_{J / \Psi}^{2}\right)], \\
\mathcal{A}_{T}\left(B_{c} \to J/\Psi D^{\ast}\right)&=& -\frac{G_{F}}{\sqrt{2}}V^{*}_{c b} V_{c d}m_{B_{c}}^{2} \left[
  \frac{a_{1}\sqrt{\lambda_1}}{1+r_{J/\Psi}}f_{D^{\ast}}r_{D^{\ast}}V^{B_{c}  J /\Psi}\left(m_{D^{\ast}}^{2}\right)\right.\non &&\left.
+  \frac{a_{2}\sqrt{\lambda_2}}{1+r_{D^{\ast}}}f_{J /\Psi} r_{J/\Psi}V^{B_{c}  D^{\ast}}\left(m_{J / \psi}^{2}\right)\right],\label{bcjpsidst}
\en
with $\lambda_1=\lambda\left(1, r_{J/\Psi}^{2}, r_{D^{\ast}}^{2}\right), \lambda_2=\lambda\left(1, r_{D^{\ast}}^{2}, r_{J /\Psi}^{2}\right)$.
As for the decay widths of the decays $B_c\to \eta_c(J/\Psi)D_s(D^{*}_s)$, they
can be obtained by performing the replacements $D\rightarrow D_s, D^{*}\rightarrow D^{*}_s, V_{cd}\rightarrow V_{cs}$ in Eqs.(\ref{bcetacd})$-$(\ref{bcjpsidst}). The calculation results are listed in Tab. \ref{etacjpsiD}.
One can find that our predictions are a little larger than most of other results, but are smaller the
the PQCD calculations. The branching ratios of the decays with $D^{(*)}_s$ involved are at least one order
larger than those of the corresponding decays with $D^{(*)}$ involved. This is because the CKM matrix element $V_{cs}$ associated with
the former is much larger than $V_{cd}$ associated with the latter. All of these decays with the ground-state S-wave  charmonia involved have large branching ratios, which lie in the range of
$10^{-4}\sim10^{-3}$ and can be detected by the present LHCb experiments.
In Ref. \cite{aajj3}, if assuming that the spectator diagram dominates and that
factorization holds, one can obtain the approximations \be
R_{D^+_s/\pi^+}&\equiv&\frac{\Gamma(B^+_c\to J/\Psi
D^+_s)}{\Gamma(B^+_c\to J/\Psi \pi^+)}
\approx\frac{\Gamma(B^+_c\to \bar D^* D^+_s)}{\Gamma(B^+_c\to \bar D^* \pi^+)},\\
R_{D^{*+}_s/D^+_s}&\equiv&\frac{\Gamma(B^+_c\to J/\Psi
D^{*+}_s)}{\Gamma(B^+_c\to J/\Psi D^+_s)}
\approx\frac{\Gamma(B^+_c\to \bar D^* D^{*+}_s)}{\Gamma(B^+_c\to
\bar D^* D^+_s)}, \en
which were measured as
$R_{D^+_s/\pi^+}=2.90\pm0.57\pm0.24$ and
$R_{D^{*+}_s/D^+_s}=2.37\pm0.56\pm0.10$ by the LHCb collaboration \cite{aajj3}, and
given as $R_{D^+_s/\pi^+}=2.76\pm0.47$ and
$R_{D^{*+}_s/D^+_s}=1.93\pm0.26$ by ATLAS \cite{atlas}. From our
calculations, these two ratios are obtained as \be
R_{D^+_s/\pi^+}&\equiv&\frac{\Gamma(B^+_c\to J/\Psi D^+_s)}{\Gamma(B^+_c\to J/\Psi \pi^+)}=3.09\pm1.05,\\
R_{D^{*+}_s/D^+_s}&\equiv&\frac{\Gamma(B^+_c\to J/\Psi D^{*+}_s)}{\Gamma(B^+_c\to J/\Psi D^+_s)}=1.48\pm0.50,
\en
where the value of $R_{D^+_s/\pi^+}$ is consistent with the measurements given by LHCb and ATLAS, while
$R_{D^{*+}_s/D^+_s}$ can explain the ATLAS result within errors.

\begin{table}
\caption{ CLFQM predictions for branching ratios $(10^{-3})$ of $ B_{c}$
 decays to final states containing a ground-state S-wave
charmonium ($\eta_c$ or $J/\Psi$) and a charmed meson. The errors are induced by the same sources as in Table \ref{swave}.}
\setlength{\tabcolsep}{1.7mm}{
\begin{center}
\begin{tabular}{c|c|c|c|c|c|c|c|c|c}
\hline\hline mode& This work &\cite{chang}& \cite{V}&\cite{Ivanov}&\cite{Naimuddin}&\cite{Colangelo}&\cite{Abd}&\cite{FU}&\cite{Rui:2014tpa} \\
\hline
$B^+_c\to \eta_{c} D^+$   &$0.22^{+0.00+0.01+0.02}_{-0.00-0.01-0.00}  $&$0.012$&$0.15$&$0.19$&$0.06$&$0.05$&$0.14$&$0.10$&$0.44$  \\
$B^{+}_{c}\to J/\psi D^+$&$0.20^{+0.00+0.00+0.03}_{-0.00-0.00-0.03}   $&$0.009$&$0.09$&$0.15$&$0.04$&$0.13$&$0.09$&$0.09$&$0.28$  \\
$B^+_c\to \eta_{c} D^{*+}$   &$0.31^{+0.01+0.00+0.01}_{-0.01-0.00-0.01}  $&$0.010$&$0.10$&$0.19$&$0.07$&$0.02$&$0.13$&$0.10$&$0.58$       \\
$B^{+}_{c}\to J/\psi D^{*+}$&$0.41^{+0.01+0.00+0.05}_{-0.01-0.00-0.01}   $&$-$&$0.28$&$0.45$&$0.18$&$0.19$&$0.28$&$0.28$&$0.67$       \\
\hline
$B^+_c\to \eta_{c} D^+_{s}$   &$6.44^{+0.11+0.73+0.94}_{-0.11-0.71-0.50}  $&$0.54$&$2.8$&$4.4$&$1.79$&$5$&$2.6$&$2.5$&$12.32$       \\
$B^{+}_{c}\to J/\psi D^+_{s}$&$6.09^{+0.11+0.36+1.15}_{-0.11-0.33-0.47}   $&$0.41$&$1.7$&$3.4$&$1.15$&$3.4$&$1.5$&$2.2$&$8.05$\\
$B^+_c\to \eta_{c} D^{*+}_{s}$   &$6.97^{+0.12+0.17+0.39}_{-0.12-0.14-0.07}     $&$0.44$&$2.7$&$3.7$&$1.49$&$0.38$&$2.4$&$2.0$&$16.54$                      \\
$B^{+}_{c}\to J/\psi D^{*+}_{s}$&$9.03^{+0.04+0.02+0.34}_{-0.04-0.02-0.32}    $&$-$&$6.7$&$9.7$&$4.4$&$5.9$&$5.5$&$6.0$&$20.45$               \\
\hline\hline
\end{tabular}\label{etacjpsiD}
\end{center}}
\end{table}

Next, we consider the decays with the P-wave charmonia involved in the final states. The P-wave charmonium can be $\chi_{c0}, \chi_{c1}$
or $h_c$. The decay widths of the decays $B_c\to \chi_{c0(1)} P(V), h_c P(V)$ are given as follows
\be
\Gamma\left(B_{c} \to \chi_{c0} P(V)\right) & =
& \frac{\left|G_{F} V_{c b} V_{u q}^{*} a_{1} f_{P(V)}
m_{B_{c}}^{2} F_{0}^{B_{c}\chi_{c0}}(m^2_{P(V)})\right|^{2}}{32 \pi
m_{B_{c}}}\left(1-r_{\chi_{c0}}^{2}\right),\\
\Gamma\left(B_{c} \to \chi_{c1} P\right) & =
& \frac{\left|G_{F} V_{c b} V_{u q}^{*} a_{1} f_{P}
m_{B_{c}}^{2} A_{0}^{B_{c} \chi_{c1}}(m^2_P)\right|^{2}}{32 \pi
m_{B_{c}}}\left(1-r_{\chi_{c1}}^{2}\right), \label{chi1p}
\en
where the subscript $q=d(s)$ in the CKM element $V_{uq}$ for the decays with $\pi,\rho(K, K^*)$ involved.
For the decays $B_c\to \chi_{c1}V$, the corresponding decay widths are the summation of the three polarizations
\be
\Gamma\left(B_{c} \to \chi_{c1} V\right) & = & \frac{|\vec{p}|}{8 \pi m_{B_{c}}^{2}}\left(
|\mathcal{A}_{L}(B_{c}\to \chi_{c1} V)|^2+2\left|\mathcal{A}_{N}\left(B_{c}\to \chi_{c1}
 V\right)\right|^{2}\right.\non &&\left.+2\left|\mathcal{A}_{T}\left(B_{c}\to \chi_{c1}V\right)\right|^{2}\right),\label{chi1v}
\en
where
\be
\mathcal{A}_{L}\left(B_{c}\to \chi_{c1} V\right) &=& \frac{G_F}{\sqrt2}V^*_{cb}V_{uq}a_1f_Vm^2_{B_c}\frac{1}{2r_{\chi_{c1}}}
\left[\frac{\lambda(1,r^2_{\chi_{c1}},r^2_V)}{1-r_{\chi_{c1}}}V^{B_c\chi_{c1}}_2(m^2_V) \right.\non && \left.
-(1-r^2_{\chi_{c1}}-r^2_{V})(1-r_{\chi_{c1}})V^{B_c\chi_{c1}}_1(m^2_V)\right],\\
\mathcal{A}_{N}\left(B^{+}_{c} \to \chi_{c1} V\right)&=& -\frac{G_{F}}{\sqrt{2}} V^{*}_{cb} V_{uq}a_{1} f_{V}
m_{B_{c}}^{2} r_{V}\left(1-r_{\chi_{c1}}\right) V_{1}^{B_{c}\chi_{c1}}\left(m_{V}^{2}\right), \\
\mathcal{A}_{T}\left(B_{c} \to \chi_{c1} V\right)&=&  -\frac{G_{F}}{\sqrt{2}} V^{*}_{cb} V_{uq} a_{1} f_{V}
m_{B_{c}}^{2} r_{V} \frac{\sqrt{\lambda\left(1, r_{\chi_{c1}}^{2}, r_{V}^{2}\right)}}{1-r_{\chi_{c1}}} A^{B_{c}\chi_{c1}}\left(m_{V}^{2}\right).
\en

It is noted that the analytic formulae of the decay widths between the decays $B_c \to h_c P(V)$ and
$B_c \to \chi_{1c} P(V)$ are similar.
Summing the branching fractions of the these decays in Tab. \ref{chi012pik}, we find that the
results of the decays with $\pi^+(\rho^+)$ involved are about one order of magnitude larger compared with those of the decays with
$K^+(K^{*+})$ involved. The difference  mainly comes from the the CKM matrix elements: the former involve a larger factor $V_{ud}\sim1$,
while the latter is associated with a smaller factor $V_{us}=\lambda\sim0.225$. Our predictions are comparable to
 most other theoretical
results, such as the QCD-motivated RQM based on the quasi-potential approach \cite{R.N}, the NRQM
\cite{Hernandez}, the RCQM \cite{M.A}.
The branching ratios of the decays $B_c\to \chi_{c0(1)} P(V)$
predicted by most works have a common property: $Br(B_c\to \chi_{c0} P(V))$ are much larger than $Br(B_c\to \chi_{c1} P(V))$.
This characteristic can be tested by the present LHCb experiments.
\begin{table}[htb]
\caption{ The CLFQM predictions for branching ratios $(10^{-3})$ of $ B_{c}$
 decays to final states containing a P-wave
charmonium  and a light pseudoscalar or vector meson. The errors are induced by the same sources as in Table \ref{swave}.}
\setlength{\tabcolsep}{0.3mm}{
\begin{center}
\begin{tabular}{c|c|c|c|c|c|c|c|c|c}
\hline\hline mode& This work &\cite{R.N}&\cite{Hernandez}&\cite{M.A}&\cite{Zong}&\cite{O.N}&\cite{Z.H}&\cite{Zhu}&\cite{Rui:2017pre}\\
\hline
$B^{+}_{c}\to \chi_{c0}\pi^{+}$&$0.66^{+0.01+0.01+0.08}_{-0.01-0.01-0.07} $&$0.21$&$0.26$&$0.55$&$0.28$&$9.8$&$0.31$&$4.2$&$1.6$ \\
$B^{+}_{c}\to\chi_{c1}\pi^+$&$0.13^{+0.00+0.00+0.00}_{-0.00-0.00-0.00}   $&$0.2$&$0.0014$&$0.068$&$0.07$&$0.089$&$0.021$&$0.05$&$0.51$  \\
$B^{+}_{c}\to h_{c}\pi^{+}$&$0.96^{+0.02+0.02+0.13}_{-0.02-0.02-0.12}    $&$0.46$&$0.53$&$1.1$&$0.5$&$16$&$0.98$&$6.2$&$0.54$   \\
\hline
$B^{+}_{c}\to\chi_{c0}\rho^+$&$1.69^{+0.00+0.02+0.20}_{-0.00-0.02-0.19} $&$0.58$&$0.67$&$1.3$&$0.72$&$33$&$0.76$&$-$&$5.8$     \\
$B^{+}_{c}\to\chi_{c1}\rho^+$&$0.43^{+0.01+0.00+0.01}_{-0.01-0.00-0.01}  $&$0.15$&$0.1$&$0.29$&$0.29$&$4.6$&$0.23$&$1.47$&$2.8$      \\
$B^{+}_{c}\to h_{c}\rho^+$&$2.42^{+0.04+0.16+0.19}_{-0.04-0.09-0.42}      $&$1.0$&$1.3$&$2.5$&$1.2$&$53$&$2.2$&$1.24$&$2.3$       \\
\hline\hline
$B^{+}_{c}\to\chi_{c0}K^+$&$0.052^{+0.001+0.001+0.006}_{-0.001-0.001-0.006}$&$0.016$&$0.02$&$0.042$&$0.0021$&$-$&$0.023$&$0.32$&$0.12$   \\
$B^{+}_{c}\to\chi_{c1}K^+$&$0.010^{+0.000+0.000+0.000}_{-0.000-0.000-0.000} $&$0.015$&$0.00011$&$0.0051$&$0.00052$&$-$&$0.0016$&$0.004$&$0.038$    \\
$B^{+}_{c}\to h_{c}K^+$&$0.075^{+0.001+0.001+0.010}_{-0.001-0.001-0.001}     $&$0.035$&$0.041$&$0.083$&$0.0038$&$-$&$0.074$&$0.47$&$0.043$     \\
\hline
$B^{+}_{c}\to\chi_{c0}K^{*+}$&$0.096^{+0.002+0.001+0.011}_{-0.002-0.001-0.011} $&$0.04$&$0.037$&$0.07$&$0.0039$&$-$&$0.045$&$-$&$0.33$     \\
$B^{+}_{c}\to\chi_{c1}K^{*+}$&$0.027^{+0.001+0.000+0.000}_{-0.001-0.000-0.000}$&$0.01$&$0.0073$&$0.018$&$0.0018$&$-$&$0.017$&$0.0707$&$0.18$      \\
$B^{+}_{c}\to h_{c}K^{*+}$&$0.13^{+0.00+0.00+0.02}_{-0.00-0.01-0.01}     $&$0.07$&$0.071$&$0.13$&$0.0068$&$-$&$0.13$&$0.0618$&$0.13$      \\
\hline\hline
\end{tabular}\label{chi012pik}
\end{center}}
\end{table}

If replacing $P(V)$ with $D(D^{*})$ in the upper decays, the branching ratios of the corresponding decays $B_c\to \chi_{c0(1)}(h_c)D(D^{*})$
can be obtained
by their decay widths:
\be
\Gamma\left(B_{c} \to \chi_{c0}D \right)  &=&
 \frac{\left(G_{F} V^{*}_{c b} V_{c d}m_{B_{c}}^{2}\right)^{2}\left(1-r_{\chi_{c0}}^{2}-r_{D}^{2}\right)}{32 \pi m_{B_{c}}}\non &&
 \times|a_{1} f_{D}
 F_{0}^{B_{c} \chi_{c0}}(m_{D}^2)+ a_{2} f_{\chi_{c0}}
F_{0}^{B_{c} D}(m_{\chi_{c0}}^2)|^2, \label{bcchi0d}\\
\Gamma\left(B_{c} \to \chi_{c0}D^{\ast} \right)  &=&
 \frac{\left(G_{F} V^{*}_{c b} V_{c d}m_{B_{c}}^{2}\right)^{2}\left(1-r_{\chi_{c0}}^{2}-r_{D^{\ast}}^{2}\right)}{32 \pi m_{B_{c}}}\non &&
 \times|a_{1} f_{D^{\ast}}F_{0}^{B_{c} \chi_{c0}}(m_{D^{*}}^2)+ a_{2} f_{\chi_{c0}}
 A_{0}^{B_{c} D^{\ast}}(m_{\chi_{c0}}^2)|^2,
 \en
 \be
 \Gamma\left(B_{c} \to \chi_{c1}D\right)  &=&
 \frac{\left(G_{F} V^{*}_{c b} V_{c d}m_{B_{c}}^{2}\right)^{2}\left(1-r_{\chi_{c1}}^{2}-r_{D}^{2}\right)}{32 \pi m_{B_{c}}}\non &&
 \times|a_{1} f_{D}
V_{0}^{B_{c} \chi_{c1}}(m_{D}^2)+ a_{2} f_{\chi_{c1}}
F_{0}^{B_{c} D}(m_{\chi_{c1}}^2)|^2. \label{chi1d}
\en

\begin{table}
\caption{CLFQM predictions for branching ratios $(10^{-3})$ of $ B_{c}$
 decays to final states containing a P-wave
charmonium ($\chi_{c0}, \chi_{c1}$ or $h_c$) and a light pseudoscalar or vector meson. The errors are induced by the same sources as in Table \ref{swave}.}
\begin{center}
\begin{tabular}{c|c|c}
\hline\hline mode& This work &\cite{hffu} \\
\hline
$B^{+}_{c}\to\chi_{c0}D^+ $&$0.075^{+0.001+0.001+0.001}_{-0.001-0.001-0.001}$&$0.033$\\
$B^{+}_{c}\to\chi_{c1} D^+$&$0.019^{+0.000+0.000+0.001}_{-0.000-0.000-0.001}$&$0.14\times10^{-3}$ \\
$B^{+}_{c}\to h_{c} D^+$&$0.075^{+0.001+0.001+0.011}_{-0.001-0.001-0.010}   $&$0.066$ \\
\hline
$B^{+}_{c}\to\chi_{c0}D^+_{s} $&$2.29^{+0.04+0.02+0.34}_{-0.04-0.07-0.23} $&$0.88$ \\
$B^{+}_{c}\to\chi_{c1} D^+_{s}$&$0.60^{+0.01+0.03+0.06}_{-0.01-0.04-0.01}$&$0.20\times10^{-2}$ \\
$B^{+}_{c}\to h_{c} D^+_{s}$&$2.24^{+0.04+0.02+0.33}_{-0.04-0.04-0.29}    $&$1.58$ \\
\hline
$B^{+}_{c}\to\chi_{c0}D^{*+} $&$0.11^{+0.00+0.00+0.01}_{-0.00-0.00-0.01}$&$0.042$    \\
$B^{+}_{c}\to\chi_{c1} D^{*+}$&$0.054^{+0.001+0.000+0.000}_{-0.001-0.000-0.000}$&$0.026$   \\
$B^{+}_{c}\to h_{c} D^{*+}$&$0.11^{+0.00+0.00+0.03}_{-0.00-0.00-0.01}    $&$0.071$  \\
\hline
$B^{+}_{c}\to\chi_{c0}D^{*+}_{s} $&$2.49^{+0.04+0.02+0.32}_{-0.04-0.33-0.30} $&$0.84$       \\
$B^{+}_{c}\to\chi_{c1} D^{*+}_{s}$&$1.23^{+0.02+0.00+0.07}_{-0.02-0.01-0.07}$&$0.49$           \\
$B^{+}_{c}\to h_{c} D^{*+}_{s}$&$2.41^{+0.04+0.02+0.34}_{-0.04-0.02-0.32}   $&$1.34$          \\
\hline\hline
\end{tabular}\label{chiD}
\end{center}
\end{table}
For the decay $B_c\to \chi_{c1} D^*$, the corresponding decay width is the summation of the three polarizations:
\be
\Gamma\left(B_{c} \to \chi_{c1} D^*\right) & = & \frac{|\vec{p}|}{8 \pi m_{B_{c}}^{2}}\left(
|\mathcal{A}_{L}(B_{c}\to \chi_{c1} D^*)|^2+2\left|\mathcal{A}_{N}\left(B_{c}\to \chi_{c1}
 D^*\right)\right|^{2}\right.\non &&\left.+2\left|\mathcal{A}_{T}\left(B_{c}\to \chi_{c1} D^*\right)\right|^{2}\right),\label{chi1ds}
\en
where the three polarization amplitudes $\mathcal{A}_{L}, \mathcal{A}_{N} $ and $ \mathcal{A}_{T}$ are given as
\be
\mathcal{A}_{L}\left(B_{c} \to  \chi_{c1} D^{\ast}\right)  &=&
\frac{G_F}{\sqrt2}V^*_{cb}V_{cd}m^2_{B_c}
\left\{-(1-r^2_{\chi_{c1}}-r^2_{D^*})\right.\non && \left.
\times\left[\frac{a_1f_{D^*}(1-r_{\chi_{c1}})}{2r_{\chi_{c1}}}V^{B_c\chi_{c1}}_1(m^2_{D^*})+\frac{a_2f_{\chi_{c1}}(1+r_{D^*})}{2r_{D^*}}A^{B_cD^*}_1(m^2_{\chi_{c1}})\right]\right.\non && \left.
+\frac{a_1\lambda_1f_{D^*}}{2r_{\chi_{c1}}(1-r_{\chi_{c1}})}V^{B_c\chi_{c1}}_2(m^2_{D^*})
+\frac{a_2\lambda_2f_{\chi_{c1}}}{2r_{D^*}(1+r_{D^*})}A^{B_cD^*}_2(m^2_{\chi_{c1}})\right\},\;\;\;\\
 \mathcal{A}_{N}\left(B_{c} \to \chi_{c1} D^{\ast}\right)&=&  -\frac{ G_{F}}{\sqrt{2}}V^{*}_{c b} V_{c d}
m_{B_{c}}^{2}[ a_{1} f_{D^{\ast}}
 r_{D^{\ast}}\left(1-r_{\chi_{c1}}\right) V_{1}^{B_{c} \chi_{c1}}\left(m_{D^{\ast}}^{2}\right)\non &&
+ a_{2} f_{\chi_{c1}} r_{\chi_{c1}}\left(1+r_{D^{\ast}}\right) A_{1}^{B_{c} D^{\ast}}\left(m_{\chi_{c1}}^{2}\right)],
\en
\be
\mathcal{A}_{T}\left(B_{c} \to \chi_{c1} D^{\ast}\right)&=& -\frac{G_{F}}{\sqrt{2}}V^{*}_{c b} V_{c d}m_{B_{c}}^{2} \left[
  \frac{a_{1}\sqrt{\lambda_1}}{1-r_{\chi_{c1}}}f_{D^{\ast}}r_{D^{\ast}}A^{B_{c}  \chi_{c1}}\left(m_{D^{\ast}}^{2}\right)\right.\non &&\left.
+  \frac{a_{2}\sqrt{\lambda_2}}{1+r_{D^{\ast}}}f_{\chi_{c1}} r_{\chi_{c1}}V^{B_{c}  D^{\ast}}\left(m_{\chi_{c1}}^{2}\right)\right],
\label{bcchi1dst}
\en
with $\lambda_1=\lambda\left(1, r_{\chi_{c1}}^{2}, r_{D^{\ast}}^{2}\right), \lambda_2=\lambda\left(1, r_{D^{\ast}}^{2}, r_{\chi_{c1}}^{2}\right)$.
As to the decay widths of the channels $B_c\to \chi_{c0}(\chi_{c1})D_s(D^{*}_s)$, they
can be obtained by performing the replacements $D\to D_s, D^{*}\rightarrow D^{*}_s, V_{cd}\rightarrow V_{cs}$ in
Eqs. (\ref{bcchi0d})$-$(\ref{bcchi1dst}).
The branching ratios of the these decays are given in
Tab. \ref{chiD}, where we also list the results given by  the Salpeter method \cite{hffu}. This method is the relativistic instantaneous
approximation of the original Bethe-Salpeter equation.

\begin{table}
\caption{CLFQM predictions for branching ratios $(\times10^{-3})$ of the decays $B_{c} \to X(3872)M$, where $M$ represents a light pseudoscalar, a
vector meson, or a charmed meson. The errors are induced by the same sources as in Table \ref{swave}.}
\begin{center}
\begin{tabular}{c|c|c|c|c}
\hline\hline mode& This work &\cite{wwang2}&\cite {Zhang:2022xqs}&\cite{Hsiao:2016pml}\\
\hline
$B^{+}_{c}\to X(3872)\pi^{+}$&$0.25^{+0.00+0.01+0.03}_{-0.00-0.01-0.01}  $&$0.17$&$0.27$&$0.06$ \\
$B^{+}_{c}\to X(3872)K^{+}$&$0.020^{+0.000+0.001+0.000}_{-0.000-0.001-0.000} $&$0.013$&$0.025$&$0.0047$ \\
$B^{+}_{c}\to X(3872)\rho^{+}$&$0.63^{+0.01+0.03+0.03}_{-0.01-0.03-0.08}$&$0.41$&$-$&$-$    \\
$B^{+}_{c}\to X(3872)K^{*+}$&$0.036^{+0.001+0.001+0.004}_{-0.001-0.002-0.002}   $  &$0.024$&$-$&$-$     \\
$B^{+}_{c}\to X(3872)D^+$&$0.033^{+0.001+0.002+0.001}_{-0.001-0.002-0.000}  $      &$-$ & &  \\
$B^{+}_{c}\to X(3872)D^{*+}$&$0.078^{+0.001+0.007+0.009}_{-0.001-0.007-0.006}  $   &$-$  & &      \\
$B^{+}_{c}\to X(3872)D^+_{s}$&$1.00^{+0.02+0.15+0.15}_{-0.02-0.15-0.10}   $ &$-$ & &     \\
$B^{+}_{c}\to X(3872)D^{*+}_{s}$&$1.78^{+0.03+0.07+0.12}_{-0.03-0.07-0.08}$ &$-$ & &           \\
\hline\hline
\end{tabular}\label{x3872}
\end{center}
\end{table}
Though the $X(3872)$ has been confirmed by many experimental collaborations, such as
CDF \cite{cdf}, D0 \cite{d0}, Babar \cite{babar0} and LHCb \cite{lhcb}, with quantum numbers $J^{PC}=1^{++}$ and isospin $I=0$, there are
still many uncertainties. Though many different exotic hadron state interpretations, such as
a loosely bound molecular state \cite{close, wong, braaten0, swanson, zhusl}, a compact tetraquark state \cite{chiu, barnea, maiani1, hogaasen},
$c\bar c g$ hybrid meson \cite{close2, li}, glueball \cite{li}, have been put forward, the first raidal excitation of $1P$ charmonium
state $\chi_{c1}(1P)$ as the most natural assignment has not been ruled out \cite{barnes, eichten, quigg}.
By assuming the $X(3872)$ as a $1^{++}$ charmonium state, we calculate the branching ratios of the decays $B^{+}_{c}\to X(3872)M$
(here, $M$ represents a light pseudoscalar, a vector meson, or a charmed meson). The analytic expressions of the
corresponding decay widths are similar to those of
the decays $B_c\to \chi_{c1} M$ listed in Eqs. (\ref{chi1p}), (\ref{chi1v}), (\ref{chi1d}) and (\ref{chi1ds}).
 In Tab. \ref{x3872}, we list the branching fractions of the decays $B_{c}\to X(3872)M$.
One can find that our predictions for the decays
$B^{+}_{c}\to X(3872)\pi^+(K^+)$ are consistent with the results given in the PQCD approach \cite{Zhang:2022xqs}, while they are much larger than
those calculated in the generalized factorization (GF) approach \cite{Hsiao:2016pml}. Certainly, some of these decays studied using
the CLFQM about 15 years ago \cite{wwang2}, the differences between our predictions and the previous calculations are induced by taking different values for some parameters.

Lastly, we turn to the branching ratios of the decays with the radially excited S-wave charmonia, such as $\eta_c(2S,3S)$
and $\psi(3S,3S)$, involved in the final states. The correspond decay widths are similar to those of the decays
$B_c\to \eta_cM, J/\Psi M$, where $M$ represents a light pseudoscalar, a vector meson, or a charmed meson ($D^{(*)},D^{(*)}_s$).
As we know, in order to compare with experiments, the ratios
\be
R_{\psi(2S)/J/\Psi}&\equiv& Br(B_c\to \psi(2S)\pi)/Br(B_c\to J/\Psi\pi),\\
R_{\eta_c(2S)/\eta_c}&\equiv& Br(B_c\to \eta_c(2S)\pi)/Br(B_c\to
\eta_c\pi), \en
are often used.
If we still employ the traditional light-front wave functions for the radially excited chuarmonia given in Eq.(\ref{betap}), we will get larger branching
ratios than most other theoretical predictions; even worse, the obtained value $R_{\psi(2S)/J/\Psi}=0.467$ is much larger than the experimental data
$R_{\psi(2S)/J/\Psi}=0.268\pm0.032\pm0.007\pm0.006$ given by PDG \cite{pdg22}.
There exists a similar case in Ref.\cite{Ke13}, so we follow the same
strategy by choosing the modified harmonic oscillator wave functions:
\be
\varphi^{\prime}_{2S}&=&4\left(\frac{\pi}{\beta^{\prime 2}}\right)^{\frac{3}{4}}
\sqrt{\frac{d p_{z}^{\prime}}{d x_{2}}} \exp
\left(-\frac{2^\delta}{2}\frac{p_{z}^{\prime 2}+p_{\perp}^{\prime 2}}{ \beta^{\prime 2}}\right)
(a_2-b_2\frac{p_{z}^{\prime 2}+p_{\perp}^{\prime 2}}{\beta^{\prime 2}}), \label{psi2s}\\
\varphi^{\prime}_{3S}&=&4\left(\frac{\pi}{\beta^{\prime 2}}\right)^{\frac{3}{4}}
\sqrt{\frac{d p_{z}^{\prime}}{d x_{2}}} \exp
\left(-\frac{3^\delta}{2}\frac{p_{z}^{\prime 2}+p_{\perp}^{\prime 2}}{ \beta^{\prime 2}}\right)
(a_3-b_3\frac{p_{z}^{\prime 2}+p_{\perp}^{\prime 2}}{\beta^{\prime 2}}
+c_3\frac{(p_{z}^{\prime 2}+p_{\perp}^{\prime 2})^2}{\beta^{\prime 4}}).\label{psi3s}
\en

In order to keep the orthogonality and normalization for the wave functions of these radially excited states, one needs to
introduce a factor $n^\delta$ into the exponential functions in these wave functions, which can be determined by fitting the data of the
corresponding decay constants. Similarly, there exists a $1/n$ exponential dependence factor in the wave functions
 of the hydrogen-like atoms, which are obtained by solving the Schr$\ddot{o}$dinger equation. In Ref. \cite{Ke13}, the authors
 supposed that the parameters shown in Eqs. (\ref{psi2s}) and (\ref{psi3s}) are the same as those for $\Upsilon(2S)$ and
 $\Upsilon(3S)$ under the heavy quark effective theory, which are given as \cite{Ke:2010vn}
\be
 a_2&=&1.88684, \;\;b_2=1.54943,\;\; \delta=1/1.82,\non
 a_3&=&2.53764,\;\; b_3=5.67431, \;\;c_3=1.85652 \;\;\;\text{(scenario I)}. \label{si}
\en

It is noted that these parameters are determined by assuming that the $\Upsilon(iS)(i=2,3)$ mesons have the same $\beta^\prime$ values
as that of  $\beta'_{\Upsilon}$ for $\Upsilon(1S)$. In fact, under this
assumption, once the
value of $\delta$ is fixed, these parameters given in Eq.(\ref{si}) can be determined using
the orthogonality and normalization for the wave functions of these ground and radially excited states.
That is to say,
if we only replace
$\beta^\prime_{\Upsilon}$ with $\beta^\prime_{J/\Psi}$, the values of parameters $a_{2,3},b_{2,3},c_3$ are not changed. We call this case as scenario
I (SI). As another possibility, we also assume here that each value of $\beta^\prime$ in the wave functions of $J/\Psi$
and $\psi(2S, 3S)$ is different but with $\delta=1/1.82$ fixed, then we can get another group of values for these parameters:
\be
 a_2&=&1.99718,\;\; b_2=1.48536, \;\;\delta=1/1.82,\non
 a_3&=&3.00375, \;\;b_3=5.54952,\;\; c_3=1.49566 \;\;\;\text{(scenario II)}, \label{s2}
\en
\begin{table}[h]
\caption{CLFQM predictions for branching ratios ($\times10^{-4}$) of the decays
$B_c\to\psi(2S)P(V)$, $\eta_c(2S)P(V)$ with $P(V)$ representing a
light pseudoscalar (vector) meson. For each decay channel, we
calculate both in scenario I (upper line) and scenario II (lower line).
The first two errors for these entries correspond to the
uncertainties from the lifetime and the decay constant
 of the initial meson $B_{c}$, respectively. The last one is from the parameter $\delta$ in the
 modified wave functions of the radially excited charmonium $\psi(2S)$ or $\eta_c(2S)$.}
\scriptsize
\setlength{\tabcolsep}{1.1mm}{
\begin{center}
\begin{tabular}{c|c|c|c|c|c|c|c|c|c|c|c|c}
\hline\hline mode& This work &\cite{Bediaga:2011cs}&\cite{Nayak:2022qaq}&\cite{Chao:1997c}&\cite{chang}&\cite{Colangelo1}&\cite{Chang:2014jca}&\cite{Zhou:2020bnm}&\cite{Ke13}&\cite{Rui:2015iia}&\cite{Rui:2016opu}&\cite{Galkin}\\
\hline
$B^{+}_{c}\to \psi(2S)\pi^+$&$4.18^{+0.07+0.02+1.42}_{-0.07-0.02-1.13} $&$3.7$&$1.39$&$0.63$&$2.2$&$2.0$&$2.66$&$1.42$&$2.97$ &$6.7$ &$4.8$ &$1.1$\\
$$&$2.52^{+0.04+0.01+0.99}_{-0.04-0.01-0.75} $&$$&$$&$$&$$&$$&$$&$$&$$ &$$&$$ &$$\\
$B^{+}_{c}\to \psi(2S) K^+$&$0.33^{+0.01+0.00+0.11}_{-0.01-0.00-0.09} $&$0.29$&$0.109$&$0.04$&$0.16$&$0.089$&$-$&$0.102$&$0.23$ &$-$ &$-$ &$0.1$\\
$$&$0.20^{+0.01+0.00+0.08}_{-0.01-0.00-0.06} $&$$&$$&$$&$$&$$&$$&$$&$$ &$$&$$ &$$\\
\hline
$B^{+}_{c}\to \psi(2S)\rho^+$&$11.93^{+0.21+0.052+4.04}_{-0.21-0.052-3.14}$&$11$&$-$&$1.6$&$6.3$&$4.8$&$6.83$&$-$&$-$ &$-$   &$-$  &$1.8$\\
$$&$7.18^{+0.13+0.04+2.84}_{-0.13-0.04-2.15}$&$$&$$&$$&$$&$$&$$&$$&$$ &$$&$$ &$$\\
$B^{+}_{c}\to \psi(2S) K^{*+}$&$0.70^{+0.012+0.024+0.21}_{-0.012-0.014-0.20} $&$0.57$&$-$&$0.081$&$0.34$&$0.27$&$41$&$-$&$-$   &$-$  &$-$ &$0.098$ \\
$$&$0.42^{+0.01+0.01+0.15}_{-0.01-0.08-0.13} $&$$&$$&$$&$$&$$&$$&$$&$$ &$$&$$ &$$\\
\hline
$B^+_c\to \eta_{c}(2S)\pi^+$   &$3.35^{+0.06+0.02+1.15}_{-0.06-0.02-0.89}$&$2.4$&$1.44$&$2.2$&$2.4$&$0.66$&$2.87$&$1.67$&$-$&$10.3$ &$14$ &$1.7$\\
$$&$0.46^{+0.01+0.01+0.31}_{-0.01-0.01-0.21}$&$$&$$&$$&$$&$$&$$&$$&$$ &$$&$$ &$$\\
$B^+_c\to \eta_{c}(2S) K^+$   &$0.27^{+0.01+0.00+0.09}_{-0.01-0.00-0.07}$&$0.18$&$0.117$&$0.16$&$0.18$&$0.049$&$-$&$0.119$&$-$&$-$   &$-$ &$0.1$  \\
$$&$0.036^{+0.001+0.000+0.025}_{-0.001-0.001-0.017}$&$$&$$&$$&$$&$$&$$&$$&$$ &$$&$$ &$$\\
\hline
$B^+_c\to \eta_{c}(2S)\rho^+$&$8.54^{+0.15+0.04+2.93}_{-0.15-0.05-2.27}$&$5.5$&$12.46$&$5.25$&$5.5$&$1.4$&$-$&$0.356$&$-$ &$-$  &$-$  &$3.6$\\
$$&$1.17^{+0.02+0.01+0.79}_{-0.02-0.02-0.53}$&$$&$$&$$&$$&$$&$$&$$&$$ &$$&$$ &$$\\
$B^+_c\to \eta_{c}(2S) K^{*+}$ &$0.49^{+0.01+0.00+0.17}_{-0.01-0.00-0.13}$&$0.26$&$0.239$&$0.25$&$0.28$&$0.0715$&$-$&$0.191$&$-$   &$-$    &$-$  &$0.2$   \\
$$&$0.067^{+0.001+0.001+0.045}_{-0.001-0.001-0.030}$&$$&$$&$$&$$&$$&$$&$$&$$ &$$&$$ &$$\\
\hline\hline
\end{tabular}\label{2spv}
\end{center}}
\end{table}
\begin{table}[h]
\caption{CLFQM predictions for branching ratios ($\times10^{-4}$) of the decays $B_c\to\psi(2S)D^+_{(s)}(D^{*+}_{(s)}), \eta_c(2S)D^+_{(s)}(D^{*+}_{(s)})$.
 For each decay channel, we calculate in SI (upper line) and SII
(lower line). The errors for these entries are the same with those in Table \ref{2spv}.}
\setlength{\tabcolsep}{1mm}{
\begin{center}
\begin{tabular}{c|c|c|c|c|c|c|c}
\hline\hline mode& This work &\cite{Zhou:2020bnm}&\cite{Ke13}&\cite{Bediaga:2011cs}&\cite{Colangelo1}&\cite{Nayak:2022qaq}&\cite{chang} \\
\hline
$B^{+}_{c}\to \psi(2S) D^+$&$0.71^{+0.01+0.00+0.23}_{-0.01-0.00-0.18} $&$0.0156$&$0.138$&$0.24$&$0.073$&$0.07(0.11)$&$0.11$  \\
$$&$0.44^{+0.00+0.00+0.17}_{-0.00-0.00-0.13} $&$$&$$&$$&$$&$$&$$  \\
$B^{+}_{c}\to \psi(2S) D^{*+}$&$0.92^{+0.16+0.05+0.27}_{-0.16-0.01-0.21}$&$1.29$&$0.42$&$-$&$0.052$&$-$&$-$       \\
$$&$0.61^{+0.01+0.00+0.20}_{-0.01-0.00-0.16}$&$$&$$&$$&$$&$$&$$  \\
$B^{+}_{c}\to \psi(2S) D^+_{s}$&$10.92^{+0.19+1.07+2.67}_{-0.19-0.97-3.85} $&$2.69$&$3.08$&$5.25$&$1.2$&$2.57(3.94)$&$4.4$     \\
$$&$6.67^{+0.12+0.84+1.79}_{-0.12-0.75-2.71} $&$$&$$&$$&$$&$$&$$  \\
$B^{+}_{c}\to \psi(2S) D^{*+}_{s}$&$19.25^{+0.33+0.21+6.16}_{-0.33-0.32-4.36}$&$27.2$&$8.85$&$-$&$1.7$&$-$&$-$                \\
$$&$12.60^{+0.22+0.16+4.53}_{-0.22-0.27-3.12}$&$$&$$&$$&$$&$$&$$  \\
$B^+_c\to \eta_{c}(2S) D^+$&$0.35^{+0.01+0.00+0.11}_{-0.01-0.00-0.10}$&$0.0364$&$-$&$0.057$&$0.022$&$0.161(0.21)$&$0.2$    \\
$$&$0.079^{+0.001+0.001+0.037}_{-0.001-0.001-0.027}$&$$&$$&$$&$$&$$&$$  \\
$B^+_c\to \eta_{c}(2S) D^{*+}$ &$0.50^{+0.01+0.00+0.15}_{-0.01-0.00-0.11}$&$0.292$&$-$&$0.21$&$0.00078$&$0.12(0.185)$&$0.11$        \\
$$&$0.13^{+0.00+0.00+0.05}_{-0.00-0.00-0.04}$&$$&$$&$$&$$&$$&$$  \\
$B^+_c\to \eta_{c}(2S) D^+_{s}$ &$10.62^{+0.19+0.67+0.87}_{-0.19-0.81-1.89}$&$4.46$&$-$&$0.67$&$0.785$&$5.81(7.4)$&$8.7$       \\
$$&$2.36^{+0.04+0.52+0.62}_{-0.04-0.60-0.34}$&$$&$$&$$&$$&$$&$$  \\
$B^+_c\to \eta_{c}(2S) D^{*+}_{s}$&$11.07^{+0.20+0.17+3.62}_{-0.20-0.28-2.45} $&$3.56$&$-$&$4.5$&$0.2$&$4.61(6.13)$&$4.4$           \\
$$&$2.76^{+0.05+0.08+1.34}_{-0.05-0.14-0.79} $&$$&$$&$$&$$&$$&$$  \\
\hline\hline
\end{tabular}\label{2sD}
\end{center}}
\end{table}
which are called as scenario II (SII).
In this work, we calculate in these two scenarios for the $B_c$ decays with $\psi(2S)$ or
$\psi(3S)$ involved in the final states.
By using these modified wave functions for $\psi(2S)$, one can obtained that $R_{\psi(2S)/J/\Psi}=0.212\pm0.071$ in SI,
which is consistent with the data. At the same time, the tensions between our predictions with other theoretical results are greatly reduced. For example, the
branching ratio $Br(B_c^+\to \eta_c(2S)\pi^+)=(7.70^{+0.11+0.03+0.84}_{-0.12-0.04-0.52})\times10^{-4}$ using the traditional light-front
wave function for $\eta_c(2S)$, while $Br(B_c^+\to \eta_c(2S)\pi^+)=(3.35^{+0.06+0.02+0.89}_{-0.06-0.02-1.20})\times10^{-4}$ by
replacing with the modified wave function in SI, which are close to the results given by Refs.\cite{Bediaga:2011cs,Chao:1997c,chang,Chang:2014jca}.
This is similar for the decay $B_c^+\to \psi(2S)\pi^+$. The branching ratios of the decays
$B_c^+\to \eta_c(2S)\pi^+$ and $B_c^+\to \psi(2S)\pi^+$ in SI are close to each other; this is also supported by most of the other
theoretical predictions shown in Table \ref{2spv}.
Furthermore, the differences of the branching ratios of the decays
$B_c\to J/\Psi({2S})P(V)$ between these two scenarios
are not large, while they are very different for the decays with $\eta_c(2S)$ involved. So one can use the decay channels
$B_c\to \eta_c({2S})P(V)$
 to check which scenario is more accurate by comparing with
the future experimental data.

In Table \ref{2sD}, we calculate the branching ratios of the decays $B_c\to\psi(2S)D_{(s)}(D^{*}_{(s)}),$ $\eta_c(2S)D_{(s)}(D^{*}_{(s)})$.
From our calculations and the numerical results, we find the following points:
\begin{enumerate}
\item
The branching ratios of
the decays with $\eta_c(2S)$ involved are more sensitive to the shape
parameter $\beta'$. For example, for the decays $B^+_c\to\eta_c(2S)D^{+}_{(s)},\eta_c(2S)D^{*+}_{(s)}$, their branching ratios
in SI are about five times larger than those in SII, while for the decays
$B^+_c\to\psi(2S)D^{+}_{(s)},\psi(2S)D^{*+}_{(s)}$, the differences of the results between these two
scenarios are less than two times.
\item
The branching ratios of the decays with $D^+_s$ or $D^{*+}_s$ involved are at least one order
larger than those of the corresponding decays with $D^+$ or $D^{*+}$ involved. It is because  the former (the latter) are
suppressed (enhanced) by the CKM matrix elements.
\item
On the whole, the predictions in SII are closer to other theoretical results than those in SI, which supports that
taking different value of the shape parameter $\beta'$ for each radially excited charmonium is more reasonable.
\end{enumerate}
\begin{table}
\caption{CLFQM predictions for branching ratios ($\times10^{-5}$) of the decays
$B_c\to\psi(3S)P(V), \eta_c(3S)P(V)$, with $P(V)$ representing a
light pseudoscalar (vector) meson. For each decay channel, we
calculate both in SI (upper line) and SII (lower line). The errors for these entries are the same with those in Table \ref{2spv}.}
\begin{center}
\begin{tabular}{c|c|c|c|c}
\hline\hline mode& This work &\cite{Nayak:2022qaq}&\cite{Ti anhong}&\cite{Rui:2016opu} \\
\hline
$B^{+}_{c}\to \psi(3S)\pi^+$&$16.15^{+0.29+0.06+17.14}_{-0.29-0.07-8.72} $&$4.7(4.8)$&$3.11$&$48$ \\
$$&$1.80^{+0.03+0.00+3.30}_{-0.03-0.00-1.29}$&$$&$$&$$\\
$B^{+}_{c}\to \psi(3S) K^+$&$1.28^{+0.02+0.01+1.36}_{-0.02-0.01-0.69}$&$0.35(0.36)$&$0.214$&$-$\\
$$&$0.14^{+0.00+0.00+0.26}_{-0.00-0.00-0.10}$&$$&$$&$$\\
$B^{+}_{c}\to \psi(3S)\rho^+$&$46.47^{+0.82+0.27+49.03}_{-0.82-0.66-24.91}$&$-$&$3.35$&$-$    \\
$$&$5.15^{+4.48+0.02+9.57}_{-4.48-0.04-3.71}$&$$&$$&$$\\
$B^{+}_{c}\to \psi(3S) K^{*+}$&$2.78^{+0.05+0.04+2.87}_{-0.05-0.02-1.51} $&$-$&$0.229$&$-$  \\
$$&$0.31^{+0.01+0.00+0.56}_{-0.01-0.00-0.22} $&$$&$$&$$\\
\hline
$B^+_c\to\eta_{c}(3S)\pi^+$&$13.50^{+0.24+0.06+14.31}_{-0.24-0.06-7.29}$&$4.7(4.8)$&$2.16$&$140$\\
$$&$0.29^{+0.01+0.00+1.14}_{-0.01-0.00-0.29}$&$$&$$&$$\\
$B^+_c\to \eta_{c}(3S) K^+$&$1.07^{+0.02+0.00+1.13}_{-0.02-0.01-0.58}$&$0.38(0.39)$&$0.153$&$-$   \\
$$&$0.024^{+0.000+0.000+0.090}_{-0.000-0.000-0.023}$&$$&$$&$$\\
$B^+_c\to \eta_{c}(3S)\rho^+$&$34.36^{+0.61+0.15+36.42}_{-0.61-0.16-18.57}$&$14.9(15.5)$&$4.29$&$-$   \\
$$&$0.76^{+0.01+0.01+2.91}_{-0.01-0.01-0.73}$&$$&$$&$$\\
$B^+_c\to \eta_{c}(3S) K^{*+}$ &$1.97^{+0.04+0.01+2.08}_{-0.04-0.01-1.06}$&$0.79(0.81)$&$0.225$&$-$      \\
$$&$0.044^{+0.001+0.000+0.017}_{-0.001-0.000-0.042}$&$$&$$&$$\\
\hline\hline
\end{tabular}\label{charm3pv}
\end{center}
\end{table}
At present, only a few papers have studied $B_c$ decays with $\psi(3S)$ or $\eta_{c}(3S)$ involved, which are listed in Tables \ref{charm3pv} and \ref{charm3d}.
Most theoretical predictions show that the branching ratios of these decays are about or less than $10^{-4}$. Meanwhile, for the decay
$B_c^+\to \eta_c(3S)\pi^+$, its branching ratio was predicted as $1.4\times10^{-3}$ in the PQCD approach \cite{Rui:2016opu}, where the authors
obtained that the branching ratios of the decays $B_c\to \eta_c(2S)\pi^+$ and  $B_c^+\to \eta_c(3S)\pi^+$ are almost equivalent. Our prediction
for the branching ratio of the decay $B_c^+\to \eta_c(2S)\pi^+$ is about 2.5 times larger than that of $B_c^+\to \eta_c(3S)\pi^+$ in SI.
Certainly, the difference between $Br(B_c^+\to \eta_c(2S)\pi^+)$ and $Br(B_c^+\to \eta_c(3S)\pi^+)$ in SII is more than one order. There exists
a similar
relation between the decays $B_c^+\to \psi(2S)\pi^+$ and $B_c^+\to \psi(3S)\pi^+$. So we suggest that the ratios
$Br(B_c^+\to \eta_c(2S)\pi^+)/Br(B_c^+\to \eta_c(3S)\pi^+)$ and $Br(B_c^+\to \psi(2S)\pi^+)/Br(B_c^+\to \psi(3S)\pi^+)$ can be measured by LHCb experiments to
distinguish which shape parameters for these radially excited charmonia are more appropriate. In Ref. \cite{Ti anhong}, the authors calculated the
branching ratios of these decays by using the improved Bethe-Salpeter method. Their results of the decays $B_c\to \psi(3S)P(V)$, where
$P(V)$ represents a light pseudoscalar (vector) meson, agree with our predictions in SI within errors. Meanwhile, there exist larger
divergences for the branching ratios of the decays $B_c\to \eta_c(3S)P(V)$ with other theoretical results.
The relativistic independent quark model (RIQM) based on a flavor-independent interaction potential was used in Ref.\cite{Nayak:2022qaq},
where two groups of results corresponding to two sets of Wilson coefficients were obtained.

\begin{table}
\caption{CLFQM predictions for branching ratios ($\times10^{-5}$) of the decays $B^+_c\to\psi(3S)D^+_{(s)}(D^{*+}_{(s)}),$ $\eta_c(3S)D^+_{(s)}(D^{*+}_{(s)})$.
 For each decay channel, we calculate in scenario I (upper line) and scenario II
(lower line). The errors for these entries are the same as those in Table \ref{2spv}.}
\begin{center}
\begin{tabular}{c|c|c|c}
\hline\hline mode& This work &\cite{Nayak:2022qaq}&\cite{Zhou:2020bnm} \\
\hline
$B^{+}_{c}\to \psi(3S) D^+$&$1.68^{+0.03+0.06+1.57}_{-0.03-0.05-0.89}$&$0.02(0.092)$&$3.62\times10^{-3}$ \\
$$&$0.25^{+0.01+0.02+0.33}_{-0.01-0.02-0.17}$&$$&$$\\
$B^{+}_{c}\to \psi(3S) D^{*+}$&$2.78^{+0.05+0.01+2.61}_{-0.05-0.00-1.36} $&$-$&$3.38$     \\
$$&$0.47^{+0.01+0.00+0.60}_{-0.01-0.00-0.27} $&$$&$$\\
$B^{+}_{c}\to \psi(3S) D^+_{s}$&$50.58^{+0.89+16.39+50.71}_{-0.89-16.54-24.73} $&$6.6(7.9)$&$3.76$   \\
$$&$7.74^{+0.14+0.96+4.44}_{-0.14-0.89-1.16} $&$$&$$\\
$B^{+}_{c}\to \psi(3S) D^{*+}_{s}$&$58.38^{+1.03+1.11+54.11}_{-1.03-0.98-29.92}$&$-$&$58.9$               \\
$$&$9.42^{+0.17+0.42+2.13}_{-0.17-0.44-5.89}$&$$&$$\\
\hline
$B^+_c\to \eta_{c}(3S) D^+$&$1.35^{+0.02+0.06+1.18}_{-0.02-0.05-0.70}$&$0.21(0.36)$&$-$  \\
$$&$0.095^{+0.002+0.015+0.133}_{-0.002-0.013-0.071}$&$$&$$\\
$B^+_c\to \eta_{c}(3S) D^{*+}$ &$1.95^{+0.03+0.00+1.74}_{-0.03-0.01-0.92}$&$0.032(0.09)$&$-$     \\
$$&$0.18^{+0.00+0.00+0.24}_{-0.00-0.00-0.10}$&$$&$$\\
$B^+_c\to \eta_{c}(3S) D^+_{s}$ &$41.11^{+0.73+1.98+2.87}_{-0.73-2.39-18.31}$&$7.72(11.5)$&$-$    \\
$$&$3.33^{+0.06+0.61+0.85}_{-0.06-0.68-1.66}$&$$&$$\\
$B^+_c\to \eta_{c}(3S) D^{*+}_{s}$&$43.19^{+0.76+1.05+37.26}_{-0.76-0.89-21.37} $&$3.2(4.6)$&$-$                  \\
$$&$3.83^{+0.07+0.28+4.90}_{-0.07-0.29-2.47} $&$$&$$\\
\hline\hline
\end{tabular}\label{charm3d}
\end{center}
\end{table}

If replacing $P(V)$ with
$D_{(s)}(D^{*}_{(s)})$, we can study the branching ratios for the decays $B_c\to\psi(3S)D_{(s)}(D^{*}_{(s)}),
\eta_c(3S)D_{(s)}(D^{*}_{(s)})$, which are listed in Table \ref{charm3d}. Just as in the case of the decays
$B_c\to\psi(2S)D_{(s)}(D^{*}_{(s)}), \eta_c(2S)D_{(s)}(D^{*}_{(s)})$, the branching ratios of the decays with
$D^{(*)}_s$ involved are much larger than those of the decays with $D^{(*)}$ involved because of the CKM factors.
In addition to the RIQ model, some of these decays are also researched by using the improved
Bethe-Salpeter method \cite{Zhou:2020bnm}, where the branching ratios of the decays $B_c\to\psi(3S)D^{*+}, \eta_c(3S)D^{*+}_{s}$ are
consistent with our predictions in SI. Whereas, the branching ratio of the decay $B^+_c\to\psi(3S)D^+$ is predicted as
$3.62\times10^{-8}$ and much smaller than our result. We predict that some of the decays with $\eta_c(3S)$ or $\psi(3S)$ involved,
such as $B_c\to \eta_c(3S)\rho,B_c\to \psi(3S)D^{*}_s$, might have larger branching ratios (up to $10^{-4}$) and may be accessible
at the High Luminosity Large Hadron Collider in the near future.

Comparing Tables \ref{swave}, \ref{2spv}, and \ref{charm3pv}, one can find that there is a hierarchy for these decays:
\be
Br(B_c\to J/\Psi P(V))>Br(B_c\to \psi(2S) P(V))>Br(B_c\to \psi(3S) P(V)),\\
Br(B_c\to \eta_c P(V))>Br(B_c\to \eta_c(2S) P(V))>Br(B_c\to \eta_c(3S)P(V)),
\en
where $P(V)$ represents a light pseudoscalar (vector) meson. This is because for the decays with the higher excited charmonia involved,
the phase spaces are tighter, and the form factors are smaller and less sensitive to the change of the momentum transfer $q^2$.

{\centering\section{Summary}\label{sum}}
In this work we study the form factors of $B_c$ decays into charmonia in the coariant light-front quark model.
 Here, the charmonia refer to the S-wave mesons, such as $J/\Psi, \eta_c$, the corresponding radially excited states, such as
$\psi(2S,3S), \eta_c(2S,3S)$, and the P-wave mesons, such as $\chi_{c0}, \chi_{c1}, h_c$ and $X(3872)$. Certainly, the form factors
of the $B_c\to D^{(*)}, D^{(*)}_s$ transitions are also considered for the purpose of the branching ratio calculation. We
find that the analytic expressions for $B_c\to S, A$ transition form factors can be obtained from those of $B_c\to P, V$ analytical expressions by some simple
replacements. The form factor $F^{B_c\eta_c}(V^{B_cJ/\Psi})$ has been calculated by many approaches,
most results of which lie in the range of $0.5\sim0.7$ $(0.5\sim1.0)$. We obtain a moderate value $F^{B_c\eta_c}=0.6 (V^{B_cJ/\Psi}=0.76)$. This can
be used to check which method is more favored by comparing to the future experimental data.  Compared with the form factors of $B_c$
transitions to these two ground-state S-wave charmonia, those of
 $B_c$ transitions to the radially excited S-wave charmonia, P-wave charmonia and charmed mesons are smaller. Except for each form factor
 at the zero recoiling point, we also calculate the corresponding one at the maximally recoiling point. Furthermore, we plot the $q^2$-dependence
 for each transition form factor.
Then we calculate
the branching ratios of 80 $B_c$ decays with a charmonium involved in each channel. We find that
the decays $B_c^+\to J/\Psi(\eta_c)\pi^+(\rho^+)$ and $B_c^+\to J/\Psi(\eta_c)D^+_s (D^{*+}_s)$ have larger branching ratios, which can
reach the order of $10^{-3}$, while most other decay channels have smaller branching ratios, which are suppressed by $1\sim3$ orders.
These predictions will be tested in the future by the LHCb experiments.
\section*{Acknowledgments}
This work is partly supported by the National Natural Science
Foundation of China under grant no. 11347030, the Program of
Science and Technology Innovation Talents in Universities of Henan
Province 14HASTIT037, and the Natural Science Foundation of Henan
Province under grant no. 232300420116. Z. Q. Zhang would like to thank Prof. Hai-Yang Cheng, Chun-Khiang Chua, and Hong-Wei Ke and Hsiang-nan Li for helpful discussions.
\appendix
\section{Some specific rules under the $p^-$ intergration}
When preforming the integraion, we need to include the zero-mode contribution. It amounts to performing the integration in a proper way in the CLFQM. Specificlly we
use the following rules given in Refs. \cite{jaus,hycheng}:
\be \hat{p}_{1 \mu}^{\prime} &\doteq &   P_{\mu}
A_{1}^{(1)}+q_{\mu} A_{2}^{(1)},\\
\hat{p}_{1 \mu}^{\prime}
\hat{p}_{1 \nu}^{\prime}  &\doteq & g_{\mu \nu} A_{1}^{(2)} +P_{\mu}
P_{\nu} A_{2}^{(2)}+\left(P_{\mu} q_{\nu}+q_{\mu} P_{\nu}\right)
A_{3}^{(2)}+q_{\mu} q_{\nu} A_{4}^{(2)},\\
Z_{2}&=&\hat{N}_{1}^{\prime}+m_{1}^{\prime 2}-m_{2}^{2}+\left(1-2
x_{1}\right) M^{\prime 2} +\left(q^{2}+q \cdot P\right)
\frac{p_{\perp}^{\prime} \cdot q_{\perp}}{q^{2}},\\
A_{1}^{(1)}&=&\frac{x_{1}}{2}, \quad A_{2}^{(1)}=
A_{1}^{(1)}-\frac{p_{\perp}^{\prime} \cdot q_{\perp}}{q^{2}},\quad A_{3}^{(2)}=A_{1}^{(1)} A_{2}^{(1)},\\
A_{4}^{(2)}&=&\left(A_{2}^{(1)}\right)^{2}-\frac{1}{q^{2}}A_{1}^{(2)},\quad A_{1}^{(2)}=-p_{\perp}^{\prime 2}-\frac{\left(p_{\perp}^{\prime}
\cdot q_{\perp}\right)^{2}}{q^{2}}, \quad A_{2}^{(2)}=\left(A_{1}^{(1)}\right)^{2}.  \en
{\centering
\end{document}